\documentclass[useAMS,usenatbib]{mn2e}
\pdfoutput=1
\usepackage{color}
\usepackage{colortbl}
\usepackage{dcolumn}
\usepackage{amsmath}
\usepackage{amssymb}
\usepackage{graphicx}
\usepackage{epsfig}
\usepackage{changebar}
\usepackage{natbib}
\usepackage{multirow}
\usepackage{subfigure}
\usepackage{txfonts}
\usepackage{relsize}
\usepackage{verbatim} 
\usepackage{natbib}
\usepackage{rotating}
\usepackage{lscape}
\usepackage{epstopdf}
\DeclareGraphicsExtensions{,.ps,.pdf,.jpg,.jpeg}
\usepackage{lastpage}
\newcommand {\NH}{NH$_3$}
\newcommand {\water}{H$_2$O}

\newcommand{\lsun}{L$_\odot$}
\newcommand{\msun}{M$_\odot$}

\newcommand{\kms}{km\,s$^{-1}$}
\newcommand{\cmthree}{cm$^{-3}$}
\newcommand{\cmtwo}{cm$^{-2}$}

\newcommand{\aap}{A\&A}
\newcommand{\araa}{ARAA}
\newcommand{\aaps}{aaps}
\newcommand{\mnras}{MNRAS}
\newcommand{\apj}{ApJ}
\newcommand{\apjs}{ApJS}
\newcommand{\apjl}{ApJL}
\newcommand{\pasp}{PASP}
\newcommand{\pasa}{PASA}
\newcommand{\aj}{AJ}
\newcommand{\nat}{Nature}

\newcommand{\COI}{$^{12}$CO}
\newcommand{\COII}{$^{13}$CO}
\newcommand{\COIII}{C$^{18}$O}
\newcommand{\SFR}{\msun\,yr$^{-1}$}

\title[The G305 star-forming complex]{The G305 star-forming complex: radio continuum and molecular line observations}
\author[L.Hindson]{L.Hindson$^{1,2,3}$\thanks{E-mail:
l.hindson@herts.ac.uk}, M.A. Thompson$^{2}$, J.S.Urquhart$^{3,7}$, A. Faimali$^{2}$, M. Johnston-Hollitt$^{1}$,  \newauthor
 J.S. Clark$^{4}$, B. Davies$^{5,6}$\\
$^{1}$\rm School of Chemical and Physical Science, Victoria University of Wellington, PO Box 600, Wellington 6140, New Zealand \\
$^{2}$\rm Centre for Astrophysics Research, Science and Technology Research Institute, University of Hertfordshire, College Lane, Hatfield, AL10 9AB, UK\\ 
\rm $^{3}$ATNF, CSIRO Astronomy and Space Science, P.O. Box 76, Epping, NSW 1710, Australia\\ 
$^{4}$\rm Department of Physics and Astronomy, The Open University, Walton Hall, Milton Keynes, MK7 6AA, UK\\
$^{5}$\rm Institute of Astronomy, University of Cambridge, Madingley Road, Cambridge, CB3 0HA, UK\\
$^{6}$\rm School of Physics \& Astronomy, University of Leeds, Woodhouse Lane, Leeds, LS2 9JT, UK\\
$^{7}$\rm Max-Planck-Institut f\"ur Radioastronomie, Auf dem H\"ugel 69, 53121 Bonn, Germany}

\begin{document}
\date{Accepted. Received ; in original form }

\pagerange{\pageref{firstpage}--\pageref{lastpage}} \pubyear{2010}

\maketitle

\label{firstpage}
\begin{abstract}
We present 109--115\,GHz (3\,mm) wide-field spectral line observations of \COI, \COII\ and \COIII\ \mbox{$J=1$--$0$} molecular emission and 5.5 and 8.8\,GHz (6 and 3\,cm) radio continuum emission towards the high-mass star forming complex known as G305. The morphology of G305 is dominated by a large evacuated cavity at the centre of the complex driven by clusters of O stars surrounded by molecular gas. Our goals are to determine the physical properties of the molecular environment and reveal the relationship between the molecular and ionised gas and star formation in G305. This is in an effort to characterise the star-forming environment and constrain the star formation history in an attempt to evaluate the impact of high-mass stars on the evolution of the G305 complex.

Analysis of CO emission in G305 reveals 156 molecular clumps with the following physical characteristics; excitation temperatures ranging from 7--25\,K, optical depths of 0.2--0.9, H$_2$ column densities of $0.1$--$4.0\times10^{22}$\,\cmtwo, clump masses ranging from $10^2$--$10^4$\,\msun\ and a total molecular mass of $>3.5\times10^5$\,\msun. The 5.5 and 8.8\,GHz radio continuum emission reveals an extended low surface brightness ionised environment within which we identify 15 large-scale features with a further eight smaller sources projected within these features. By comparing to mid infrared emission and archival data, we identify nine HII regions, seven compact HII regions, one \mbox{UCHII} region, four extended regions. The total integrated flux of the radio continuum emission at 5.5\,GHz is $\sim180$\,Jy corresponding to a Lyman continuum output of $2.4\times10^{50}$\,photons\,s$^{-1}$. We compare the ionised and molecular environment with optically identified high-mass stars and ongoing star formation, identified from the literature. Analysis of this dataset reveals a star formation rate of 0.008--0.016\,\SFR\ and efficiency of 7--12\%, allows us to probe the star formation history of the region and discuss the impact of high-mass stars on the evolution of G305. 

\end{abstract}


\section{Introduction}
High-mass stars (M\,$>8$\,\msun) are predominantly observed in young ($\oldle 20$\,Myr), massive  clusters (M\,$\oldge 10^3$\,\msun) located within giant molecular clouds (GMCs). Despite their rarity and short life times, high-mass stars are responsible for injecting a significant amount of energy into the interstellar medium (ISM) by means of stellar winds, ionising radiation and supernovae. The impact of such feedback mechanisms from high-mass stars on the ISM and star formation is still a topic of debate but is thought to play an important role in driving the evolution of GMCs. Feedback can be both constructive and destructive by sweeping up and clearing out parsec-scale cavities (or bubbles) in the ISM \citep{Deharveng2010} resulting in dense gas shells or destroying and dispersing the molecular environment. This means that feedback from high-mass stars should have the potential to both inhibit or enhance subsequent star formation and profoundly influence the star formation history (SFH) of GMCs \citep{Zinnecker2007}. 

The G305 GMC complex is located within the Scutum-Crux arm of the Milky Way at $l\approx305.4\degr$, $b\approx0.1\degr$ at a distance of $\sim 3.8$\,kpc and is one of the most massive and luminous star-forming regions in the Galaxy \citep{Murray2010, Clark2004}. The G305 complex consists of a number of molecular clouds that surround a cavity centered on the open clusters Danks\,1 and 2 (Fig.~\ref{im:ObsMopra}, triangles) and the Wolf Rayet star WR\,48a. The feedback from the high-mass stars in and around Danks\,1 and 2 appears to be responsible for driving the expansion of a large ($\sim 15\times 6$\,pc) cavity into the surrounding molecular gas. Previous observations of G305, summarised by \cite{Clark2004}, have revealed the presence of ongoing star formation surrounding the central cavity (see Fig.\,2 in \citealt{Clark2004}), which suggests that the overall morphology of the complex is consistent with numerous sites of ongoing high-mass star formation that may have been triggered by interaction between the winds and ionising radiation of Danks\,1 and 2 and the natal molecular clouds. These conditions make G305 an excellent laboratory in which to study high-mass star formation and the impact of feedback from high-mass stars on the evolution of GMCs.

In this work we present single dish observations of \COI, \COII\ and \COIII\ in the \mbox{$J=$1--0} rotational transition to explore the global distribution and properties of molecular gas in G305. CO is the most abundant molecule in the ISM, after H$_2$, and the low-energy \mbox{$J=1$--0} rotational transition is an excellent tracer of the cool ($\sim 10$\,K) low density gas ($\sim10^2$\,\cmthree) within molecular clouds \citep{Evans1999}. To explore the ionised environment we present wide-area interferometric observations of the large-scale 5.5 and 8.8\,GHz radio continuum which is generated by the thermal Bremsstrahlung mechanism in the ionised gas that surrounds high-mass stars. Previous observations of the radio continuum emission towards G305 have been limited to low resolution (4--5\arcmin) single dish observations \citep{Goss1970,Condon1993} that are unable to resolve the radio substructure on the scale of compact HII regions. Higher resolution (1-20\arcsec) interferometric radio continuum observations exist but suffer from low sensitivity and significant spatial filtering \citep{Retallack1979,Walsh1998,Hindson2012} and so are unable to identify extended features. The data set presented in the following section allows the investigation of the molecular environment in which high-mass star formation is ongoing and allows us to trace the distribution and investigate the impact of ionising radiation from high-mass stars. By searching the literature for tracers of ongoing star formation, we are able to investigate the relationship between the ionised and molecular environment with star formation and thus broadly trace the SFH of the region. 

This paper is organised as follows; in Section~\ref{Sect:obs} we present the observations, data reduction and ancillary data utilised in this study, Section~\ref{Sect:results} presents the results and analysis of the molecular and ionised observations, in Section~\ref{Sect:discussion} we present our discussion of the results including the relationship between the ionised and molecular gas and star formation, the role of triggering and SFH of the region. We end this paper with a summary in Section~\ref{Sect:summary}.

\section{Observations \& data reduction procedures}\label{Sect:obs}
\subsection{Mopra mm spectral line Observations}
CO observations of the G305 complex were obtained during two observing runs in September of 2010 and 2011 using the 3\,mm receiver on the Mopra 22-m millimetre-wave telescope operated by the Australia Telescope National Facility (ATNF). 

The back end of Mopra is the University of New South Wales Mopra spectrometer (MOPS) which consists of four 2.2\,GHz bands that overlap to provide 8\,GHz of continuous bandwidth. The narrow or ``zoom'' mode of MOPS was used which allows 16 intermediate frequencies (IFs) to be observed simultaneously. Each of the 16 zoom ``windows'' has a bandwidth of 137.5\,MHz over 4096 channels giving a velocity resolution at 110\,GHz of $\sim 0.1$\,\kms\ per channel. The velocities presented in this work are in the kinematic local standard of rest (LSR) frame centred on $V_{\rm LSR} \approx -37$\,\kms, the systematic velocity of G305 \citep{Hindson2010}. We centred the 8\,GHz band at 112\,GHz and placed the zoom windows of MOPS to cover three CO isotopologues; \COI, \COII\ and \COIII\ in the \mbox{$J=1$--$0$} rotational transition (Table~\ref{Tab:MopsLines2}). The remaining zoom windows were set to the rest frequency of lines of interest such as HC$_3$N (\mbox{12--11}), CH$_3$CN (\mbox{6--5}) and C$^{17}$O (\mbox{1--0}) but no emission $>3\sigma$, where $\sigma$ is the RMS background noise, was detected.

\begin{figure*}
\begin{center}
\includegraphics[trim=10 0 10 30, width=0.9\textwidth]{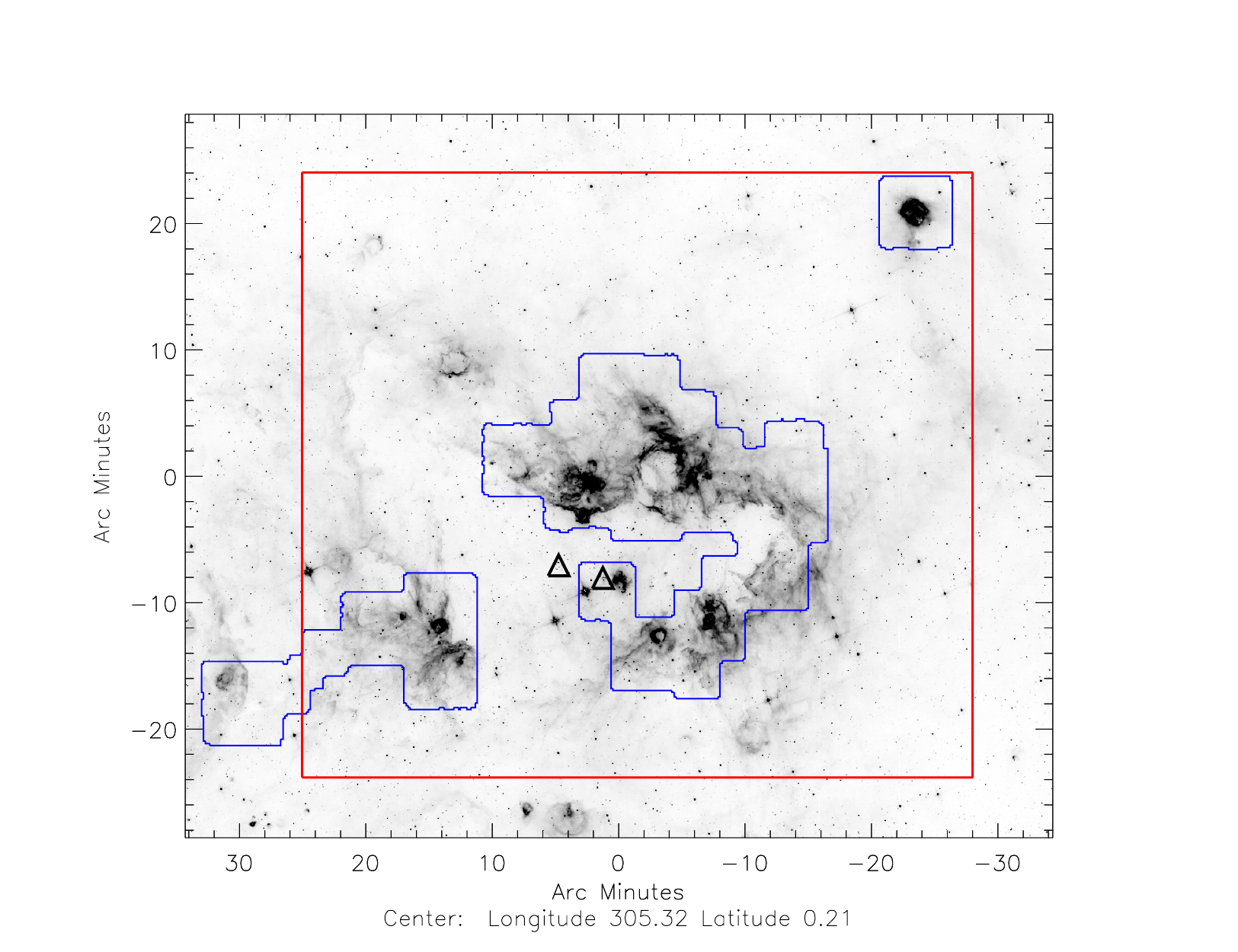} 
\caption[Mopra molecular observations scan area]{The area observed in our Mopra 109--115\,GHz spectral line (blue outline) and ATCA 5.5 and 8.8\,GHz radio continuum (red outline) programmes overlaid onto a 5.8\,$\umu$m GLIMPSE grey scale image. Black triangles indicate the positions of Danks\,1 and Danks\,2 at $l=305.34$ $b=+0.07$ and $l=305.40$ $b=+0.09$ respectively. }
\label{im:ObsMopra}
\end{center}
\end{figure*}

\begin{table}
\caption[CO observation characteristics]{Molecular transitions detected in the Mopra 109-115\,GHz observations and the corresponding central frequency, velocity resolution and size of the Mopra beam.}
\begin{center}
  \begin{tabular}{lccc}
   \hline Line & Frequency & Resolution & FWHW\\
       ($J=1$--0)    & (GHz) & (ms$^{-1}$) & (\arcsec) \\
 	\hline
\COIII	&	109.738	&	92.0	&	$36\pm 3$	\\
\COII	&	110.221	&	91.7	&	$36\pm 3$	\\
\COI	&	115.258	&	87.6	&	$33\pm 2$	\\
\hline
\end{tabular}
\end{center}
\label{Tab:MopsLines2}
\end{table}

These observations took advantage of Mopra's on-the-fly (OTF) observing mode. In this mode the telescope beam is scanned along a line on the sky at a constant rate whilst recording spectra every few seconds. The mapping coverage was based upon previous untargeted observations of \NH\ \citep{Hindson2010} and \COI\ (\mbox{$J=$1--0}) taken during a test phase of an increased OTF mapping speed. Using these observations as a pathfinder a number of sub-maps, ranging from 3\arcmin\ to 7\arcmin\ squares, were centred on the \NH\ and \COI\ emission peaks (Fig.~\ref{im:ObsMopra}). Each of these sub-maps overlaps its neighbour by 56\arcsec\ and each scan row is separated by 9\arcsec\ to ensure Nyquist sampling at the highest frequency. Using the fastest scan rate available of two seconds per point and including calibration and other overheads resulted in two orthogonal maps taking between two and four hours. Changeable weather conditions resulted in atmospheric water vapour between 15 and 25\,mm and system temperatures ($T_{\rm sys}$) between 250--350\,K for \COII\ and \COIII\ and 500--600\,K for \COI. Antenna pointing checks performed every two hours showed that the average pointing accuracy was better than 10\arcsec\ rms. The observed antenna temperatures ($T^*_{\rm A}$) were corrected for atmospheric absorption, ohmic losses and rearward spill over, by measuring an ambient load when performing $T_{\rm sys}$ measurements every 10--15 minutes. Taking measurements of an emission free off source reference position before each scan row allowed us to remove the contribution of the atmosphere by comparing to the hot load.

\subsection{Mopra data reduction}
The reduction of Mopra data is performed with two packages called {\sc LIVEDATA} and {\sc GRIDZILLA} provided by the ATNF, both are AIPS++ packages written by Mark Calabretta\footnote{http://www.atnf.csiro.au/computing/software/livedata.html} for the Parkes radio telescope and adapted for Mopra. Raw data files are first processed in {\sc LIVEDATA}, which applies user specified flagging to bad channels followed by a bandpass calibration for each scan row using the off-source reference position. This is followed by fitting a first order polynomial to the spectral baseline using line-free channels and applying a position and time stamp to each spectrum. The data is then passed to {\sc GRIDZILLA}, which calculates the pixel value from the spectral values of multiple files using the weighted mean estimation. {\sc GRIDZILLA} then resamples the data onto a common pixel scale ($l$, $b$ and $v$) and performs interpolation in velocity space to convert measured topocentric frequency channels into LSR velocity. We convert the instrument dependent antenna temperature ($T^*_{\rm A}$) to the main beam brightness temperature ($T_{\rm b}$) using $T_{\rm b}=T^*_{\rm A}/\eta_{\rm b}$, where $\eta_{\rm b}$ is the frequency-dependent beam efficiency. The Mopra beam efficiency between 86--115\,GHz was characterised by \cite{Ladd2005} and for CO (\mbox{$J=$1--0}) observations the main beam efficiency is $\eta_{\rm b} =  0.42$. The final \COI, \COII, and \COIII\ data cubes have a brightness temperature sensitivity of approximately $0.64$, $0.45$ and $0.35$\,K per 0.1\,\kms\ channel respectively. The RMS noise across these maps varies by $\sim 10\%$ due to the changing atmospheric conditions, elevations and receiver performance during observations.

\subsection{ATCA centimetre continuum observations}

Observations of the radio continuum towards G305 were made using the Australia Telescope Compact Array (ATCA), which is located at the Paul Wild Observatory, Narrabri, New south Wales, Australia\footnote{The Australia Telescope Compact Array is part of the Australia Telescope National Facility, which is funded by the Commonwealth of 
Australia for operation as a National Facility managed by CSIRO.}. We made use of six array configurations; east-west (\mbox{E-W}) 6A, 1.5A and 750B long baseline arrays and H214, H168 and H75 short baseline hybrid arrays. This combination of array configurations provides spatial sensitivity to both small scales ($\sim 1$\arcsec) presented in \cite{Hindson2012} and large scales ($\sim10$\arcmin) that we present here. In this study we are interested in the large-scale radio continuum emission associated with HII regions and extended low surface brightness ionised gas and so we exclude the longest baseline provided by the 6$^{\rm th}$ antenna in the data reduction described in the following section. The dates of observations and synthesised beams of each array can be found in Table~\ref{Observing_table}.

\begin{table}
\caption{Observational parameters for ATCA radio continuum observations excluding the 6$^{\rm th}$ antenna.}
\begin{center}
\begin{tabular}{ccccc}
   \hline Array & Date & Synthesised & Baseline  \\
   		Config  &	dd/mm/yy   &	Beam (\arcsec) & Range (m) \\
   			
   	\hline
  	
   	   		6A		&	06/06/09	&	$3.2\times3.6$	 	& 628  -- 2923	\\
   			1.5A	&	28/07/09	&	$4.8\times5.4 $ 	& 153 -- 1469	\\
   			750B	&	26/02/10	&	$9.3\times10.5 $	& 168 -- 765	\\
   			H75 	&	10/07/09	&	$87.0\times87.0 $	& 31 -- 82	\\
   			H168 	&	10/05/09	&	$40.8\times40.8 $& 61 -- 192	\\
   			H214 	&	23/05/09	&	$33.6\times33.6 $& 92 -- 247	\\

 \hline
   \end{tabular}
 \label{Observing_table}
 \end{center}
 \end{table}

The observations were made simultaneously at two IF bands centred at 5.5 and 8.8\,GHz and utilised the wide-band continuum mode of the Compact Array Broadband Backend \citep{Wilson2011}. This resulted in a bandwidth of 2\,GHz with 2048$\times$1\,MHz channels in each IF. At the longest baseline of 3\,km we obtain a maximum resolution of $3.2\times 3.6 \arcsec$ and by using a joint deconvolution approach \citep{Cornwell1988, Ekers1979} the shortest baseline of 9\,m provides sensitivity to emission up to $\sim\, 10$ and $\sim\, 5\arcmin$ at 5.5 and 8.8\,GHz respectively. 

\mbox{E-W} array observations were made over a twelve-hour period and hybrid array configurations over a six-hour period in a number of short snap-shots to provide good hour angle coverage. To correct for fluctuations in the phase and amplitude caused by atmospheric and instrumental effects the total scan time is split into blocks of 15 minutes of on source integration sandwiched between 2-minute observations of the phase calibrator PMN J1352--63. For absolute calibration of the flux density, the standard flux calibrator PKS B1934--638 was observed once during the observation for approximately 10 minutes. To calibrate the bandpass the bright point source PKS B1253--055 was also observed during the observations. In order to map the entire G305 region, at Nyquist sampling, the 8.8\,GHz band required a mosaic of 357 individual points in a hexagonal pattern (Fig.~\ref{im:ObsMopra}) this resulted in over sampling at 5.5\,GHz. With a scan rate of 2s per point (spaced over a range of hour angles in order to improve $u$-$v$ coverage) we could thus complete a map in approximately one hour and observe each pointing centre $\sim\,7$--11 times for each hybrid and \mbox{E-W} configuration respectively to give a total on source integration time on each point of $\sim\,1$ minute.

\subsection{ATCA data reduction}\label{Sect:ATCADataReduction}
The steps taken to image the large-scale radio continuum emission follows the standard joint deconvolution approach, with the inclusion of single-dish data, described in the {\sc Miriad} user manual \citep{Sault1995}. In this case all fields are handled simultaneously by the imaging and deconvolution software, which results in a simplified and faster imaging process that is particularly beneficial for low signal-to-noise ratio mosaics and extended emission \citep{Cornwell1988}.

The calibrated $u$-$v$ data set was first Fourier transformed into the image plane using the task {\sc Invert} with super-uniform weighting applied to reduce side-lobes caused by gaps in the $u$-$v$ coverage and the over density of $u$-$v$ points near the centre of the $u$-$v$ plane. A $5\times5\arcsec$ Gaussian tapering function was applied down-weight the data at the outer edge of the $u$-$v$ plane and suppress small-scale side-lobes. To take advantage of the available 2\,GHz bandwidth multi-frequency synthesis was applied to improve the $u$-$v$ coverage and account for possible spectral variation of the flux within the band. Total power (zero spacing) data at 5.5\,GHz is provided by the Parkes-MIT-NRAO survey (PMN) \citep{Condon1993} at 4.85\,GHz with a resolution of 4.92\arcmin. Unfortunately, no single dish data exists at 8.8\,GHz and total flux estimates had to be scaled assuming an optically thin spectral energy distribution (SED) with spectral index of $\alpha=-0.1$. This assumption is valid for thermal Bremsstrahlung emission associated with HII regions, where the SED turns over from optically thick to thin above 5\,GHz \citep{Kurtz2005}.  

The final 5.5 and 8.8\,GHz maps have a resolution of $10.2\times 8.2$\arcsec\ and $8.7\times 7.6$\arcsec\ and sensitivity of approximately 0.30 and 0.35\,mJy\,beam$^{-1}$ respectively. This is higher by a factor of 10 than the expected theoretical thermal noise of 0.03\,mJy\,beam$^{-1}$. The brightest source in the 5.5\,GHz map is $\sim 1500$\,mJy\,beam$^{-1}$ giving a dynamic range of $\sim 5000$. An integrated flux of 180 and 170\,Jy is estimated for the 5.5 and 8.8\,GHz maps  respectively in good agreement with previous estimates (193\,Jy at 5\,GHz; \citealt{Clark2004}) suggesting that the flux distribution has been reliably recovered.

\subsection{Ancillary Data}\label{Sect:ancillary}
To complement the observations presented above we utilise mid-infrared, radio, \water\ and methanol maser data to broaden our view of the G305 complex and aide in the description of the morphology and identify high-mass stars and ongoing star formation. 

We define high-mass stars in G305 as stars $\gtrsim8$\,\msun\ and $\gtrsim10^3$\,\lsun\ that have reached the main sequence and had time to clear their immediate surroundings of dust and gas and so be visible at optical and near-IR wavelengths. The two largest concentrations of such high-mass stars are found in the two open clusters Danks\,1 and 2 that formed in the center of the G305 complex within the last 6\,Myr. These two clusters contain a confirmed population of 20 spectroscopically identified high-mass stars (see Table\,2 of \citealt{Davies2012}), however there exists at least two further populations of high-mass stars within G305. The first of these is found in the three high-mass stars associated with the centre of the HII region designated G305.254+0.204 (\citealt{Leistra2005}; L05-A1, A2 and A3) in the (Galactic) north-west of the complex. The second population is a diffuse group of at least eight WR stars located towards the central cavity of G305 with projected distances from Danks\,1 and 2 of $\sim2$--$25$\,pc \citep{Mauerhan2011,Shara2009}. The origin of this diffuse population is unclear, they may have formed in-situ, however given their close proximity to Danks\,1 and 2, apparent isolation from other stellar sources and the relatively young age of the clusters it is plausible that they are runaway stars \citep{Mauerhan2011,Baume2009, Lundstrom1984}.Finally, early OB pre-MS stars are also suspected to be located in the ionised bubble designated PMN 1308-6215 (G304.93+0.56) to the far north-west of the region \citep{Clark2004}. Low resolution radio continuum and H$\alpha$ observations \citep{Clark2004} have detected 12 HII regions. Seven of these HII regions are located around the periphery of the central cavity (see Fig. 2 and Table 1 in \citealt{Clark2004}). These HII regions indicate that there is a younger generation of high-mass stars that, unlike Danks\,1 and 2, have not had sufficient time to completely clear their surroundings. Finally, in \cite{Hindson2012} we identified six UC\,HII regions, which directly trace young ($<10^5$\,yr; \citealt{Comeron1996, Davies2011}) high-mass stars. 

Ongoing star formation is deeply embedded within dust and gas and can be identified by observations of the mid-infrared emission associated with YSOs and star formation tracers such as maser emission. The Red MSX Source (RMS) survey\footnote{http://www.ast.leeds.ac.uk/RMS/} \citep{Urquhart2008} aims to locate massive young stellar objects (MYSOs) through colour selection in the Midcourse Space Experiment (MSX) point source catalogue and extensive follow-up observations \citep{Urquhart2007A,Urquhart2009,Urquhart2011B}. By searching the RMS database we are able to identify eight high-mass YSOs within G305 with bolometric luminosities ranging from $1.3\times10^3$ to $4.9\times10^4$\,\lsun. A number of masers are known to form within the environment surrounding ongoing star formation \citep{Elitzur1992}, we employ two such masers; 22\,GHz \water\ masers and 6.7\,GHz Class II methanol masers. Class II methanol masers are known to exclusively trace high-mass star-forming regions \citep{Urquhart2013} whilst \water\ masers form in both low and high-mass star forming regions. Both maser species only exist within star forming environments on timescales of a few 10$^5$\,yr \citep{Elitzur1992}. We utilise the Methanol Multi Beam (MMB) \citep{Green2009,Green2012,Caswell2009} to identify 14, 6.7\,GHz methanol masers and 15 \water\ masers identified in \cite{Hindson2010}. 

\section{Results \& Analysis}\label{Sect:results}

\subsection{Molecular gas towards G305}\label{Sect:COResults}

\begin{figure}
\begin{center}
\includegraphics[trim=0 10 0 10,width=0.57\textwidth]{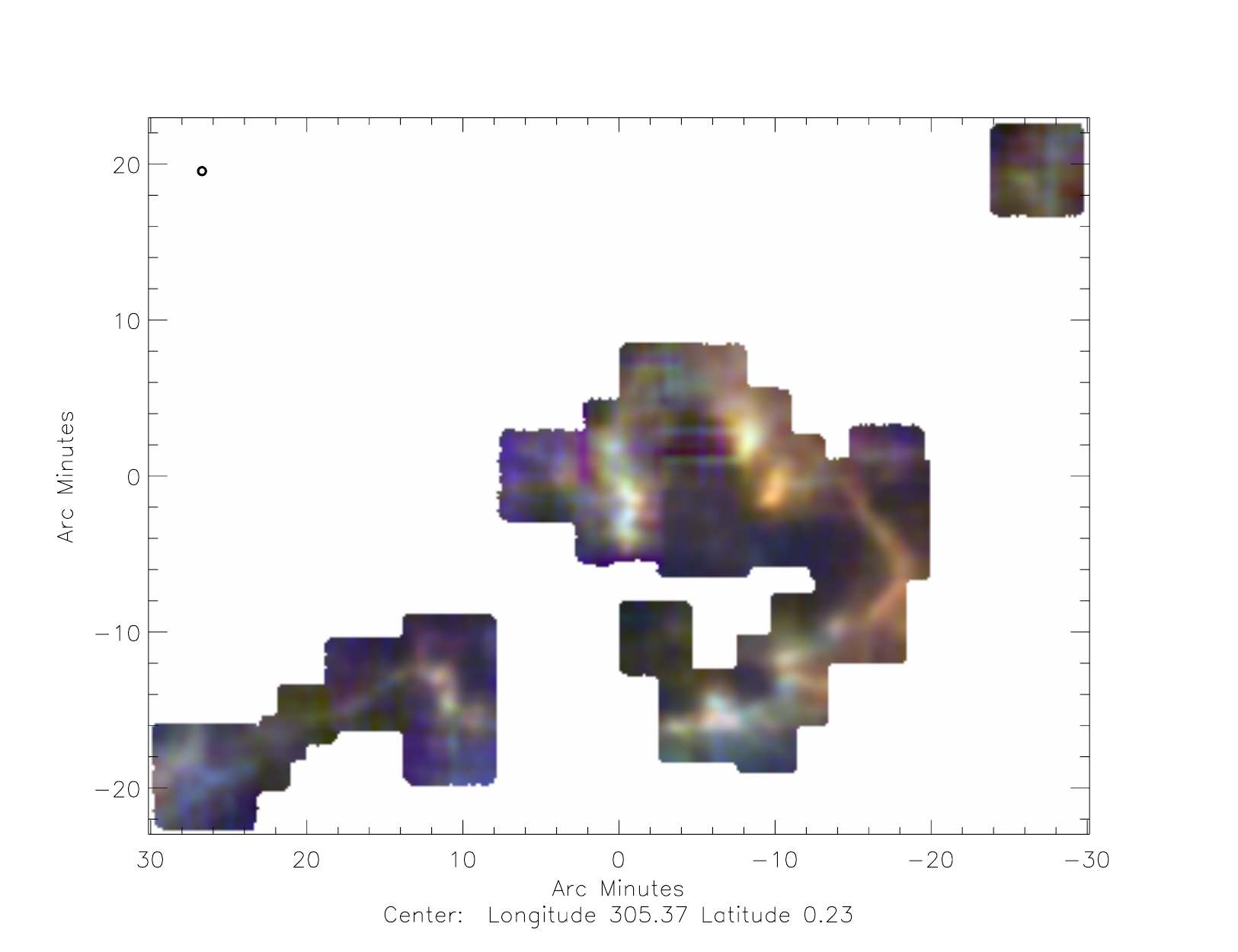}
\includegraphics[trim=0 10 0 40,width=0.57\textwidth]{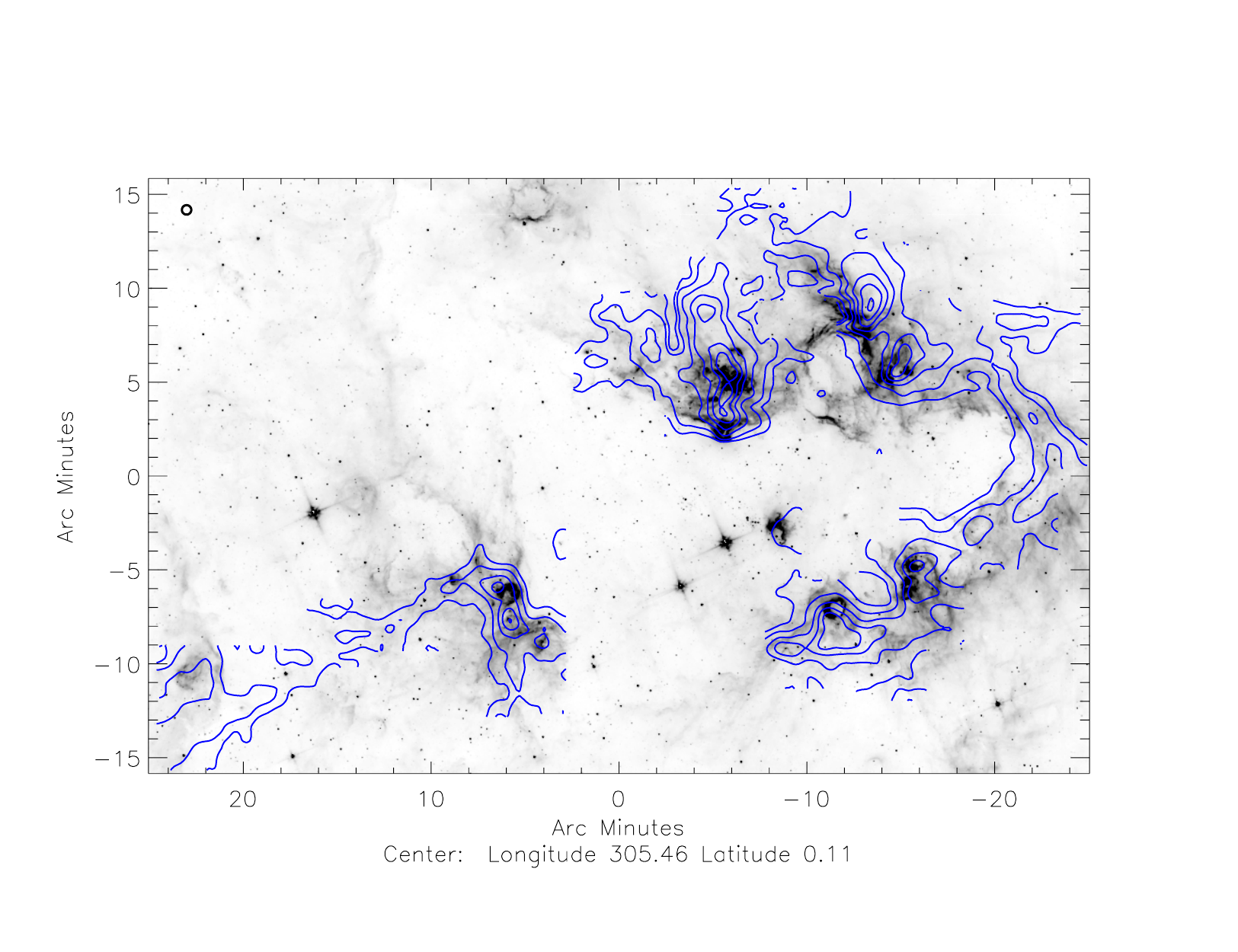}
\end{center}
\caption{Top: Three colour composite image of the \COI, \COII\ and \COIII\ $J=1$--0 emission towards G305 in red, green and blue respectively. Each image has been integrated between a velocity range of $-50.6$ and $-24.9$\,\kms. The Mopra beam is shown in the top-left by a 36\arcsec\ black circle. Bottom: \COII\ contours in blue are presented over a Glimpse 5.8\,$\umu$m grey scale background. Contours begin at 7 K\,\kms\ and then increment by 5\,K\,\kms\ from 10 to 35\,K\,\kms. Note that the bottom figure covers a smaller field-of-view, excluding the isolated emission to the north-east.}
\label{im:CO}
\end{figure}

The observations of CO towards G305 reveal a complex and clumpy molecular environment with a range of spatial scales of both interconnected and isolated molecular gas typical of GMCs (Fig.~\ref{im:CO}). To describe these various scales of emission we define the emission in terms of clouds (3--20\,pc), clumps (0.5--3\,pc) and cores ($\oldleq 0.5$\,pc). Molecular gas is concentrated in four large clouds, corresponding to large molecular clouds detected in \NH\ \citep{Hindson2010}, to the (Galactic) north, south and east of the central cavity. In the top panel of Fig.~\ref{im:CO} we present a three colour composite image of \COI, \COII\ and \COIII\ in red, green and blue respectively. The \COI\ emission has the lowest excitation level and so traces the lowest density gas whilst \COII\ and \COIII\ reveal denser gas. Figure~\ref{im:CO} reveals a number of dense clouds embedded within low-density gas, which connects the northern and southern lobes in the west of the complex.

\begin{figure}
\begin{center}
\includegraphics[width=0.5\textwidth]{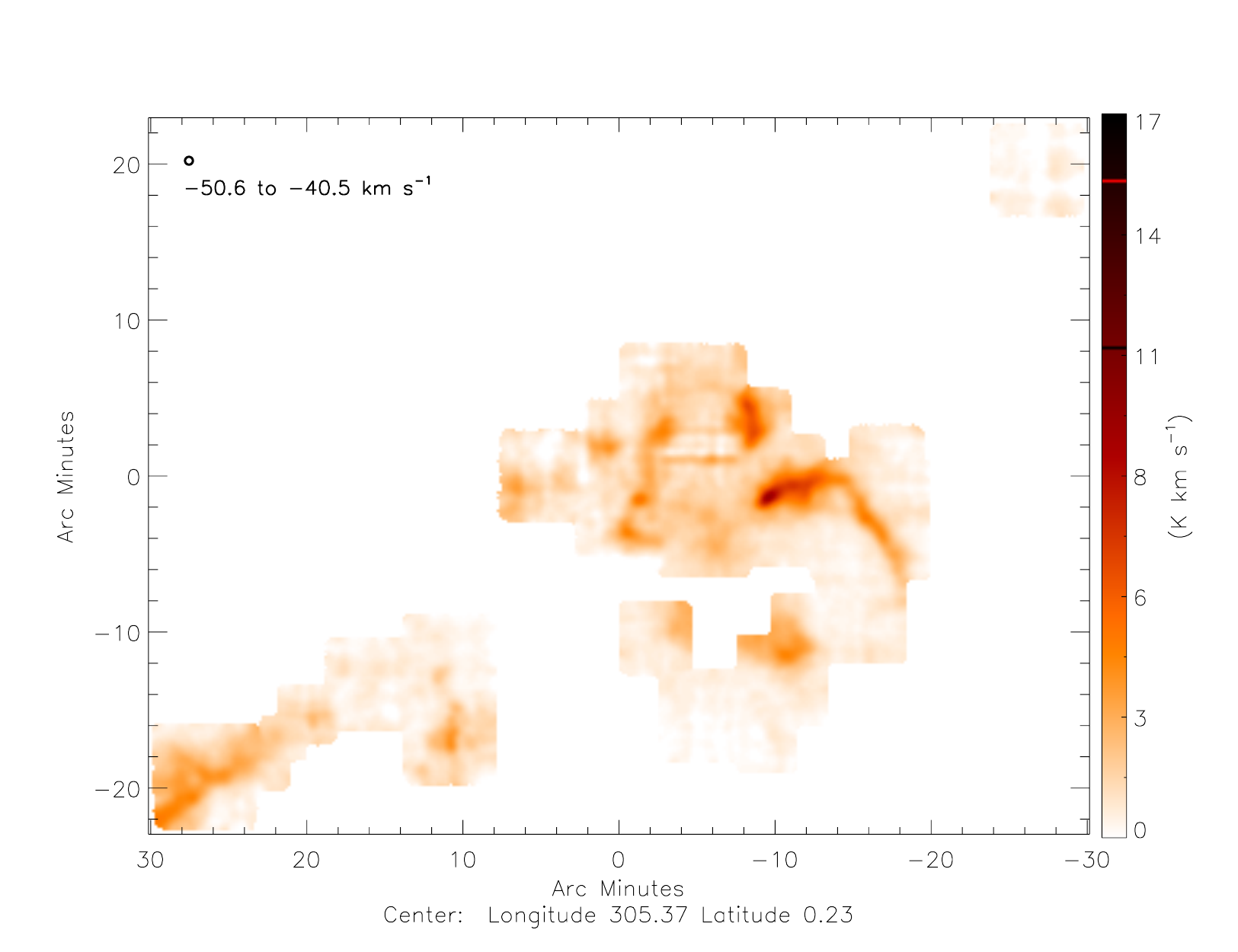}
\includegraphics[width=0.5\textwidth]{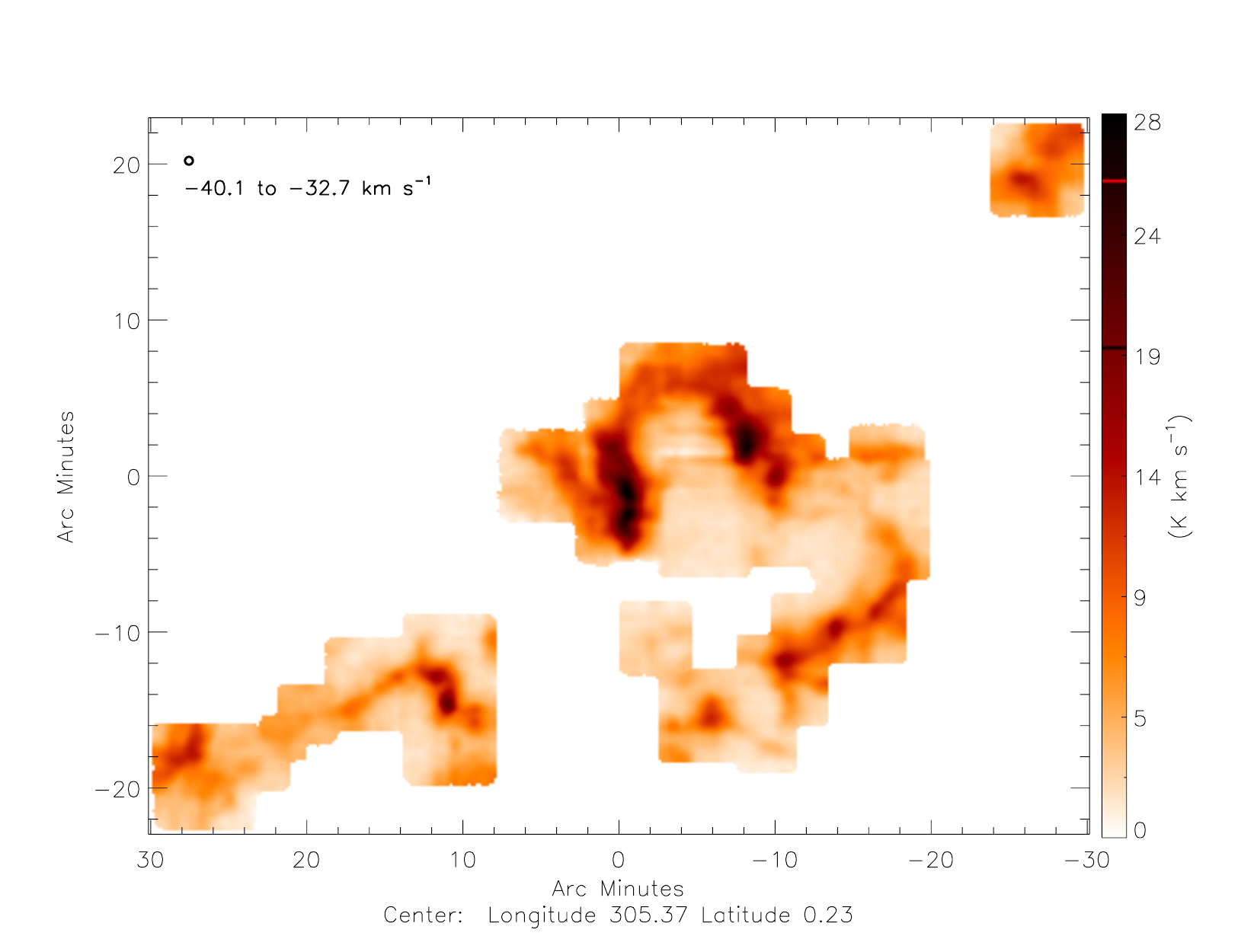}
\includegraphics[width=0.5\textwidth]{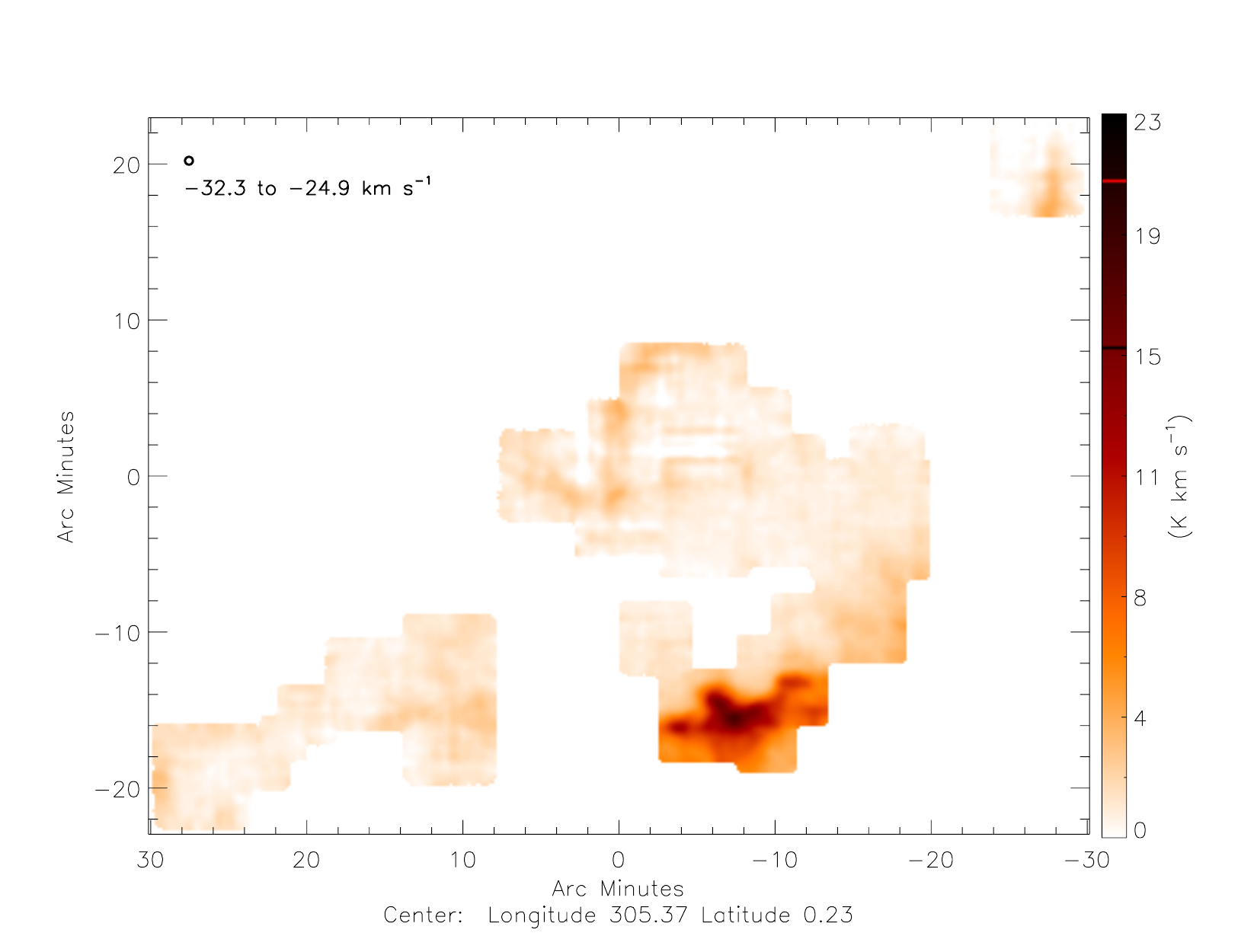}
\end{center}
\caption{\COII\ (\mbox{$J=$1--0}) integrated images of the G305 complex between $-50.6$ and $-24.9$\,\kms. Top: integrated between $-50.6$ and $-40.5$\,\kms. Middle: integrated between $-40.1$ and $-32.7$\,\kms. Bottom: integrated between $-32.3$ and $-24.9$\,\kms. }
\label{im:COVel}
\end{figure} 

We utilise the GLIMPSE 5.8\,$\umu$m mid-infrared band to highlight the photo-dominated region (PDR) which reveals the thin boundary between the ionised and neutral gas in G305 where UV photons are able to penetrate the surface of the molecular gas and excite polycyclic aromatic hydrocarbons (PAHs) into emission \citep{Rathborne2002}. Comparison between CO and 5.8\,$\umu$m emission allows us to speculate as to the relationship between the molecular gas and the PDR. Figure~\ref{im:CO} reveals that the  molecular material is coincident with the PDR located at the interface between the molecular and ionised gas surrounding the central cavity of G305 (Fig.~\ref{im:CO} bottom panel). The central cavity is bound to the north by two large molecular clouds, to the south by a lobe of dense molecular gas and to the west by a tenuous strip of gas. The cores of the molecular emission within these clouds lie behind the leading edge of the PDR indicating that the interior of the molecular clouds are being effectively shielded from ionising radiation. A gradient in the molecular emission brightness temperature is clearly visible in a number of regions sloping away from the PDR and ionising sources. This suggests that the molecular material close to the cavity has been swept up and compressed into a dense layer. Conversely there are a number of regions in which the molecular material appears to be in the process of being dispersed or destroyed such as between the two molecular clouds in the north, several sites along the southern lobe and most noticeably toward the east. We discuss the morphology of the molecular gas in more detail and present more detailed figures in Section~\ref{Sect:Overview}.

The mean central velocity of the CO emission presented here and \NH\ emission presented in \cite{Hindson2010} are in good agreement at $-36.6$ and $-37.4$\,\kms\ respectively confirming they trace the same material. However, the higher spatial and velocity resolution of the CO observations resolve multiple clumps within the bounds of the \NH\ clouds as expected. We argue that the G305 complex is most likely flattened in the plane of the sky rather than spherically symmetric. First, there is very little molecular emission projected towards the centre of the complex at any velocity, which excludes the spherical shell hypothesis. Second from the channel map (Fig.~\ref{im:COVel}) we can see that the velocity distribution of molecular gas is not spherical. The clouds of molecular gas to the north and south of the central cavity are at approximately $-41$ and $-30$\,\kms, respectively. The northern clouds of gas are approaching $>10$\,\kms\ faster than the southern clouds (Fig.~\ref{im:COVel}). This velocity gradient across the complex could be driven by expansion or shearing motion and indicates that G305 is not being viewed straight on but projected at an angle. To explore the morphology further we consider the PDR. If we assume the complex is a flattened sphere viewed face on with no inclination we would expect to see a clear boundary between the PAH and molecular emission. If the orientation of G305 is at some inclination angle we would expect the PAH and molecular emission to be superimposed upon each other. We find both instances exist within G305; the north-east region has PAH emission projected onto the face of the molecular emission whilst the north-west has a clear boundary and the south-west exhibits a mixture of the two. We conclude that the molecular gas in G305 is distributed in a complex three-dimensional structure with non-uniform orientation.

The bulk of molecular emission is found around the central cavity but we also detect a number of isolated clumps of molecular emission located away from the cavity boundary. This includes the isolated mid-infrared core in the far north-west (G304.93+0.55), which is associated with CO and faint ($T_{\rm b}\approx 0.15$\,K) \NH\ (1,1) emission. Towards the centre of the cavity faint \COI\ and \COII\ emission (clumps 82 and 119; $T_{\rm b}=4.6$ and 3.3\,K respectively) is associated with an isolated, bright and compact ``blob'' of 5.8\,$\umu$m emission. There is no \NH\ emission detected towards this region but a \water\ maser indicates there is ongoing star formation. The projection of this blob is unknown, either the physical proximity on the sky is the same as the projected distance from Danks\,2 ($<1$\,pc) or it is located on the near or far side of the cavity. We discuss this further in Section~\ref{Sect:Overview}

We detect a $\sim0.3\degr\times 0.04\degr$ ($\sim 20\times2.5$\,pc) filamentary structure in \COI\ and \COII\ to the far east of G305 extending away from the cavity boundary. With a central velocity of $-37.5$\,\kms\ and a gradient of $\sim 0.05$\,\kms\,pc$^{-1}$ leading away from the cavity boundary (Fig.~\ref{im:CO}: bottom) we suggest that the filament is a single coherent structure that is associated with G305 and not the chance alignment of a number of separate clumps along the line of sight. Similar large cohesive filaments have been discovered in the Galaxy \citep{Hill2011,Arzoumanian2011} such as the ``Nessie'' nebula \citep{Jackson2010} an infrared dark cloud $\sim80$\,pc in length and 0.5\,pc in width. Exactly how such large features form is a topic of debate but they seem to be associated with a large number of star-forming regions and so may be play a significant role in star formation.

\subsection{Molecular clump identification and physical properties}\label{Sect:MolProp}

To assign CO emission into discrete clumps we apply the automated clump finding algorithm {\sc CLUMPFIND} \citep{Williams1994} using the {\sc CUPID} package (part of the {\sc STARLINK} project software\footnote{http://starlink.jach.hawaii.edu/starlink}). There are a large number of clump finding algorithms available (e.g.\ {\sc GAUSSCLUMPS, FELLWALKER, DUCHAMP, dendogram analysis}), we chose {\sc CLUMPFIND} over these because it makes no assumption as to the clump profile and is widely used \citep{Rathborne2009, Buckle2010}.

{\sc CLUMPFIND} works in a ``top-down'' approach by finding peaks within the data and then contouring down to a user defined base level detecting new peaks along the way and assigning pixels associated with more than one peak to a specific clump using the ``friends-of-friends'' algorithm \citep{Williams1994}. The clump catalogue of CO emission presented in this paper is generated from the \COII\ \mbox{$J=1$--0} data cubes. The \COII\ cube is used because it is generally optically thin unlike \COI\ and has a higher signal-to-noise ratio (SNR) than \COIII. To account for the background noise and variations due to changing opacity, elevation level and system performance, {\sc CLUMPFIND} was applied to a \COII\ cube with the background subtracted using the {\sc FINDBACK} algorithm. The output of the clump finding process is a mask that can be applied to the CO data cubes to extract the physical properties over clumps in $l$, $b$ and $v$. Recent simulations suggest that extraction of clump properties in the position-position-velocity (PPV) dimension using CLUMPFIND accurately estimates the clump properties in the position-position-position (PPP) dimension \citep{Ward2012}. 

Two parameters are responsible for the majority of control in {\sc CLUMPFIND}; the minimum emission level to begin clump finding ($T_{\rm low}$) and the step width that defines the spacing of the contours ($\Delta T$). The recommended parameters for {\sc CLUMPFIND} are $\Delta T=2\sigma$ and $T_{\rm low} = 2\sigma$ \citep{Williams1994}. We found that applying these parameters to the native 0.1\,\kms\ velocity resolution \COII\ data cube tends to artificially fragment clumps into a main clump surrounded by unphysical low line-width fragments occupying only a few pixels each. This may be due to the low dipole moment of the CO \mbox{$J=$1--0} transition resulting in it being easily excited into emission in low-density diffuse regions which leads to confusion in the lower S/N outskirts of the clumps (e.g. \citealt{Rathborne2009, Pineda2009}). Inspecting the integrated velocity-weighted dispersion of the \COII\ data cube shows that the bulk of the emission has a line-width in excess of 0.5\,\kms\ and so we chose to bin our data cubes to this velocity resolution. At a resolution of 0.5\,\kms\ the clump fragmenting effect is largely eliminated and the brightness temperature noise in the CO maps is reduced by a factor of $\sqrt{5}$ to $\sim 0.2$\,K per 0.5\,\kms\ channel. The minimum emission level chosen to begin clump finding was $T_{\rm low}=3\sigma$ and the step width that defines the spacing of the contours $\Delta T = 3\sigma$. As we wish to only recover clumps and avoid regions of low surface brightness emission, we set a requirement of a $10\sigma$ detection at the peak, and clumps below this threshold are omitted from further analysis. Additionally, clumps were rejected if they contained any voxels ($l$, $b$, $v$ pixels) that touched the edge of the data cube, consisted of fewer than 16 voxels or were smaller than the beam size and velocity resolution (36\arcsec\ and 0.5\,\kms). The clumps were then checked by eye and using manual aperture photometry to ensure that only meaningful structures were included in the final catalogue. We note that these criteria impose a completion limit for the catalogue for clumps with diameters less than 0.6\,pc (This is the maximum diameter the smallest clump can have given the 16 voxel limit) and peak brightness temperature less than $\sim1.9$\,K. We identify 156 molecular clumps towards G305 the properties reported by {\sc CLUMPFIND} can be found in Table~\ref{Tab:COCFResults10}. In the following subsections, we present the method used to derive the physical properties of the identified molecular clumps.

\begin{table*}
 \caption[\COII\ emission properties]{The first 10 clumps and observed properties of the 156 detected $^{13}$CO clumps reported by {\sc CLUMPFIND}. Note that the radius is the effective radius. This table can be found in its entirety in the appendix.}
\centering
  \begin{tabular}{cccccccccccc}
   \hline 
	Clump & Clump & \multicolumn{3}{c}{Peak} & \multicolumn{4}{c}{Dimension}  &  Peak \\\cline{3-5} \cline{6-9}
	Number  & Name  &        \multicolumn{2}{c}{Galactic}  &$V_{\rm LSR}$         & $\Delta l$ & $\Delta b$ &$R_{\rm eff}$& $\Delta V$ &          $T_{\rm b}$       \\
	       &        &        $(l)$ & $(b)$  & (\kms)& (\arcsec) & (\arcsec) & (pc)& (\kms)  & (K)\\ 
\hline
1	&	G305.24+0.26	&	305.24	&	0.26	&	-38.7	&	123.4	&	157.46	&	2.84	&	5.66	&	15.38	\\
2	&	G304.94+0.55	&	304.94	&	0.55	&	-35	&	97.55	&	112.86	&	1.98	&	1.94	&	14.29	\\
3	&	G305.56+0.02	&	305.56	&	0.02	&	-39.6	&	112.39	&	83.02	&	1.97	&	3.04	&	13.55	\\
4	&	G305.26-0.03	&	305.26	&	-0.03	&	-32.7	&	136	&	184.61	&	3	&	4.27	&	13.17	\\
5	&	G304.93+0.54	&	304.93	&	0.54	&	-34.6	&	103.69	&	121.68	&	2.11	&	2.19	&	12.69	\\
6	&	G305.37+0.21	&	305.37	&	0.21	&	-35	&	103.83	&	86.84	&	1.88	&	4.14	&	12.14	\\
7	&	G305.36+0.17	&	305.36	&	0.17	&	-39.1	&	126.77	&	81.12	&	2.23	&	4.73	&	11.52	\\
8	&	G305.19+0.01	&	305.19	&	0.01	&	-31.8	&	104.46	&	83.39	&	1.84	&	3.49	&	11.29	\\
9	&	G305.55-0.01	&	305.55	&	-0.01	&	-38.2	&	84.13	&	101.53	&	1.87	&	3.28	&	11.07	\\
10	&	G305.36+0.19	&	305.36	&	0.19	&	-35	&	101.76	&	80.13	&	1.86	&	2.83	&	10.98	\\
\hline
 \end{tabular}
 \label{Tab:COCFResults10}
 \end{table*}	

\subsubsection{Excitation temperature}
Assuming the \COI\ emission is in local thermodynamic equilibrium (LTE) and optically thick, the observed brightness temperature ($T_{\rm b}$) and excitation temperature ($T_{\rm ex}$) are linked \citep{Rohlfs2004}. The excitation temperature is calculated for each voxel in the PPV cubes (i.e.\ ($l$, $b$, $v$) position) defined by clump finding using: 

\begin{equation}
T_{\rm ex}(l,b,v)=5.53\frac{1}{{\rm ln}\left  ( 1+\frac{5.53}{T_{\rm{12}}(l,b,v)+0.837}\right )}    \; [\rm K ]
\end{equation}

\noindent where $T_{\rm{12}}$ is the \COI\ brightness temperature expressed in units of K and the  equation includes the subtraction of the cosmic microwave background (2.73\,K). The average \COI\ excitation temperature ranges from $7.05\pm0.28$ to $24.57\pm0.09$\,K with a mean of 13.20\,K. Under the assumption of LTE the \COI\ and \COII\ excitation temperature can be assumed identical, as the energy levels of the two isotopomers are approximately the same. Since CO is usually thermalised within molecular clouds due to its low dipole moment this is a reasonable assumption. However, this assumption breaks down in the more diffuse envelopes of molecular clouds, where the optically thick \COI\ can remain thermalised due to radiative trapping, while the optically thin \COII\ is radiatively excited. In this case, the \COII\ excitation temperature would be lower than the \COI\ excitation temperature however; our selection criteria for clumps should nullify this issue. The assumption that the beam is uniformly filled with emission is implicit in this calculation but is unlikely to hold leading to the excitation temperature being underestimated due to beam dilution.

\subsubsection{Optical depth}
By using the \COI\ excitation temperature as a proxy for the \COII\ excitation temperature we are able to derive the \COII\ optical depth via:

\begin{equation}
\tau_{13}(l,b,v)=-{\rm ln}\left (1-\frac{0.189T_{13}(l,b,v)}{\left ( e^{\frac{5.3}{T_{\rm ex}(l,b,v)}}-1\right )^{-1}-0.16}  \right )
\end{equation}

\noindent where $\tau_{13}$ is the \COII\ optical depth and $T_{13}$ is the \COII\ brightness temperature. The optical depth was evaluated at each voxel ($l$, $b$, $v$) where a \COI\ excitation temperature is available. The line center optical depth was computed at each pixel ($l$, $b$) by taking the peak optical depth along the line of sight. The mean of the line center optical depths over the clump ($\tau$) was then determined and is reported in Table~\ref{Tab:COProp}. All \COII\ emission is found to be optically thin with a range of $0.18\pm 0.02$ to $0.86\pm0.07$ and average of 0.36. The error associated with the optical depth is dependent on the excitation temperature and the temperature scale.

\subsubsection{Column densities}
From the optical depths and excitation temperature the \COII\ column density ($N_{^{13}{\rm CO}}$) may be derived at each pixel ($l$, $b$) using:

\begin{equation}\label{Eq:NCO}
N_{^{13}{\rm CO}}(l,b)=2.6\times10^{14}\int \frac{T_{\rm ex}\left ( l,b,v \right )\tau_{13}(l,b,v)}{1-e^{\frac{-5.3}{T_{\rm ex}\left ( l,b,v \right )}}}\,d\nu \; [\rm cm^{-2} ]
\end{equation}

\noindent where the integration is performed over the velocity in \kms. 
 The composition of molecular clouds is dominated by H$_2$ and so to derive the molecular cloud masses from CO observations we must first make an assumption as to the abundance ratio between CO and H$_2$. The abundance ratio between \COII\ and \COI\ has been shown to have a gradient with distance from the Galactic center ($D_{\rm GC}$) that follows the relationship $^{12}{\rm CO}\,{:}\,{^{13}{\rm CO}}=6.21D_{\rm GC}+18.71$ \citep{Milam2005}. For G305 $D_{\rm GC}=6.5$\,kpc thus the abundance ratio between \COI\ and \COII\ is $^{12}{\rm CO}\,{:}\,{^{13}{\rm CO}}=59$. The abundance ratio between \COI\ and H$_2$ is $^{12}{\rm CO}\,{:}\,{\rm H_2}=8 \times 10^{-5}$ \citep{Langer1990, Blake1987}. Assuming a constant abundance ratio between \COII\ and H$_2$ of $1.36\times10^{-6}$ results in clump averaged H$_2$ column densities ranging from $0.13\pm0.02\times10^{22}$ to $4.04\pm0.19\times10^{22}$\,cm$^{-2}$ with a mean of  $0.9\times10^{22}$\,cm$^{-2}$. The error associated with the column density originates from the uncertainty in the excitation temperature, opacity and the abundance ratio. The assumption that the abundance ratio is constant across the complex is implicit but does not hold for a number of reasons that we briefly describe. The CO abundance declines steeply with decreasing extinction (A$_{\rm V}$) due to photodissociation by the Galactic radiation field at A$_{\rm V} < 3$ \citep{Glover2010}. In contrast, self-shielded H$_2$ can exist at A$_{\rm V}$ as low as 0.2 \citep{Wolfire2010}. As a result, the $^{13}{\rm CO}\,{:}\,{\rm H_2}$ abundance is likely to be a factor of 2 lower in the envelopes of molecular clumps than in their denser cores. In addition CO depletion is known to occur due to freeze out onto dust at densities greater than $10^4$\,\cmthree\ \citep{Bacmann2002} which could potentially render some of the molecular mass inside a molecular clump invisible to CO observations. Such high densities exist within G305 \citep{Hindson2010} but remain unresolved by the Mopra beam, we therefore caution that the CO column densities may underestimate the H$_2$ column density and mass towards the densest clumps. Finally, isotopomer-selective photodissociation whereby the \COII\ dissociates faster than \COI\ in the PDR is likely to occur \citep{Dishoeck1988}, lowering the abundance ratio between \COII\ and \COI\ at the clump periphery.

\subsubsection{Mass}
From the column density and area, we derive the mass of a clump from the following equation:

\begin{equation}
M=0.35\,D^2\int_{l,b,v}\frac{T_{\rm ex}(l,b,v)\tau_{13}(l,b,v)}{1-e^{\frac{-5.3}{T_{\rm ex}(l,b,v)}}}\,dv\,dl\,db \;[\rm M_\odot]
\end{equation}

\noindent where $D$ is the distance to G305 in kpc and the integral is performed over the velocity in \kms\ and extent $l$ and $b$ in arc-minutes. This results in clump averaged masses ranging from $0.1\pm0.1\times10^3$ to $23.0\pm1.1\times10^3$\,\msun\ with a mean of  $2.0\times10^3$\,\msun\ and total mass in clumps of $3.1\times10^5$\,\msun. We evaluate the total mass of molecular gas in G305, that is including the diffuse envelopes, by applying the above analysis to all the emission $>3\sigma$ and thus determine the combined clump and diffuse gas mass traced by CO to be $>3.5\times10^5$\,\msun. Due to our incomplete sampling of the region this is a lower limit.

\subsubsection{Virial mass}
The virial mass of a molecular cloud is defined as the mass for which the cloud is in virial equilibrium, i.e.\ when the internal kinetic energy, $K$, equals half the gravitational energy, $U(2K+U=0)$. The stability of molecular clumps against collapse can be tested by comparing the gas mass to the virial mass, which we derive using the standard equation (e.g.\ \citealt{Evans1999}):

\begin{equation}\label{Eq:Virial}
M_{\rm vir}\simeq210R\langle\ \Delta V^{2} \rangle \; [\rm M_\odot ]
\end{equation}

\noindent where $R$ is the clump radius in pc and $\Delta V$ is the full-width at half maximum (FWHM) line width in \kms. The virial parameter, $\alpha_{\rm vir}$, of a molecular cloud is the ratio of its virial mass ($M_{\rm vir}$) to its mass ($M$). It describes the ratio of the internal supporting energy to the gravitational energy such that $\alpha_{\rm vir}=M_{\rm vir}/M$. Therefore, for $M>M_{\rm vir}\, (\alpha_{\rm vir} <1), \, 2K+U<0$ and the molecular cloud is gravitationally bound, potentially unstable and may collapse. For $M<M_{\rm vir}\, (\alpha_{\rm vir} >1), \, 2K+U>0$ and the molecular cloud is not gravitationally bound and is in a stable or expanding state unless confined by external pressure. This is valid under the assumption of a simple spherical cloud with a density distribution of $\rho \propto 1/r$, ignoring magnetic fields and bulk motions of the gas \citep{MacLaren1988}. The virial mass varies by a factor of three under different density distributions. We find that the virial parameter ranges from $0.3\pm0.2$ to $11.7\pm6.0$ with a mean of 1.8 which suggests that most of the molecular clumps are not gravitationally bound. We identify 20 clumps with virial masses more than 3 times greater than the LTE derived mass in Section 3.2.4. These 20 clumps are faint and compact, which leads to the low mass estimates but have broad velocity dispersions. This leads to large virial mass estimates compared to the LTE derived mass, which either indicates that these sources are bright examples of gravitationally unbound gas or highly perturbed clumps.

\subsubsection{Number and surface density}
The mean number density of particles (H$_2$ and He) in the molecular clumps is estimated assuming spherical symmetry via: 

\begin{equation}
n({\rm H_2+He})=15.1\times M \times \left ( \frac{4}{3}\pi R^3 \right )^{-1}  \; [\rm cm^{-3}]
\end{equation}

Where $R$ is the physical radius in pc, $M$ is the mass in solar mass units. In this way the number density is found to be between $0.3\pm0.2\times10^3$ and $4.5\pm2.2\times10^3$\,\cmthree\ with a mean of $1.5\times10^3$\,\cmthree. The surface mass density $\sum_{\rm c}$ of the molecular clouds is calculated by simply dividing the mass by the area:

\begin{equation}
\Sigma_{\rm c}=MA^{-1} \; [\rm M_\odot\,pc^{-2}]
\end{equation}

We find surface densities ranging from $0.3\pm0.1\times10^2$ to $9.1\pm2.9\times10^2$\,\msun\,pc$^{-2}$ with a mean of $2.0\times10^2$\,\msun\,pc$^{-2}$. We convert to units of g\,cm$^{-2}$ to compare to theoretical studies (e.g \citealt{Krumholz2008}) using a conversion factor of $2.08\times10^{-4}$ and find values ranging from 0.01 to 0.2\,g\,cm$^{-2}$ with a mean value of 0.04, we discuss the implications of this in Section~\ref{Sect:SFH}. We present the derived physical properties of the first 10 molecular clumps in Table~\ref{Tab:COProp} and the range and average properties in Table~\ref{Tab:COAvgProp}.

\begin{table*}
 \caption[CO emission physical properties]{Physical properties of the first 10 of 156 $^{13}$CO clumps. This table can be found in appendix. }
\begin{center}
  \begin{tabular}{cccccccccc}
   \hline 
	Clump & Clump   & $\tau$ & $T_{\rm ex}$ & $N_{\rm H_{2}}$ & $n_{\rm H_2}$& $\sum_{\rm c}$ & $M$ & $M_{\rm vir}$ & $\alpha_{\rm vir}$ \\
	No.  & Name  & (\COII)    &  (K)     &  $10^{22}$(cm$^{-2}$)  &$10^3$(\cmthree)&$10^2$(\msun\,pc$^{-2}$) & $10^3$(\msun) & $10^3$(\msun)&  \\        
\hline
1	&	G305.24+0.26	&	    0.38 $\pm$     0.02	&	   16.93 $\pm$     0.13	&	    4.04 $\pm$     0.19	&	     3.6 $\pm$      1.7	&	     9.1 $\pm$      2.9	&	    23.0 $\pm$      1.1	&	    19.0 $\pm$      9.5	&	    0.83 $\pm$     0.41	\\
2	&	G304.94+0.55	&	    0.32 $\pm$     0.01	&	   24.57 $\pm$     0.09	&	    2.30 $\pm$     0.09	&	     3.0 $\pm$      1.4	&	     5.2 $\pm$      1.6	&	     6.3 $\pm$      0.2	&	     1.6 $\pm$      0.8	&	    0.25 $\pm$     0.12	\\
3	&	G305.56+0.02	&	    0.30 $\pm$     0.01	&	   17.90 $\pm$     0.13	&	    1.99 $\pm$     0.09	&	     2.6 $\pm$      1.2	&	     4.5 $\pm$      1.4	&	     5.5 $\pm$      0.3	&	     3.8 $\pm$      1.9	&	    0.70 $\pm$     0.35	\\
4	&	G305.26-0.03	&	    0.50 $\pm$     0.02	&	   16.70 $\pm$     0.13	&	    3.50 $\pm$     0.15	&	     3.0 $\pm$      1.4	&	     7.8 $\pm$      2.5	&	    22.0 $\pm$      1.0	&	    11.0 $\pm$      5.7	&	    0.51 $\pm$     0.26	\\
5	&	G304.93+0.54	&	    0.43 $\pm$     0.01	&	   20.26 $\pm$     0.11	&	    2.11 $\pm$     0.08	&	     2.5 $\pm$      1.2	&	     4.7 $\pm$      1.5	&	     6.6 $\pm$      0.2	&	     2.1 $\pm$      1.1	&	    0.32 $\pm$     0.16	\\
6	&	G305.37+0.21	&	    0.39 $\pm$     0.01	&	   18.16 $\pm$     0.12	&	    2.41 $\pm$     0.09	&	     3.3 $\pm$      1.5	&	     5.4 $\pm$      1.7	&	     6.0 $\pm$      0.2	&	     6.7 $\pm$      3.4	&	    1.13 $\pm$     0.57	\\
7	&	G305.36+0.17	&	    0.32 $\pm$     0.02	&	   14.58 $\pm$     0.15	&	    1.76 $\pm$     0.11	&	     2.0 $\pm$      1.0	&	     3.9 $\pm$      1.3	&	     6.2 $\pm$      0.4	&	    10.0 $\pm$      5.2	&	    1.69 $\pm$     0.85	\\
8	&	G305.19+0.01	&	    0.52 $\pm$     0.02	&	   13.52 $\pm$     0.15	&	    1.91 $\pm$     0.11	&	     2.6 $\pm$      1.3	&	     4.3 $\pm$      1.4	&	     4.5 $\pm$      0.3	&	     4.7 $\pm$      2.3	&	    1.03 $\pm$     0.52	\\
9	&	G305.55-0.01	&	    0.35 $\pm$     0.02	&	   15.56 $\pm$     0.14	&	    1.79 $\pm$     0.09	&	     2.4 $\pm$      1.2	&	     4.0 $\pm$      1.3	&	     4.4 $\pm$      0.2	&	     4.2 $\pm$      2.1	&	    0.96 $\pm$     0.48	\\
10	&	G305.36+0.19	&	    0.38 $\pm$     0.02	&	   15.38 $\pm$     0.15	&	    1.76 $\pm$     0.08	&	     2.4 $\pm$      1.1	&	     3.9 $\pm$      1.3	&	     4.3 $\pm$      0.2	&	     3.1 $\pm$      1.6	&	    0.73 $\pm$     0.37	\\
\hline
 \end{tabular}
  \end{center}
 \label{Tab:COProp}
 \end{table*}

\begin{table}
 \caption{Global minimum, maximum and mean properties of the 156 molecular clumps in G305.}
\begin{center}
  \begin{tabular}{lccc}
   \hline 
	Parameter & Min & Max & Mean  \\
\hline
$T_{\rm ex}$ (K)	&	    7.05	&	   24.57	&	   13.20	\\
$\tau_{\rm ^{13}CO}$	&	    0.18	&	    0.86	&	    0.36	\\
$R$ (pc)	&	    0.68	&	    3.00	&	    1.43	\\
$\Delta V$ (\kms)	&	    0.97	&	    5.66	&	    2.24	\\
$V_{\rm LSR}$ (\kms)	&	  $-47.80$	&	  $-17.60$	&	  $-37.44$	\\
$N_{\rm H_2} 10^{22}$(cm$^{-2}$)	&	    0.13	&	    4.04	&	    0.65	\\
Mass $10^3$(\msun)	&	    0.1	&	   23.0	&	    2.0	\\
$\sum_{\rm c} 10^2$(\msun\,pc$^{-2}$)	&	    0.3	&	    9.1	&	    2.0		\\
$M_{\rm vir} 10^3$(\msun)	&	    0.2	&	   19.0	&	    2.1	\\
$\alpha_{\rm vir}$	&	    0.3	&	   11.7	&	    1.8	\\
$n_{\rm H_2} 10^{3}$(cm$^{-3}$)	&	    0.3	&	    4.5	&	    1.5	\\
\hline
 \end{tabular}
 \end{center}
  \label{Tab:COAvgProp}
 \end{table}	

\subsection{Large-scale radio continuum emission}\label{Sect:LargeClass}
The morphology of the radio continuum emission is complex ranging from extended low surface brightness features to compact bright emission. To remove side lobes and aid in the description of the large-scale radio emission we have used a Gaussian kernel to smooth from a resolution of $10.2\times 8.2$\arcsec\ to $30\times30$\arcsec\ (Fig~.\ref{im:LargeScale}). The extended radio continuum emission associated with G305 is concentrated around the rim of the central cavity and correlates with 5.8\,$\umu$m PAH emission highlighting the PDR. The central cavity to the north, south and west is bound by diffuse, low surface brightness emission. In contrast, we find little radio continuum emission within the central cavity and, with the exception of a small region of emission, the east of the complex. The compact radio continuum emission is concentrated towards six bright sources, which are projected within the extended emission. We also detect two radio continuum sources isolated away from the central cavity and PDR boundary in the far north-west (Fig.~\ref{fig:Large13}) and north-east (Fig.~\ref{fig:Large11}). In the following section, we describe the steps taken to classify these radio continuum sources.

\subsection{Radio continuum source classification and properties}\label{Sect:radioclass}

\begin{figure}
\includegraphics[trim=40 10 30 50,width=0.5\textwidth]{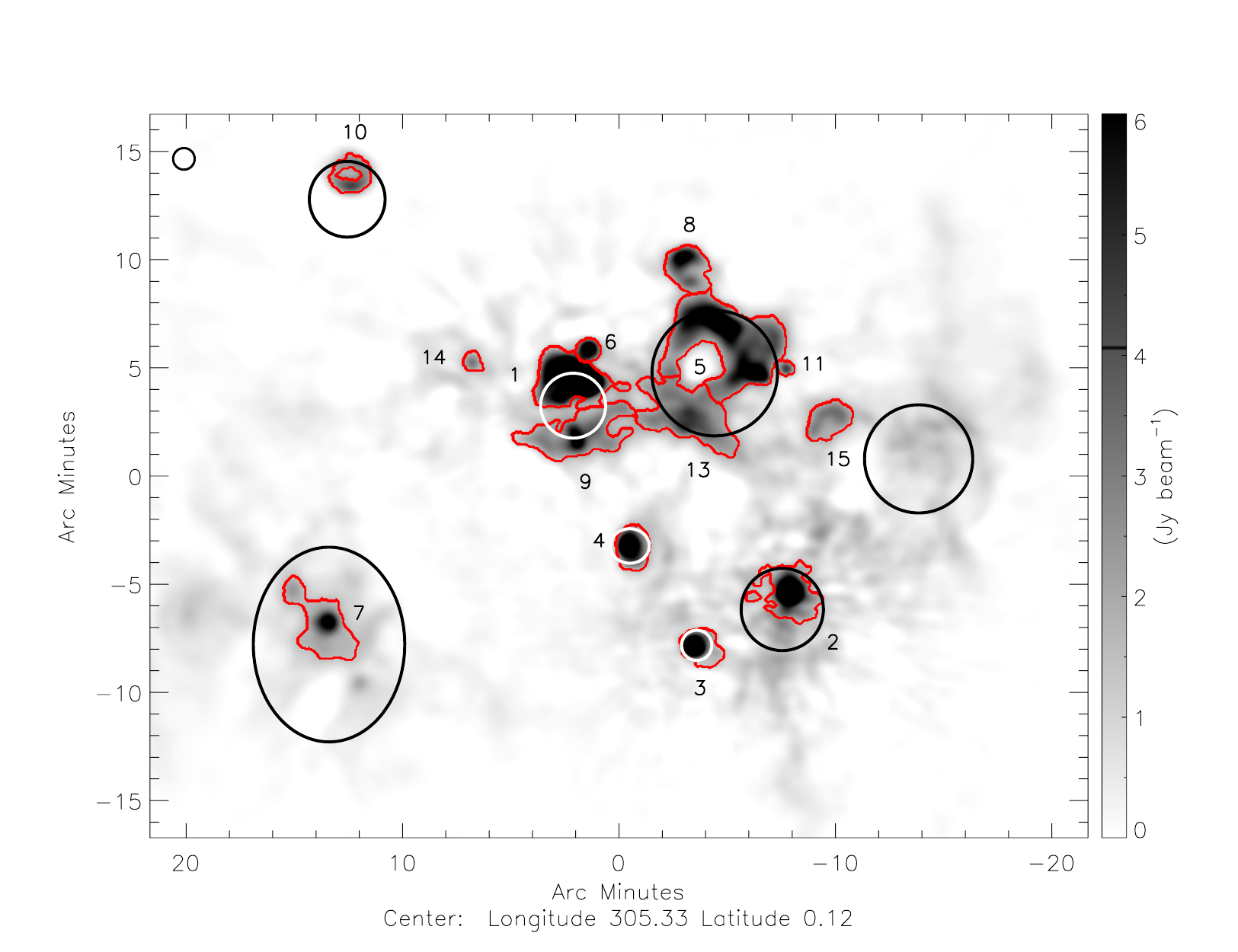} 
\includegraphics[trim=40 10 50 20,width=0.5\textwidth]{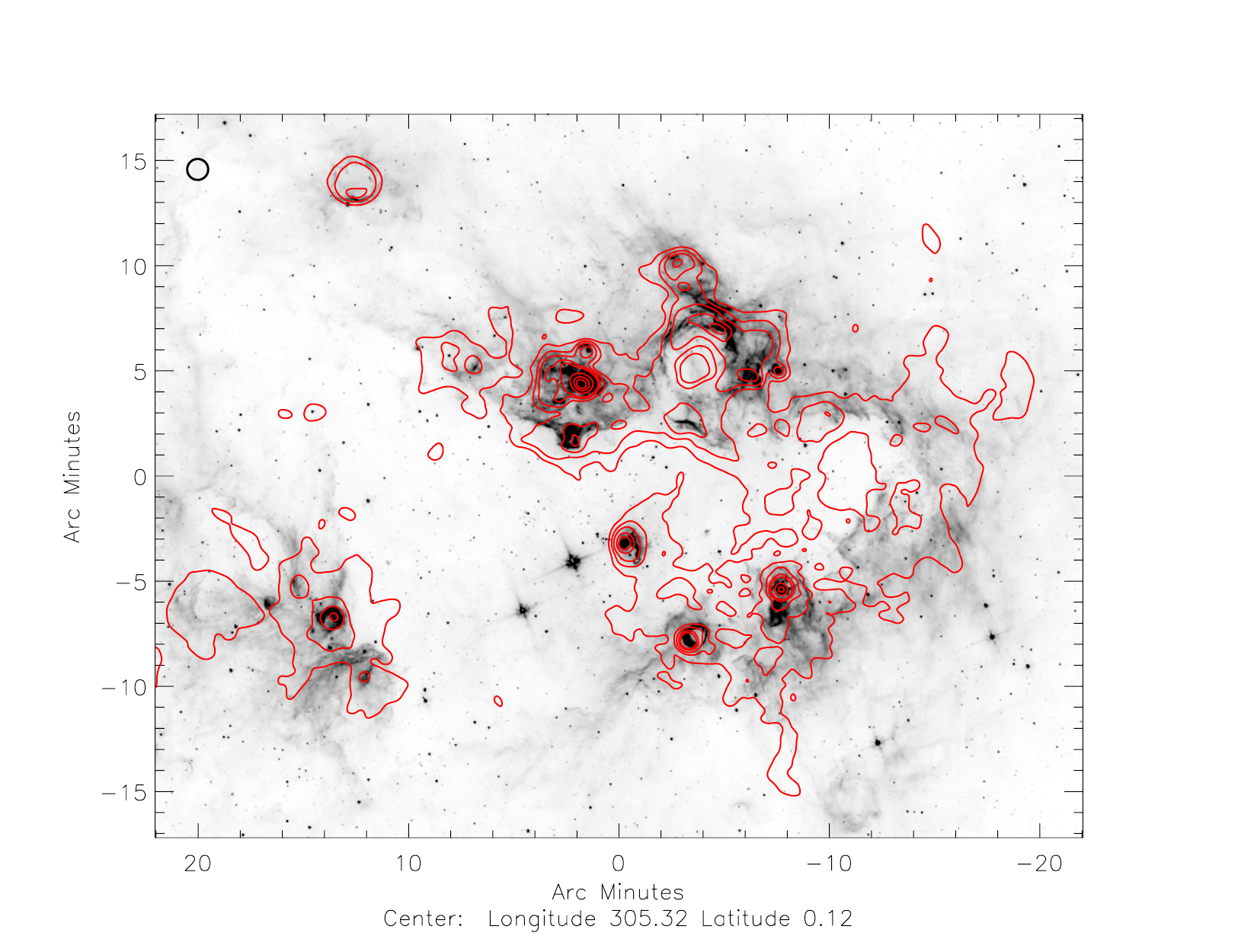}
\caption[G305 large-scale 5.5\,GHz radio continuum]{Top: The 5.5\,GHz large-scale radio continuum map with the sources identified by {\sc CLUMPFIND} outlined in red. The corresponding source numbers are presented as well as ellipses indicating previously identified HII regions \citep{Clark2004}. Bottom: The 5.5\,GHz radio emission contours (red) are presented over a 5.8\,$\umu$m GLIMPSE image. Contours begin and increment by 0.15\,Jy\,beam$^{-1}$ up to 0.75\,Jy\,beam$^{-1}$ and then proceed in steps of 1\,Jy\,beam$^{-1}$ up to 5.75\,Jy\,beam$^{-1}$. Both these images have been smoothed by a $30\times30\arcsec$ Gaussian (top-left). Source 12 in the far north-west has been cropped to improve the detail visible around the cavity but can be seen in Fig.~\ref{fig:Large13}.}
\label{im:LargeScale}
\end{figure}

Defining discrete sources in extended and complex radio emission is non-trivial. To characterise the different scales of emission presented in Fig.~\ref{im:LargeScale} we first apply the thresholding {\sc CLUMPFIND} algorithm (T$_{\rm low}=3\sigma$ and $\Delta T=3\sigma$) to a smoothed 5.5\,GHz image (Fig.~\ref{im:LargeScale}; top). This results in the radio continuum emission being separated into 15 broad but distinct features (Fig.~\ref{im:LargeScale}; red outline, Table~\ref{Tab:LargeSources}). Comparison between these 15 radio features with the HII regions reported by \cite{Caswell1987} (Fig.~\ref{im:LargeScale}; black ellipses) reveals that all the radio emission, with exception of sources\,8, 14 and 15, is associated with previously identified HII regions. We clearly resolve substructure within these HII regions and identify multiple emission features. We also note that one HII region detected by \cite{Caswell1987} ($305.097+0.138$) does not appear to be associated with an HII region but rather identifies where the central cavity is breaking out into the wider ISM.

\begin{figure*}
\subfigure[Sources 1, 1-1, 6, 9, 9-1 and 9-2$^*$  ]{
\includegraphics[width=0.3\textwidth]{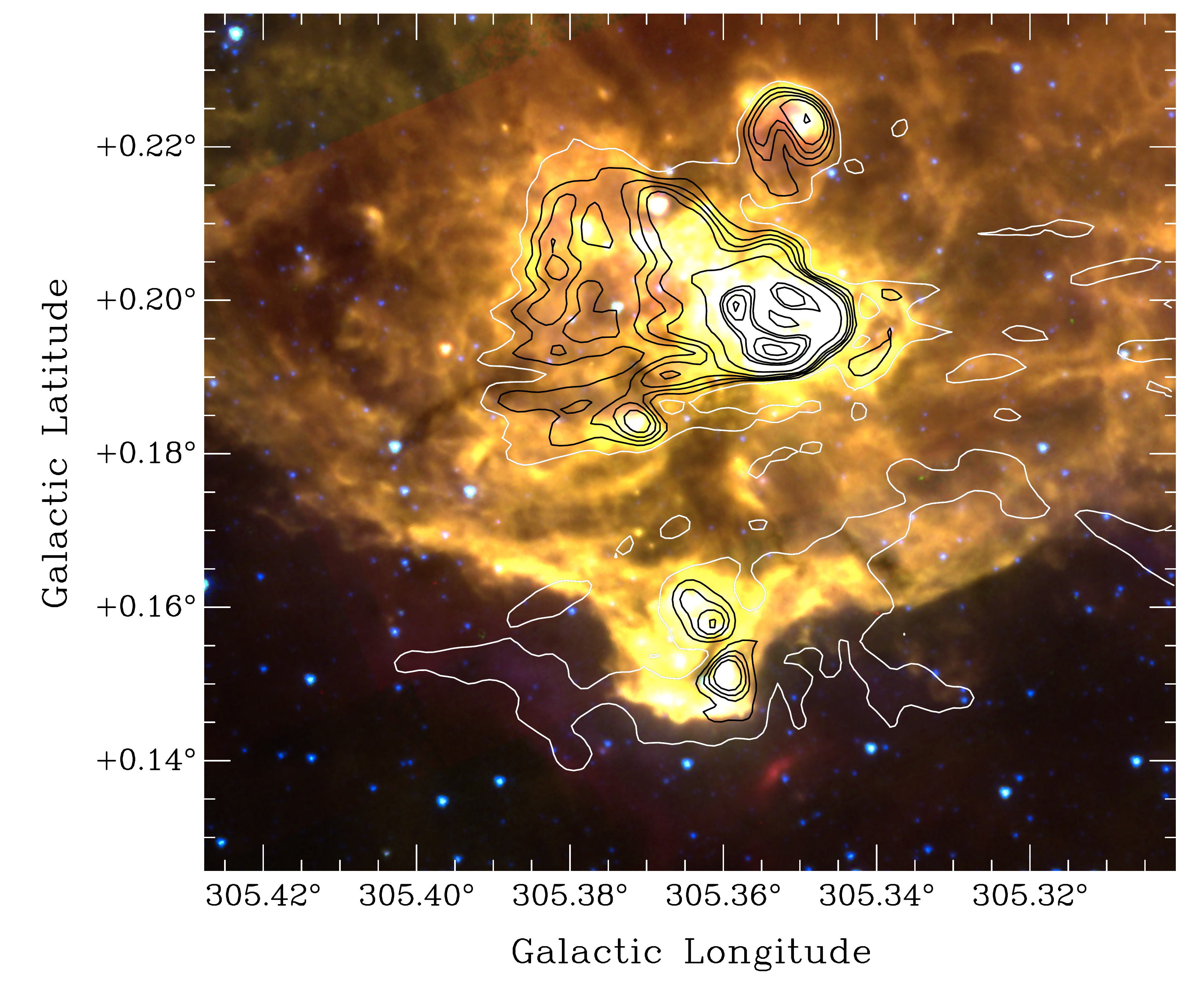}
   \label{fig:Large1}
 }
 \subfigure[Source 2$^*$, 2-1$^*$]{
\includegraphics[width=0.3\textwidth]{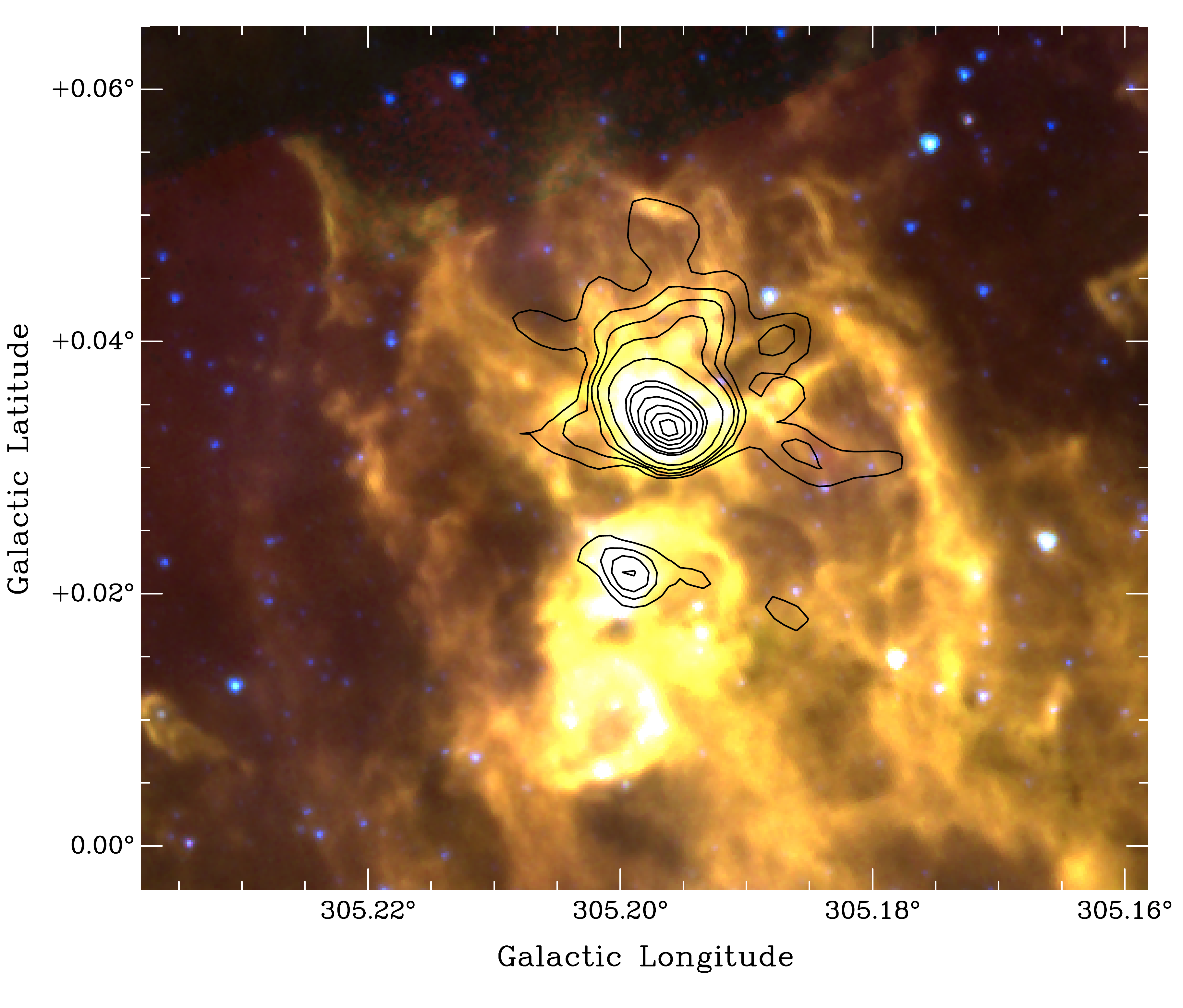}
   \label{fig:Large2}
 }
\subfigure[Source 3$^*$]{
\includegraphics[width=0.31\textwidth]{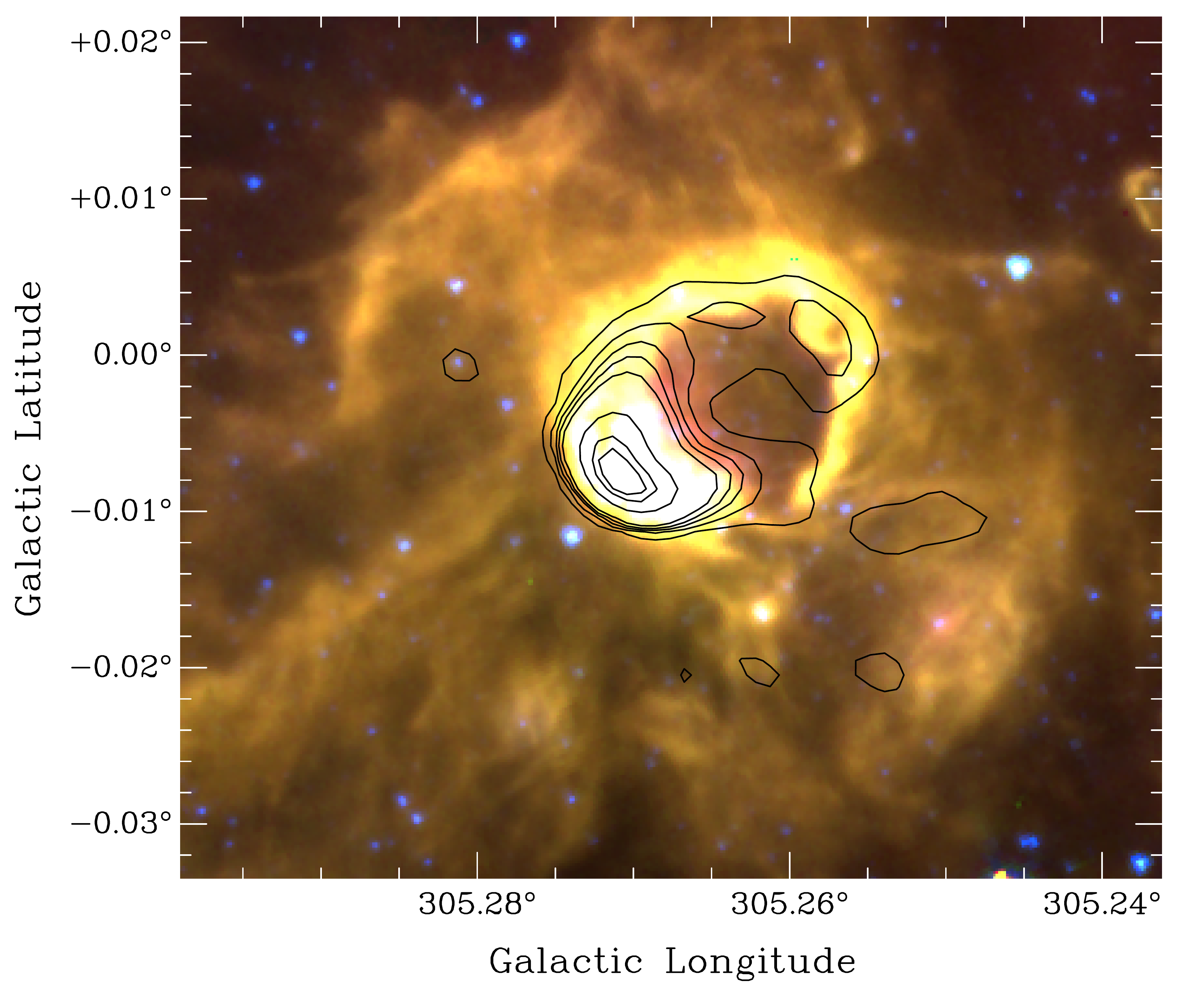}
   \label{fig:Large3}
 }
 \subfigure[Source 4]{
\includegraphics[width=0.3\textwidth]{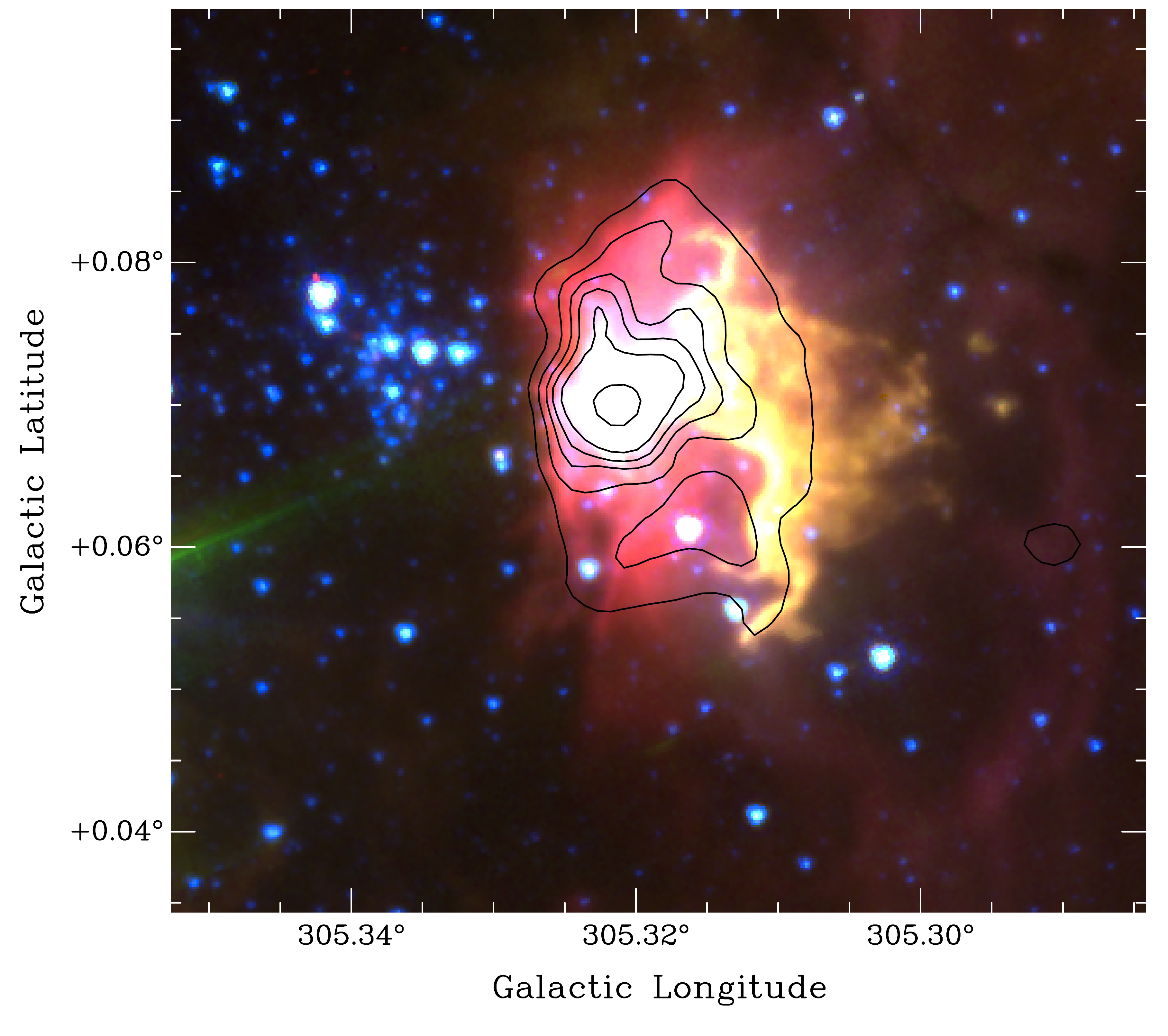}
   \label{fig:Large4}
 }
 \subfigure[Sources 5, 5-1, 5-2, 8 and 11]{
\includegraphics[width=0.3\textwidth]{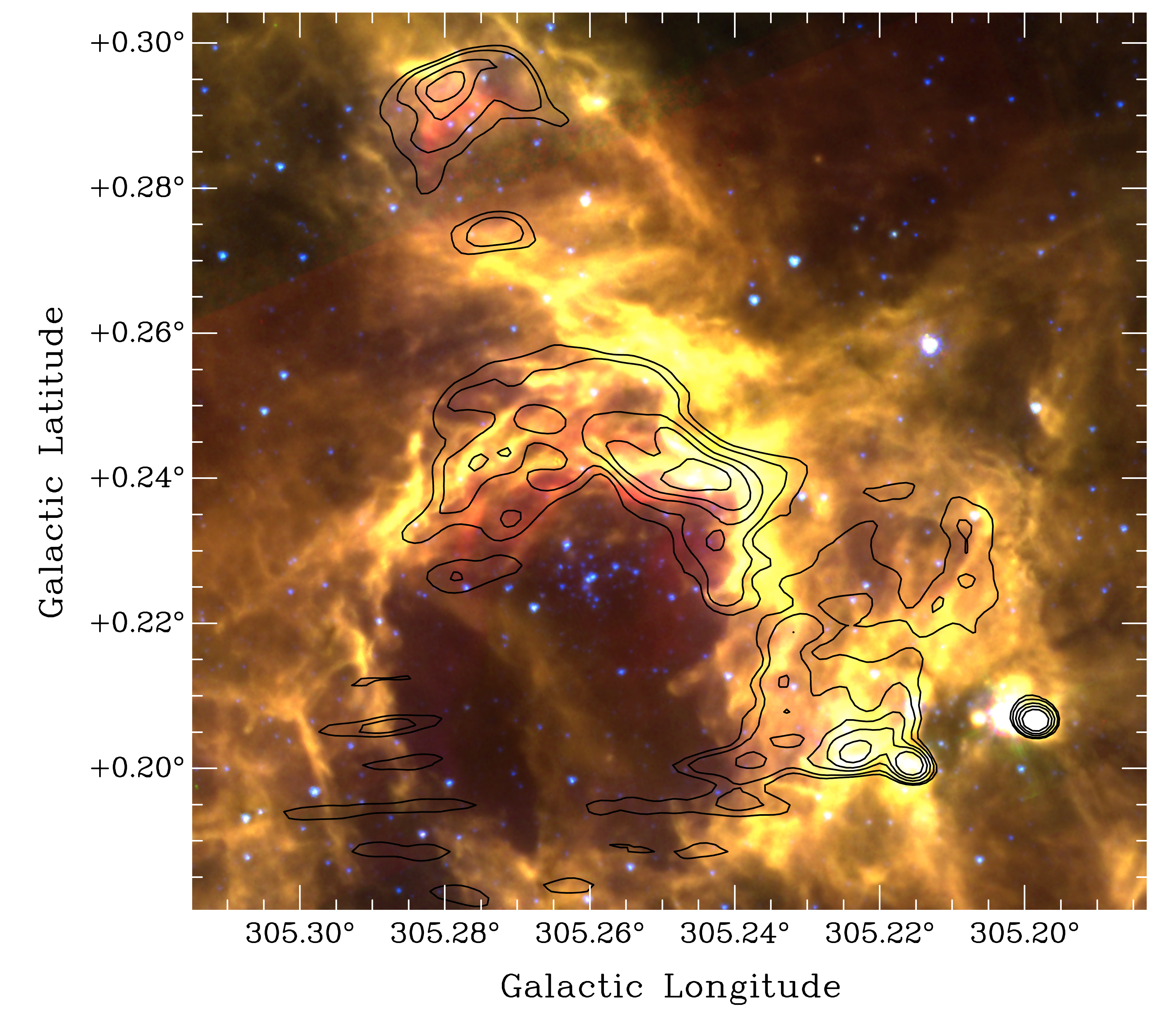}
   \label{fig:Large5}
 }
 \subfigure[Source 7, 7-1 and 7-2]{
\includegraphics[width=0.3\textwidth]{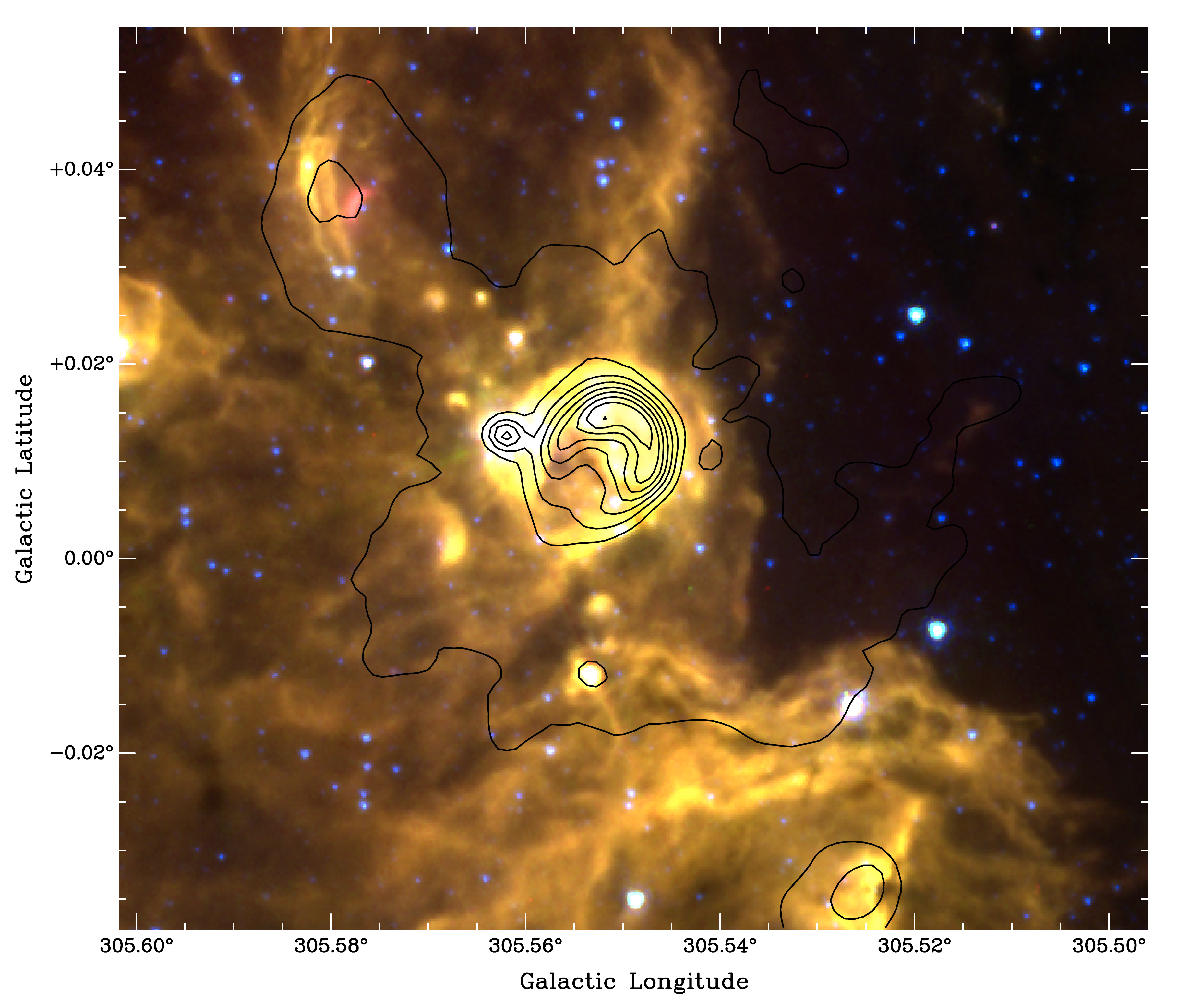}
   \label{fig:Large8}
 }
  \subfigure[Source\,10]{
\includegraphics[width=0.3\textwidth]{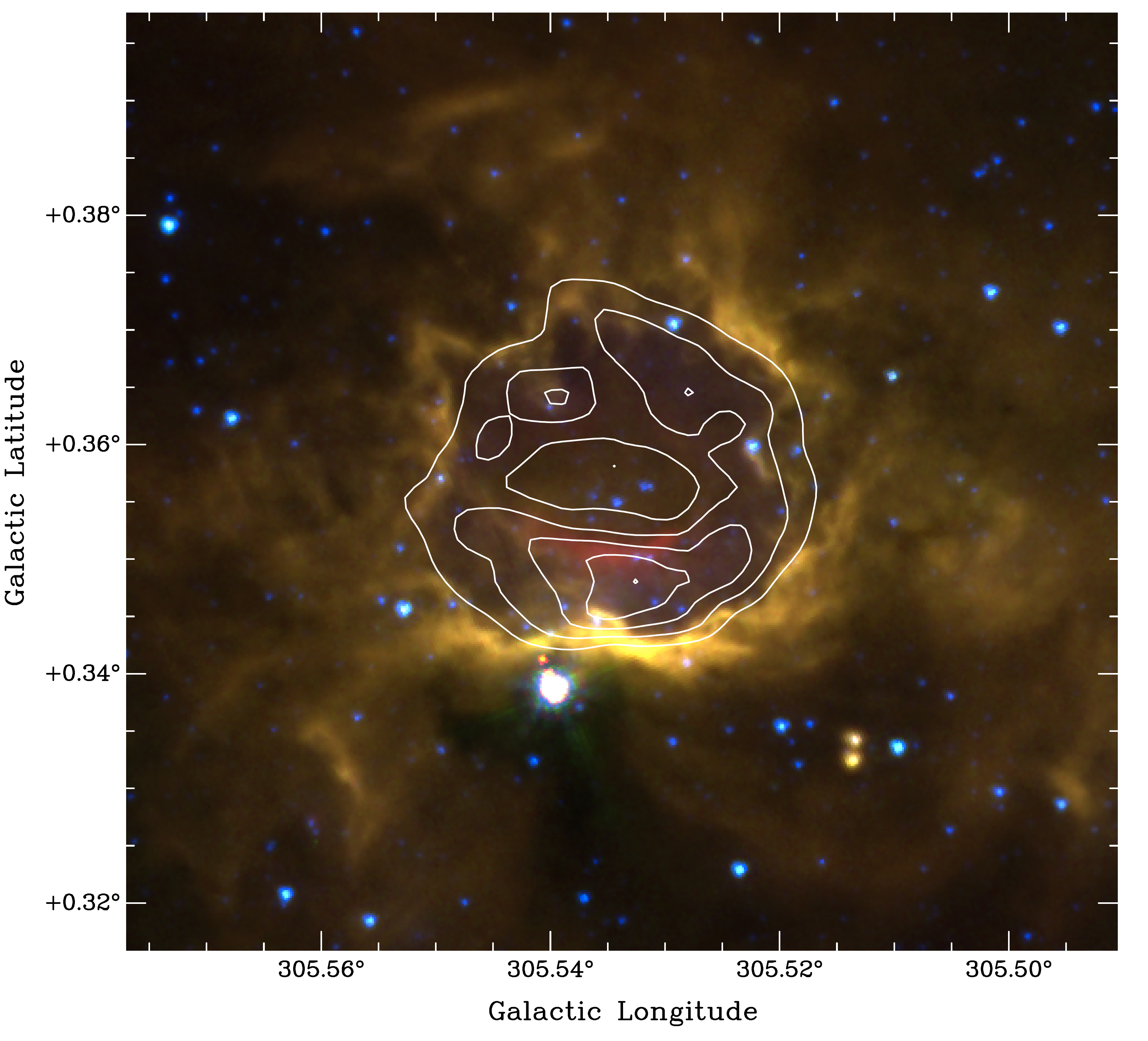}
   \label{fig:Large11}
 } 
 \subfigure[Source\,12]{
\includegraphics[width=0.3\textwidth]{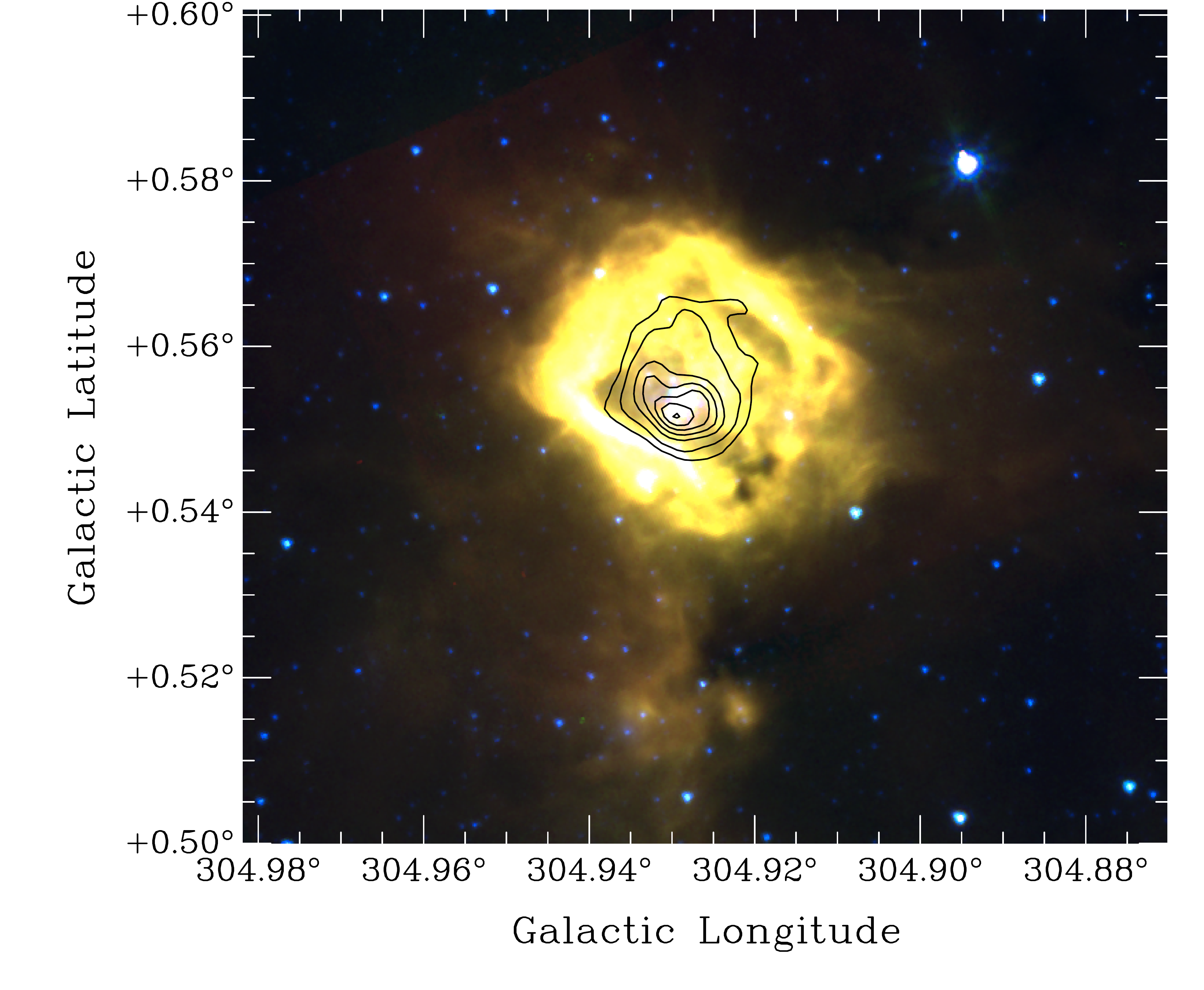}
   \label{fig:Large13}
 }
 \subfigure[Source\,14]{
\includegraphics[width=0.3\textwidth]{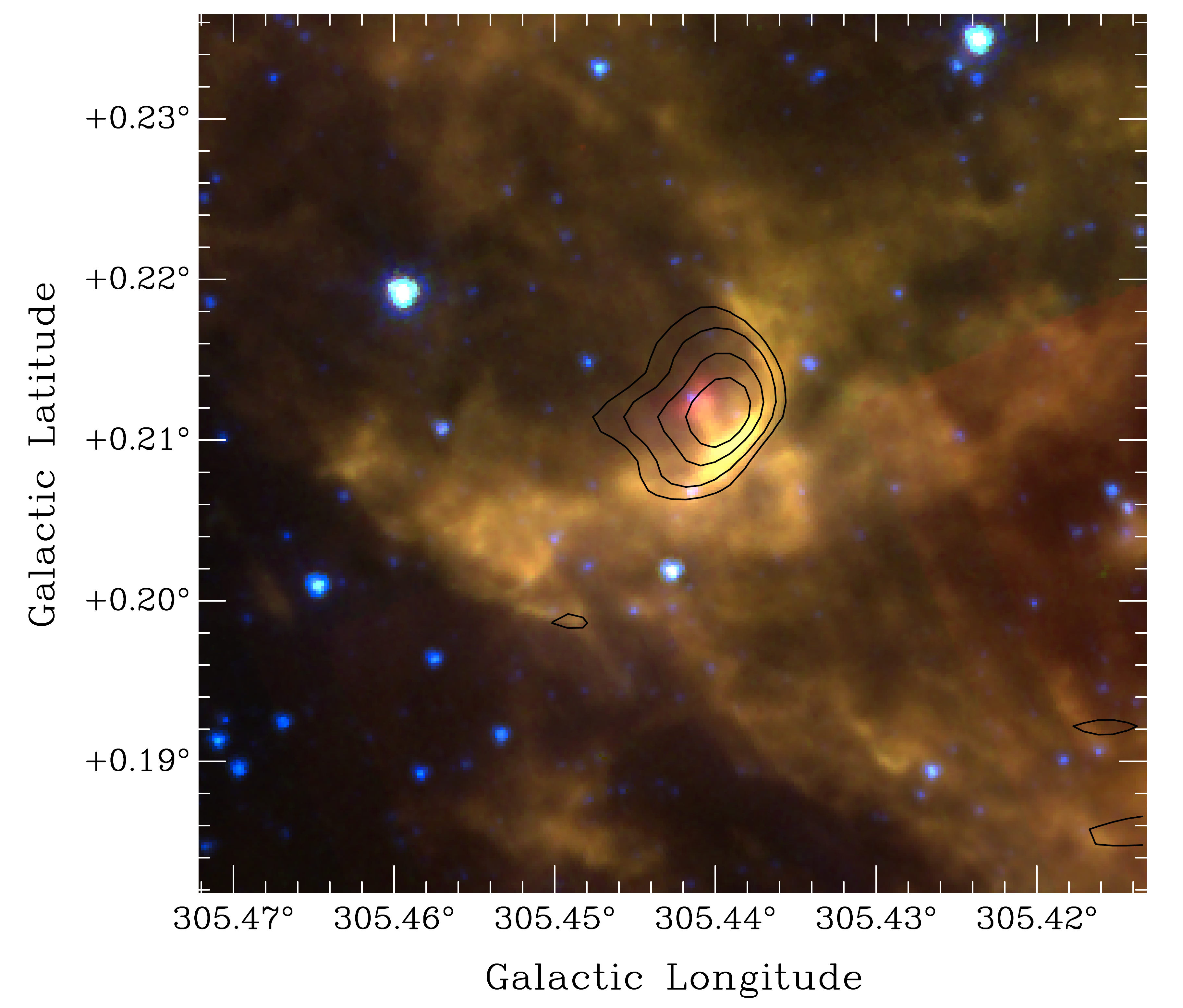}
   \label{fig:Large15}
 }
   \subfigure[Source\,15]{
 \includegraphics[width=0.3\textwidth]{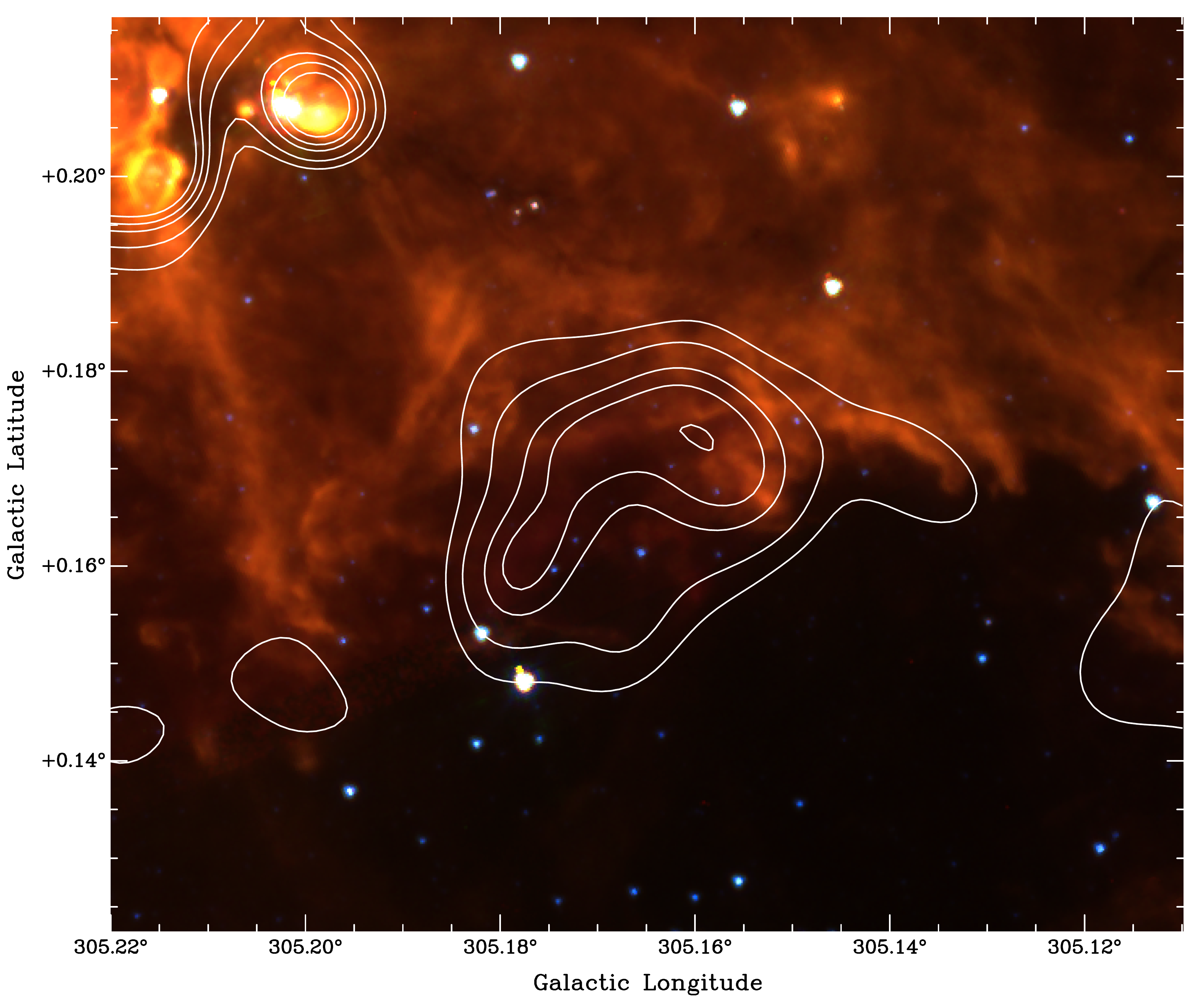}
   \label{fig:Large16}
 }

\caption[Large-scale radio emission three-colour images]{The large-scale 5.5\,GHz radio contours are presented over a three-colour images GLIMPSE image using the 4.5, 5.8 and 8.0\,$\umu$m bands (red, blue and green respectively). Contour levels vary between maps due to artefacts, for images marked with an asterisk contours begin at 0.04 and increment by 0.02\,Jy\,beam$^{-1}$ sources without an asterisk begin at 0.01 and increment by 0.01\,Jy\,beam$^{-1}$. Contours are black or white depending on the contrast of the background.}
\label{im:largescale3}
\end{figure*}

The next step is to characterise the substructure within these 15 broad radio features. This is achieved by searching for obviously isolated compact radio features by eye in the full resolution 5.5\,GHz mosaic (see Fig.~\ref{im:largescale3}). The dimensions of the substructure are defined by carefully fitting apertures around the emission for pixels that are $3\sigma$ above the average flux of the associated large-scale feature. In this way we identify 8 smaller radio sources that are either embedded or projected against the large-scale radio emission. We therefore identify 23 radio features in the 5.5\,GHz image. 

To aid in the classification of the observed radio continuum emission we extract images from the GLIMPSE survey \citep{Benjamin2003, Churchwell2009} using the 4.5, 5.8 and 8.0\,$\umu$m IRAC bands in blue, green and red, respectively and overlay the 5.5\,GHz radio continuum emission (Fig.~\ref{im:largescale3}). Due to the varied and complex nature of the radio continuum emission we consider each source individually and classify the emission based on features such as the morphology, coincidence with mid-infrared emission, size and age \citep{Deharveng2010}. We classify radio emission as an HII region if it shows a bubble, or spherical morphology (50\% of HII regions have such a morphology \citealt{Anderson2011}) that is associated with mid-infrared emission, strong in the 4.5 and 5.8\,$\umu$m bands, indicative of heated dust that is known to surround HII regions, and the feature has a radius of $>0.5$\,pc. Compact HII regions are much younger ($<0.4$\,Myr) and smaller ($<0.5$\,pc) but are also associated with bright compact mid-infrared emission indicating heated dust. Extended radio emission is classified by its association with the border of the PDR with no obvious core or spherical morphology. Finally, sources that do not fall into any of these categories are classified as unknown sources. Following these criteria we present the source type in column 3 of Table~\ref{Tab:LargeSources}. We identify 9 HII regions, 7 compact HII regions, a single \mbox{UC\,HII region} (previously identified in \citealt{Hindson2012}) and 4 extended emission features. The remaining two sources (8 and 14) do not fall into the HII region or extended classification but appear to be associated with an ionisation front or shocked region. Further classification would require improved $u$-$v$ coverage to reduce artefacts and allow physical property estimates and/or follow-up observations of radio recombination line emission \citep{Anderson2011} to obtain turbulent and thermal energies in the HII regions and derive electron temperatures at comparable resolution.

\subsubsection{Lyman continuum flux}\label{Sect:Lyman}
From the integrated 5.5\,GHz emission we derive the total ionising photon flux of the Lyman continuum ($N_{\rm{Ly}}$), a property that is independent of source geometry, using the modified equation (7) presented in \cite{Carpenter1990}
\begin{equation}\label{Eq:Lyman}
N_{\rm{Ly}} =7.7\times10^{43}S_{\rm int}D^2\nu^{0.1} \; [\rm s^{-1}]
\end{equation}

\noindent where $N_{\rm{Ly}}$ is the total number of Lyman photons emitted per second, $S_{\rm int}$ is the integrated radio flux (mJy), $D$ is the distance to the source ($3.8\pm0.6$\,kpc) and $\nu$ is the frequency of the observation (GHz). Under the assumption of optically thin emission we find Lyman continuum fluxes from $4.68\times10^{49}$ to $8.7\times10^{46}$\,photon\,s$^{-1}$. The spectral type and mass of the ionising star responsible for the derived Lyman continuum is estimated by comparing the Lyman continuum flux to the derived value of the total number of ionising photons generated by high-mass stars tabulated by \citet{Martins2005}. This is under the assumption that the radio emission observed is caused by a single zero age main sequence (ZAMS) star, no UV flux is absorbed by dust and that the HII region is ionisation bounded. Under the same assumptions we derive the corresponding mass if a single star and ten stars that would be required to drive the estimated Lyman continuum flux using table\,1 from \cite{Davies2011}. Whilst the most high-mass star will dominate the Lyman flux if there are multiple stars that make a significant contribution to the ionising flux the spectral type will be later than we have estimated. Conversely, if there is significant absorption by dust and/or the nebular is not ionisation bounded the spectral type and mass may be earlier and lower than our estimate respectively. This leads to an error in the spectral classification of approximately half a spectral type (e.g. \citealt{Wood1989A}). 

\subsubsection{Dynamical age}

We estimate the dynamical age ($t_{\rm dyn}$) of the detected HII regions by using the simple model of an HII region expanding into a uniform density ISM composed of molecular hydrogen described by \citep{Dyson1997,Hosokawa2005}:

\begin{equation}
t_{\rm dyn}(R)=\frac{4R_{\rm s}}{7c_{\rm s}}\left [ \left ( \frac{R}{R_{\rm s}} \right )^{7/4}-1 \right ] \; [\rm s]
\end{equation}

\noindent where $R$ is the physical radius of the HII region, $c_{\rm s}$ is the sound velocity in the ionised gas ($c_{\rm s} = 10$\,\kms), $N_{\rm Ly}$ is the number of ionising photons per unit time (s$^{-1}$) and $R_{\rm s}$ is the radius of a Str$\rm\ddot{o}$mgren sphere given by:

\begin{equation}\label{Eq:Stromgren}
R_{\rm{s}}\approx \left (\frac{3N_{\rm{Ly}}}{4\pi \alpha_{\rm{H}} n_{\rm{e}}^{2}}\right )^{1/3} \; [\rm cm]
\end{equation}

\noindent where $n_{\rm e}$ is the electron density (\cmthree) and $\alpha_{\rm{H}}$ describes the hydrogen recombination rate ($\alpha_{\rm H}=2.7\times 10^{-13}$\,\cmthree\,s$^{-1}$). In order to derive the Stromgren radius we must assume an initial electron density of the ambient medium that surrounds the HII region. This posses a problem because the electron density around HII regions varies and is not constant over time but varies from $10^2$--$10^7$ and follows a $n_{\rm e}\propto D^{-1}$ relationship \citep{Kim2001, Garay1999}; as the HII region evolves and expands the electron density drops. The density of ionised gas that surrounds \mbox{UC\,HII} regions, the earliest directly observable stage of high-mass star formation, has been reported as $\sim 1\times10^5$\,\cmthree\ \citep{Wood1989A} but may be as high as  $1\times10^7$\,\cmthree\ \citep{Depree1995} although this is likely to be a strong upper limit \citep{Xie1996}. Without prior knowledge of the ambient density of the HII regions and how it has evolved we assume a representative density of $1\times10^{5}$\,\cmthree\ for classical HII regions. The ambient density surrounding compact and ultra compact HII regions is likely to be higher due to their younger age and the expansion of the nearby classical HII regions into the molecular gas, which is likely to raise the ambient density by a factor of 10-100 \citep{Hosokawa2005,Kim2001}. We therefore assume a higher ambient density of $10^7$\,\cmthree\ for compact and ultra compact HII regions. In addition to the unknown ambient electron density, uncertainty in the dynamical age estimate stems from the difficulty in defining the physical radius of the radio emission because of its association with extended emission. We find dynamical ages of compact HII regions ranging from 0.2 to 0.8\,Myr, classical HII regions from 0.4 to 4.0\,Myr.

\begin{table*}
 \caption[Large-scale radio sources]{Identifiers and observed properties of the 16 detected large-scale radio sources. The classification of the source (see Section~\ref{Sect:LargeClass} for details) is presented in column 3 as; HII - HII region, C - compact HII region, UCHII – ultracompact HII region, E - extended and U - unknown. }
\centering
  \begin{tabular}{ccccccccccccccc}
   \hline 
   Source  & Source	 & Source & \multicolumn{3}{c}{Dimensions}  & \multicolumn{4}{c}{Observed flux density}  & $N_{\rm Ly}$ & Spectral &\multicolumn{2}{c}{Mass} & Dynamical \\ 
	       Index   & Type  & Name &\multicolumn{3}{c}{Observed} & \multicolumn{2}{c}{Peak}  &\multicolumn{2}{c}{Integrated} & Log &Type & \multicolumn{2}{c}{No. Stars} & Age \\ \cline{4-5} \cline{6-10}
	       	       &     &     & maj & min & $R$ &  \multicolumn{2}{c}{(mJy beam$^{-1}$)}       & \multicolumn {2}{c}{(Jy)}  & (s$^{-1}$) & &$\times1$ & $\times10$&(Myr)\\
	        &      &     & ('')   & ('')  & (pc)  & \textit{f}$_{5.5}$ & \textit{f}$_{8.8}$ & \textit{f}$_{5.5}$ & \textit{f}$_{8.8}$  &  & &(\msun) &(\msun) &   \\
	        \hline
       1 & HII & G305.353+0.193 & 270.6 & 181.5 & 2.04 & 1136.7 & 923.2 & 35.2 & 28.5 & 49.67 & O5 & 100  &  30	&     1.8 \\
       1--1 &  C & G305.370+0.185 & 13.2 & 9.9 & 0.09 & 146.0 & 110.9 & 0.3 & 0.3 & 47.61 & B0 & 20  &  15  &      0.2 \\
       2 & HII & G305.195+0.033 & 211.2 & 178.2 & 1.79 & 1526.5 & 1286.3 & 15.0 & 12.6 & 49.30 & O6 & 50  &  25  &      1.7 \\
       2--1 &  C & G305.199+0.021 & 13.2 & 16.5 & 0.11 & 116.9 & 81.6 & 0.3 & 0.2 & 47.54 & B0 & 20  &  15  &      0.4 \\
       3 & HII & G305.270+-0.007 & 118.8 & 108.9 & 1.05 & 729.6 & 592.7 & 6.5 & 6.3 & 48.93 & O6.5 & 40  &  20  &      0.8 \\
       4 &  HII & G305.320+0.070 & 92.4 & 128.7 & 1.00 & 251.5 & 188.9 & 5.9 & 5.3 & 48.89 & O6.5 & 35  &  20  &      0.8 \\
       5 & HII & G305.254+0.204 & 359.7 & 330.0 & 3.17 & 168.1 & 117.2 & 29.8 & 17.9 & 49.59 & O5.5 & 90  &  30  &      4.0 \\
       5--1 & C & G305.223+0.202 & 26.4 & 13.2 & 0.16 & 132.1 & 91.3 & 0.6 & 0.5 & 47.87 & O9.5 &  20  &  15  &     0.6 \\
       5--2 & C & G305.215+0.200 & 19.8 & 16.5 & 0.14 & 168.1 & 117.2 & 0.5 & 0.4 & 47.84 & B0 &20  &  15  &       0.5 \\
       6 & HII & G305.348+0.223 & 69.3 & 62.7 & 0.61 & 215.5 & 146.0 & 2.4 & 1.6 & 48.50 & O8 &  30  &  20  &      0.4 \\
       7 & E & G305.551+0.014 & 194.7 & 240.9 & 2.00 & 90.9 & 71.0 & 6.7 & 4.1 & 48.94 & O6.5 & 40  &  20  & 0     - \\
       7--1 & HII & G305.551+0.014 & 46.2 & 46.2 & 0.33 & 90.9 & 71.0 & 1.5 & 1.3 & 48.28 & O8.5 &  25  &  20  &    1.6 \\
       7--2 & UC\,HII & G305.561+0.013 & 6.6 & 6.6 & 0.05 & 55.8 & 39.9 & 0.1 & 0.1 & 46.94 & B0.5 &15  &  12  &   0.1 \\
       8 & U & G305.278+0.295 & 125.4 & 132.0 & 1.19 & 98.0 & 72.2 & 5.2 & 4.4 & 48.84 & O6.5 &  35  &  20  &    - \\
       9 & E & G305.358+0.152 & 343.2 & 148.5 & 2.08 & 195.4 & 156.5 & 10.1 & 5.8 & 49.13 & O6 & 50  &  25  &      - \\
       9--1 & C & G305.358+0.152 & 16.5 & 16.5 & 0.13 & 195.4 & 156.5 & 0.5 & 0.4 & 47.80 & B0 & 20  &  15  &      0.4 \\
       9--2 & C & G305.361+0.158 & 19.8 & 26.4 & 0.19 & 120.0 & 82.9 & 0.5 & 0.4 & 47.84 & O9.5 &  20  &  15  &     0.8 \\
      10 & HII & G305.532+0.348 & 115.5 & 105.6 & 1.02 & 51.2 & 42.9 & 3.0 & 3.1 & 48.59 & O7.5 &   30  &  20  &    1.0 \\
      11 & C & G305.197+0.206 & 16.5 & 13.2 & 0.11 & 194.5 & 142.3 & 0.4 & 0.4 & 47.74 & B0 &  20  &  15  &     0.3 \\
      12 & HII & G304.929+0.552 & 59.4 & 59.4 & 0.55 & 62.4 & 60.1 & 0.8 & 1.1 & 48.05 & O9.5 &  25  &  20  &     0.5 \\
      13 & E & G305.277+0.161 & 277.2 & 237.6 & 2.36 & 56.7 & 38.6 & 12.0 & 7.1 & 49.20 & O6 &  50  &  25  &     - \\
      14 & U & G305.439+0.211 & 62.7 & 56.1 & 0.55 & 52.3 & 31.7 & 0.7 & 0.4 & 47.98 & O9.5 &   20  &  15  &    - \\
      15 & E & G305.169+0.174 & 128.7 & 108.9 & 1.09 & 45.6 & 30.9 & 3.2 & 2.9 & 48.63 & O7 &  30  &  20  &     - \\
\hline
 \end{tabular}
 \label{Tab:LargeSources}
 \end{table*}

\subsection{The total ionising flux of G305}\label{Sect:Flux}
Comparison between the 5.5\,GHz integrated flux presented here (180\,Jy) and the single dish 5\,GHz flux (193\,Jy; \citealt{Clark2004}) suggests that our observations reliably reproduce the total flux of the complex. We are therefore able to broadly explore the relative contribution of high-mass stars in the region to the observed integrated radio flux.

The integrated 5.5\,GHz radio flux and location of HII regions presented in Fig.~\ref{im:LargeScale} and Table~\ref{Tab:LargeSources} reveals that $\sim 60$\% (103.2\,Jy) of the total 5.5\,GHz flux (180\,Jy) in G305 originates from classical and compact HII regions embedded around the rim of the central cavity. This would suggest that in addition to the $\sim 30$ high-mass stars detected via spectroscopic observations in G305 \citep{Davies2012} there is a population of at least $17$ high-mass stars (Table~\ref{Tab:LargeSources}) embedded around the periphery of the central cavity. If we include the five \mbox{UC\,HII} regions detected in \citep{Hindson2012} this brings the total number of high-mass stars within G305 to $>51$. This is under the assumption that the observed radio continuum features are powered by a single high-mass star, however it is possible that the observed radio continuum is produced by multiple stars (Table~\ref{Tab:LargeSources} column 14). If this is the case then it is possible that the are significant clusters of high-mass stars embedded around the central cavity with cluster masses on the order of $10^3$\,\msun. We find little evidence of radio emission originating from the highest concentration of high-mass stars found in Danks\,1 and 2. However, just because no detectable diffuse radio emission appears to be associated with Danks\,1 and 2 does not mean they are not significant sources of ionising radiation. What this result reveals is that there is little gas left within the central cavity for Danks\,1 and 2 to ionise and the radiation is free to escape. 

Ideally we would compare the observed ionising flux inferred from our 5.5\,GHz observations with the ionising flux expected from the high-mass stars identified in G305. However, the ionising flux produced by high-mass stars is difficult to determine accurately without modelling each star individually and the complex nature of the 5.5\,GHz emission makes it difficult to isolate the ionising radiation associated with individual groups of high-mass stars. Such modelling is far beyond the scope of this study but we estimate the ionising flux of Danks\,1 and 2 by realising that in practice the ionising flux of a cluster is dominated by the most high-mass stars. In the case of G305 this is given by the ionising flux of the three WNLh stars, D1-1, D1-2 and D1-3 (see Table\,2 in \citealt{Davies2012}). The ionising flux associated with these high-mass stars is estimated by utilising the Lyman flux estimates for similar WNLh stars in the Arches cluster scaled to the luminosities of these objects \citep{Martins2008}. The amount of ionising radiation emitted by the three WNLh stars in Danks\,1 alone is comparable to the total ionising flux within the entire G305 region estimated here \citep{Davies2012,Mauerhan2011}. This implies that G305 suffers a significant leakage of UV photons ($>50$\%) into the surrounding ISM. Hence, estimating the stellar contents and star formation rates of G305 and similar regions using radio fluxes \citep{Kennicutt1994, Kennicutt1998} alone will result in significant underestimates for all but the youngest complexes, where feedback has yet to disrupt the molecular material.

\section{Discussion}\label{Sect:discussion}
The radio spectral line and continuum observations presented in the previous sections allow us to trace the distribution of the molecular and ionised gas, on similar spatial scales, across the extent of the G305 complex. We have located 156 molecular clumps associated with G305 which have physical properties characteristic of other well known GMCs \citep{Roman2010,Rahman2010,Roman2010}. We have identified HII regions and extended ionised gas from which we have inferred the young embedded main-sequence massive stellar component of G305. Using these results in combination with tracers of ongoing star formation and identified high-mass stars, we are able to compare the molecular and ionised environment with star formation. We are therefore in a position to broadly trace the spatial and temporal distribution of high-mass star formation across the G305 complex and so infer the SFH of the region. In order to describe the various features in G305 we use source, clump and cloud to indicate the radio continuum sources in Table~\ref{Tab:LargeSources}, CO clumps in Table~\ref{Tab:COCFResults10} and the large scale \NH\ clouds from \cite{Hindson2010}.

\begin{figure*}
\centering

\subfigure[north-east]{
\includegraphics[width=0.49\textwidth]{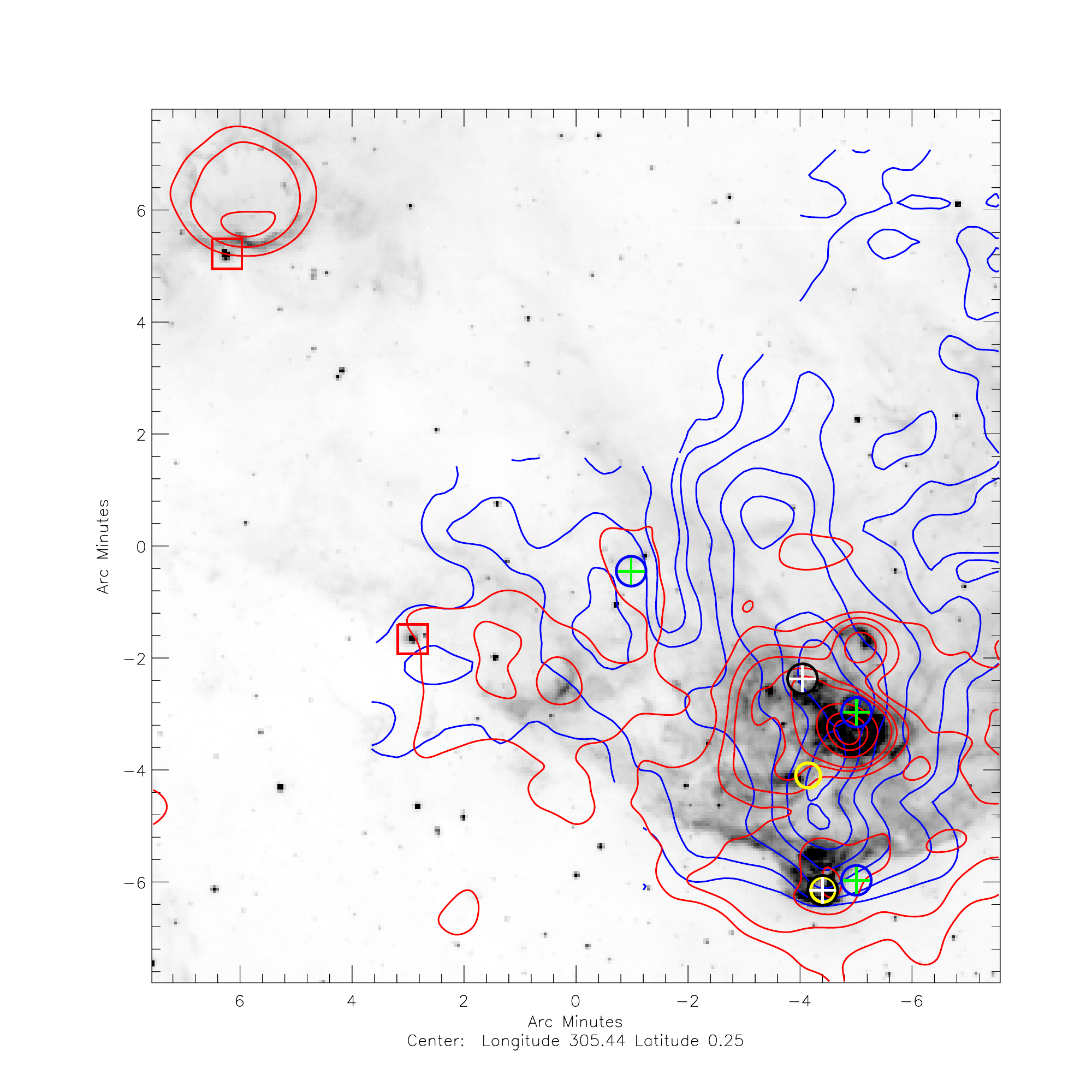}
   \label{fig:NE}
 }
 \subfigure[north-west]{
\includegraphics[width=0.49\textwidth]{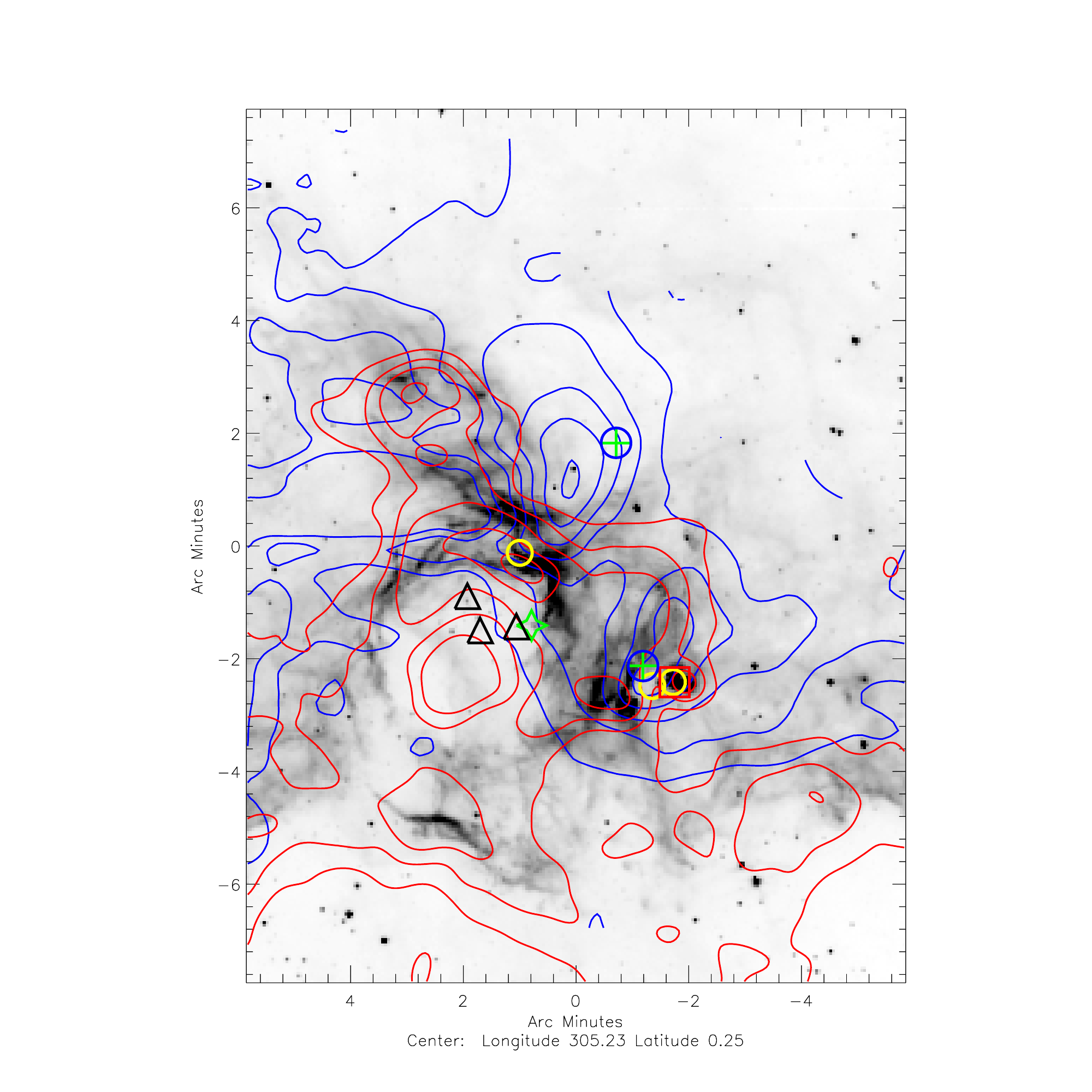}
   \label{fig:NW}
 }
   \subfigure[east]{
 \includegraphics[trim=0 60 0 60,width=0.49\textwidth]{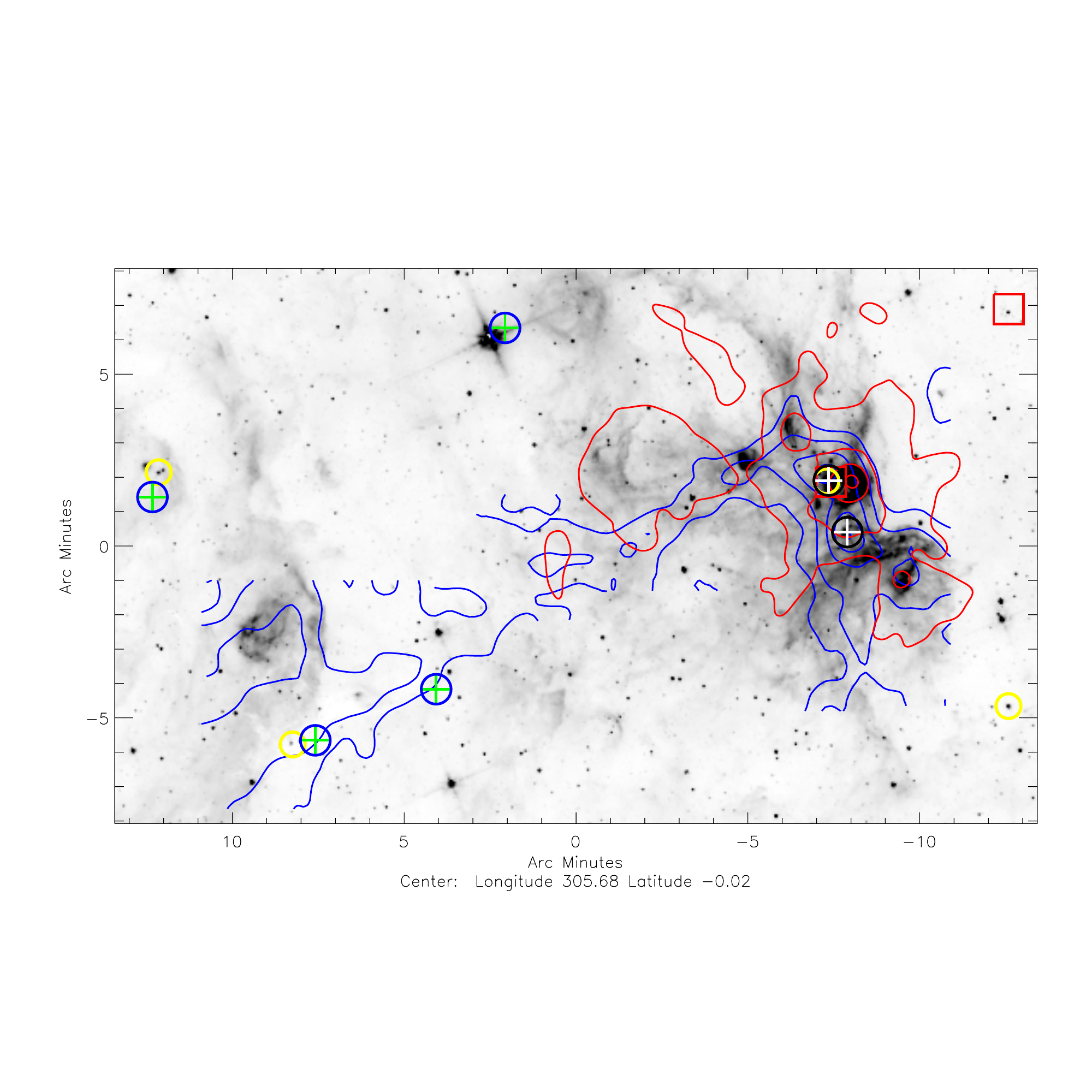}
   \label{fig:E}
 }
  \subfigure[south-west]{
 \includegraphics[trim=0 60 0 100, width=0.49\textwidth]{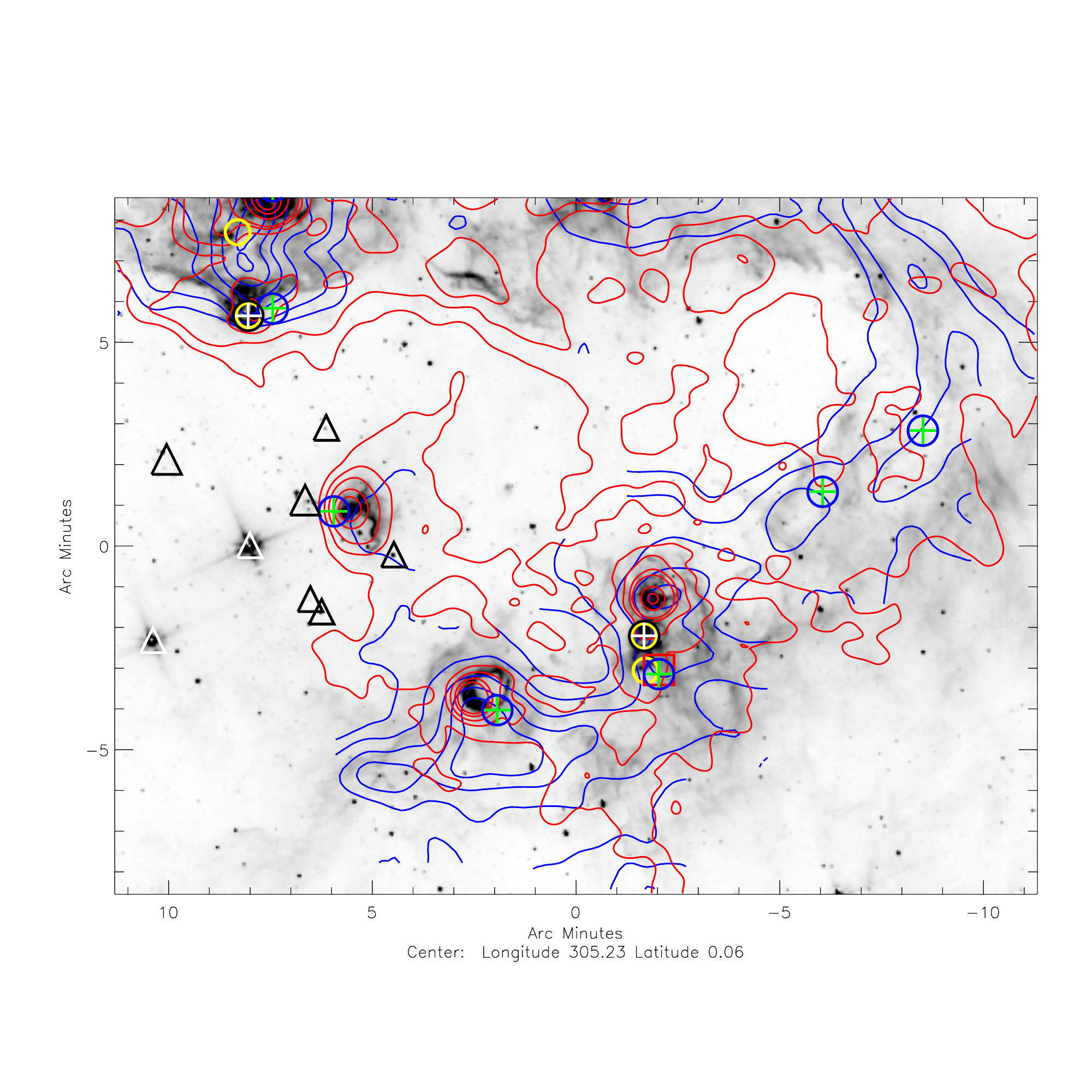}
   \label{fig:SW}
 }
\caption[Recent and ongoing tracers of star formation towards G305 sub-regions]{GLIMPSE 5.8\,$\umu$m grey scale background with 5.5\,GHz radio continuum and molecular gas traced by \COII\ presented as red and blue contours respectively with identical contour levels to Fig.~\ref{im:LargeScale} and Fig.~\ref{im:CO} bottom panels. The positions of identified high-mass stars are shown by black and white triangles depending on contrast. Note that Danks\,1 and 2 contain multiple high-mass stars which all fall within the area of a single triangle symbol. Black circles with white crosses indicate \mbox{UC\,HII} regions. Masers are shown as circles; Methanol = yellow, \water\ = blue (centred on green crosses) and red boxes indicate YSOs detected in the RMS survey.}
\label{Fig:comb}
\end{figure*}

\subsection{The molecular and ionised gas morphology and star formation}\label{Sect:Overview}

In Figure~\ref{Fig:comb} we present the molecular and ionised gas (blue and red contours respectively), optically identified high-mass stars and ongoing star formation towards subregions of G305 following the same approach as earlier work \citep{Hindson2010}. The molecular and ionised gas is clearly correlated; with diffuse ionised gas confined within the cavity to the north, south and west by the molecular gas. The HII regions embedded within the molecular clumps surrounding the cavity coincide with the peak CO emission. We find that the molecular clumps in G305 are being irradiated by strong UV radiation from both Danks\,1 and 2 and to a greater extent by HII regions embedded around the rim of the central cavity.

High-mass stars in G305 (Fig.~\ref{Fig:comb}; triangles) are concentrated towards Danks\,1 and 2 in the centre of G305 and a small cluster associated with the G305.254+0.204 HII region in the north west (Fig.~\ref{fig:NW}). The ongoing star formation towards G305 is indicated by: eight RMS YSOs (Fig.~\ref{Fig:comb}; red squares), 14 methanol masers and 15 \water\ masers (Fig.~\ref{Fig:comb}; yellow circles and blue circles centred on green crosses respectively) and five UC\,HII regions (Fig.~\ref{Fig:comb}; white crosses centered within black circles). This star formation in G305 is occurring at a projected distance from Danks\,2 of $\sim2$--35\,pc. The closest sign of ongoing star formation is an \water\ maser to the west of Danks\,2 that peaks at a velocity of $-42.7$\,\kms\ (\water\ maser\,8; see Table\,5 in \citealt{Hindson2010}). The broad velocity range over which \water\ masers emit makes it difficult to identify the corresponding molecular emission associated with the maser. We find that the most likely molecular clumps are 82 and 119 which are found at a velocity of $-45.1$ and $-40.1$\,\kms\ respectively. These two clumps are at $-7.7$ and $-2.7$\,\kms\ from the central velocity of G305 ($-37.4$\,\kms) and coupled with their low density (0.7 and 0.9\,\cmthree\  respectively) we propose that this site of star formation is unlikely to have survived in such close proximity to Danks\,1 and 2 and so is more likely to be located on the near side of the cavity. More than half ($\sim60$\%) of the ongoing star formation is located in a narrow ($\sim 2$\,pc) molecular region surrounding the central cavity at a projected distance from Danks\,2 of between 7 and 17\,pc. Star formation is concentrated around embedded HII regions and bright mid-infrared emission indicating that star formation is taking place in close proximity to high-mass stars still embedded in the molecular environment. The remainder of star formation is taking place towards the east of the complex, and unlike around the periphery of the cavity, it is dispersed over a wide area. In contrast, there is a distinct lack of ongoing star formation extending towards the west of the complex. We now describe each of the sub regions shown in Figure~\ref{Fig:comb} in more detail.

To the north-east (Fig.~\ref{fig:NE}) of Danks\,1 and 2 the molecular and ionised peak emission are superimposed. The ionised gas extends to the south into the central cavity whilst the molecular emission terminates with the 5.8\,$\umu$m emission highlighting an ionised boundary layer. We identify a density gradient in the molecular emission with the column density increasing away from the cavity boundary suggesting that the expansion of the central cavity is sweeping up the molecular material. Two HII regions (sources 1 and 6) are embedded within the molecular gas and are surrounded by signs of ongoing star formation. At the molecular, PAH and ionised gas interface towards the south of this region we detect two compact HII regions (sources 9--1 and 9--2), an \mbox{UC\,HII} region (source\,8 from \citealt{Hindson2012}), methanol and \water\ maser. Further to the north-east we find an HII region (source\,10) with a bubble morphology which may have triggered the YSO found towards its brightened edge \citep{Thompson2012}.

The HII region G305.254+0.204 in the north-west (source\,5, Fig.~\ref{fig:NW}) has cleared out a cavity ($R\approx 1.5$\,pc) in the surrounding molecular gas. The stars responsible for this HII region are most likely L05-A1, A2 and A3 \citep{Leistra2005}, a group of massive OB stars in the G$305.2+0.2$ cluster \citep{Dutra2003}. We identify a well-defined boundary between the ionised and molecular gas with a clear gradient in the molecular gas density increasing from the boundary. We interpret this as the result of expansion of the G305.254+0.204 HII region into the molecular gas which is compressing the molecular material at the interface. It is along this dense boundary that we find the highest concentration of star formation in this subregion. In addition to this apparently constructive feedback, we also see evidence for destructive feedback; the two largest molecular clouds in this sub region (clumps 1 and 11) have been split in two. The ionised gas associated with the expanding HII region appears to be driving a wedge between clump 1 and 11 extending through the molecular gas. There are no signs of star formation between these clumps suggesting the expansion of the HII region has terminated star formation.

To the east of the complex (Fig.~\ref{fig:E}) the powerful radiation and winds of high-mass stars towards the centre of the complex have completely blown out of the molecular confines of the cavity and dispersed into the wider ISM, leaving only a small pocket of over-dense gas near to the PDR (clump\,3). This clump is associated with the HII region source 7--1, which is in the process of splitting and dispersing the remaining gas and is associated with an \mbox{UC\,HII} region (source 7--2). This implies two things, first that in the past the molecular material in G305 was not homogenous, secondly some fraction of the high energy radiation originating from high-mass stars in the central cavity is free to escape the complex as we suggested in Section~\ref{Sect:Flux}. Ongoing star formation is concentrated around this HII region but also extends along the filament presented in Section~\ref{Sect:COResults} 

In the south-west of the complex (Fig.~\ref{fig:SW}) we identify a lobe of molecular gas that extends to the north-west and connects the south-west and north-west subregions. The south-west region is host to significant extended low surface brightness radio emission particularly towards the west. Two HII regions (sources 2 and 3) are detected in this region located towards the peaks of the molecular gas within the southern lobe. Ongoing star formation is concentrated towards these HII regions but two \water\ masers to the north-west of the HII regions reveals that low levels of star formation also extends to the north-west. It can clearly be seen that the molecular gas has been swept up at the cavity boundary and dispersed between these two HII regions. The source of the diffuse ionising radiation is most likely high-mass stars within the central cavity (Fig.~\ref{fig:SW}; black and white triangles) and demonstrates that the feedback from this population of older high-mass stars is still affecting the region.

\subsection{Triggering}\label{Sect:Triggering}
When studying GMCs, HII regions and ongoing star formation, the question of triggered star formation often arises. High-mass star(s) are thought to be able to promote star formation in nearby molecular material by driving winds and shocks into the gas (collect and collapse) and overrunning and compressing pre-existing condensations (radiative driven implosion). Subsequent gravitational instability and/or increase in external pressure collectively lead to triggered star formation \citep{Elmegreen1998,Zinnecker2007}. The result of triggered star formation should be a spatial ordering of sequentially younger generations of stars from the central high-mass star(s) that instigated the triggering cascade. However, proving triggering is the source of a given stellar population conclusively is problematic at best. Typical methods involve ``not disproving'' the hypothesis, often by assembling a consistent time line for the different phases of star formation and searching for morphological signposts such as star formation on the borders of expanding shells \citep{Deharveng2005,Urquhart2007, Thompson2012}. 

First, we consider the distribution of the four epochs of high-mass stars in G305 indicated by optically visible high-mass stars, classical, compact and UC HII regions. The oldest generation of high-mass stars (3--6\,Myr) is found in Danks\,1 and 2, a co-spatial distribution of visible high-mass stars within the centre of that complex and the cluster G$305.2+0.2$. The second and third generation is indicated by classical HII regions (0.4--1.8\,Myr) and compact HII regions (0.1--0.8\,Myr) surrounding the central cavity and embedded within molecular gas (Fig.~\ref{Fig:comb}). The fourth generation of high-mass stars is found around these HII regions in the form of \mbox{UC\,HII} regions and also 6.7\,GHz methanol masers ($<0.1$\,Myr). These four epochs of high-mass star formation are clearly separated both spatially and temporally, which suggests that high-mass star formation may have taken place in a number of distinct bursts rather than constantly over the lifetime of the complex. We note that two HII regions (G305.254+0.204 and G304.93+0.56) are isolated away from the main cavity. These HII regions could not of been affected by feedback from Danks\,1 and 2 which suggests that high-mass stars have also formed spontaneously within the complex. 

Next, we look to the morphology of the ionised and molecular gas and the location of star formation tracers within G305. Figure~\ref{Fig:comb} clearly shows that the ongoing star formation is almost exclusively associated with the ionised and molecular gas interface surrounding HII regions and the central cavity. This is where we would expect the expansion of HII regions to be triggering star formation through the collect and collapse and radiative driven implosion process. Such a distribution of the molecular and ionised gas and star formation is highly suggestive of triggered star formation via the collect and collapse process as proposed by \citet{Elmegreen1977}. A number of star-forming regions show similar distributions of star formation, HII regions and ionised and molecular gas to G305 such as NGC3603 \citep{Nurn2002}, 30\,Dor \citep{Walborn1997,Walborn2002} and to a lesser extent Carina \citep{Brooks2001}. In all these cases, the authors suggest that the distribution of star formation reflects the propagation of triggered star formation through the molecular cloud. However, the morphology and distribution of stars is not considered sufficient evidence to confirm triggering. The central problem of triggered star formation is the difficulty in identifying the origin of the discovered star formation. When attempting to identify star formation that has been triggered one must first exclude the possibility that the star(s) would have formed spontaneously without the influence of a trigger. Star formation located around an HII region for instance could have formed spontaneously and simply been uncovered by the expansion of the HII region. Additional evidence is required to support the morphological argument and confirm that a particular star's formation was induced and not spontaneous.  

The best model for triggered star formation in G305, given its age, size and morphology is the collect and collapse process. To explore the timescales involved in triggered star formation via the collect and collapse process we apply the theoretical approach of \cite{Dale2007} and \cite{Whitworth1994} and calculate the time it would take for a swept up shell of molecular gas to gain sufficient mass to become gravitationally unstable and fragment called the ``fragmentation timescale'' ($t_{\rm frag}$). In order for a region to be consistent with the collect and collapse scenario, the time from the formation of the ionising star(s) and turning on of the associated HII region to the gravitational instability and fragmentation of the swept up shell, plus the age of the triggered star formation population must be comparable to the total age of the triggering source. The fragmentation timescale is calculated by assuming a simple model of the gravitational stability of a uniform shocked shell driven by an HII region expanding in a medium of uniform density composed of pure hydrogen such that: 

\begin{equation}\label{Eq:tfrag}
t_{\rm frag}\sim 1.6\frac{a^{7/11}}{0.2}\frac{N_{\rm Ly}^{-1/11}}{10^{49}}\frac{n_3^{-5/11}}{10^3} \; [\rm Myr]
\end{equation}
 
\noindent where $a$ is the isothermal sound speed inside of the shocked layer in units of 0.2\,\kms, $N_{\rm Ly}$ is the Lyman continuum photon rate in s$^{-1}$ and $n_3$ is the initial gas atomic number density in units of $10^3$\,\cmthree. The fragmentation timescale is strongly dependent on the value of the natal ambient atomic density and to a lesser extent the Lyman continuum photon rate, unfortunately neither of these values are easily derived for dynamically evolving GMCs such as G305. The atomic number density assumed in \cite{Dale2007} is 200\,\cmthree, in Section~\ref{Sect:MolProp} we determine an average current molecular density of 1500\,\cmthree\ implying that the assumed atomic density may be low by a factor of 2.5. As we demonstrated in Section~\ref{Sect:Flux} the contribution to the Lyman continuum rate from Danks\,1 and 2 and high-mass stars within the central cavity is difficult to estimate. To obtain a rough estimate of the fragmentation timescale of Danks\,1 and 2 we note that in Danks\,1 the three WNLh stars have a Lyman continuum output proportional to the total Lyman continuum flux of the whole complex and so assume $N_{\rm Ly}\approx3.2\times10^{50}$ \citep{Davies2012,Clark2004}. Therefore, assuming the simple model of \cite{Dale2007} $a=0.2$\,\kms\ and $n_3=200$\,\cmthree\ the fragmentation timescale required for star formation to be triggered by Danks\,1 and 2 is $\sim 2.4\pm0.8$\,Myr. Comparing this to the dynamical age of the HII regions around the periphery of the central cavity (0.4--4.0\,Myr; Table~\ref{Tab:LargeSources}) and the age of Danks\,1 and 2 ($\sim 1.5^{+1.5}_{-0.5}$ and $\sim 3^{+3}_{-1}$\,Myr respectively) it is unlikely that the oldest HII regions on the periphery of the cavity could have been triggered by Danks\,1 and 2 via the collect and collapse process. However, it is possible that the younger episodes of star formation on the order of $10^5$\,yrs old could of been triggered by Danks\,1 and 2.

One of the strongest morphological cases for triggering in G305 is towards the G305.254+0.204 HII region (Fig.~\ref{fig:NW}) in the north-west. A number of authors \citep{Clark2004,Leistra2005,Longmore2007} have suggested that the over density of star formation tracers on the periphery of the HII region is indicative of triggered star formation. With an integrated Lyman continuum flux of $3.9\times10^{49}$\,s$^{-1}$ and using the same assumptions as the previous fragmentation timescale calculation for the ambient density and sound speed ($n_3=200$\,\cmthree\ and $a_3=0.2$\,kms) we find a fragmentation timescale of $t_{\rm frag}=2.9\pm 1.5$\,Myr. Considering the dynamical age of the G305.254+0.204 HII region is 4.0\,Myr and the star formation tracers detected along the periphery are $<0.6$\,Myr it is plausible that the expanding HII region could be responsible for triggering star formation. 

.  

Constraining the timescale involved in the collect and collapse process further would require more detailed modelling of G305, which would be problematic due to the clumpy, non-static and large-scale structure of the region. The high-mass stars in G305 have formed and evolved in an expanding, increasingly more massive giant shell of molecular material. This type of dynamically changing environment and its interaction with an evolving HII region is beyond the scope of current models. Given the inherent uncertainty in the calculation of the dynamical age and fragmentation timescale we suggest that these results are an indication that some star formation may have been triggered within G305.

\subsection{Star formation rate}\label{Sect:SFR}
The star formation rate (SFR) describes the rate at which mass in a GMC is converted into stars and along with the star formation efficiency (SFE) provides a fundamental physical parameter for characterising the evolution of star-forming regions and galaxies (see, e.g.\ \citealt{Kennicutt1998,Calzetti2009,Kennicutt2012} for reviews of SFR indicators). We have estimated the SFR of G305 from the oldest and most recent episodes of high-mass star formation; the central clusters Danks\,1 and 2 and extrapolation of the initial mass function (IMF) based on the most high-mass stars associated with \mbox{UC\,HII} regions (see Section\,5.4, \citealt{Hindson2012}). The radio continuum observations presented in this work allow us to apply the same method to determine the SFR of the compact and classical HII regions and so broadly trace how the SFR has evolved. 

In \cite{Hindson2012} we extrapolate the IMF from estimates of the most high-mass stars associated with \mbox{UC\,HII} regions in G305 and use this to determine a lower limit to the recent ($<10^5$ yr) SFR of 0.002--0.004\,\msun\,yr$^{-1}$. We discuss the limitations of this approach in detail in \cite{Hindson2012}. To briefly summarise this method is dependant on the chosen IMF, in this case Salpeter, and is only constrained by the massive stellar component that we detect. Thus we are forced to extrapolate the number of low mass stars and so the total and average mass determined by the IMF fit is inevitably associated with at most an order of magnitude error. We perform these steps to derive the SFR based upon the older population of high-mass stars indicated by compact and classical HII regions presented in this study (Table~\ref{Tab:LargeSources}). Extrapolation of the IMF reveals $1.95\times10^3$ and $6.34\times10^3$ stars or YSOs associated with compact and classical HII regions respectively. This corresponds to a total stellar mass of $0.67\times10^4$ and $2.17\times10^4$\,\msun\ for stars associated with the compact and classical HII regions, comparable to the combined mass of Danks\,1 and 2 (\citealt{Davies2012}; $1.10\times10^4$\,\msun). To determine the SFR across the epochs of star formation in G305 in a consistent manner we assume that the stars in each epoch have formed over a period of 0.5\,Myr \citep{Hindson2012}. We find a SFR based on compact and classical HII regions of $0.013$ and $0.043$\,\msun\,yr$^{-1}$ respectively (Table~\ref{Tab:SFR}). The main caveat is that this method is sensitive to the mass derived for the most high-mass stars which is fixed by the ionising flux derived from the radio continuum flux in Section~\ref{Sect:Lyman}. Specifically, the error associated with the number of stars derived from classical HII regions is difficult to quantify due to the complex morphology of the radio emission and possibility of photon leakage and so we underestimate the low mass YSO contribution to the IMF. The compact and \mbox{UC\,HII} regions may not be completely sampled and flux estimates may be underestimated due to dust absorption. We therefore suggest that the SFR derived above are accurate to approximately a factor of 10.

If we include the YSOs detected in \cite{Faimali2012} ($0.8\times10^4$\,\msun) the combined mass in stars for the entire G305 complex is $4.89\times10^4$\,\msun, these stars have formed over the last 3--6\,Myr resulting in an average SFR over the lifetime of the complex of 0.008--0.016\,\SFR. This result is comparable to the SFR of Carina (\citealt{Povich2011}; $>0.008$\,\SFR) and M17 (\citealt{Chomiuk2011}; $0.005$\,\SFR), however unlike Carina in which the SFR appears relatively constant we find evidence a variable SFR in G305. These results are also in good agreement with the SFR derived from the dense gas detected by Hi-Gal of 0.006--0.02\,\SFR\ \citep{Faimali2012}. A number of other methods are commonly used for determining the SFR in molecular clouds; we explored a number of these presented in Table\,3 of \cite{Faimali2012} we discuss and refine three of these methods below.

\begin{table}
 \caption[]{The SFR for the four epochs of high-mass star formation in G305. For Danks\,1 and 2 we use the mass estimates from \cite{Davies2012} and for the HII regions we derive the mass in stars by by fitting the IMF to the most high-mass star (Table~\ref{Tab:LargeSources}). To determine the SFR we assume that star formation took place over 0.5\,Myr.}
\centering
  \begin{tabular}{lccc}
  \hline
  Tracer & Epoch & Total Stellar  & SFR\\
         & 	(Myr)	 & Mass 10$^4$(\msun)   & (\msun\,yr$^{-1}$) \\
         \hline
 Danks\,1 \& 2	&	3.0--1.5	&	1.10	&	0.022	\\
 Classical HII	&	0.4--4.0	&	2.17	&	0.043	\\
 Compact HII	&	0.2--0.8	&	0.67	&	0.013	\\
 UC HII			&	$<0.1$		&	$>0.15$	&	$>0.003$	\\
   \hline 
 \end{tabular}
 \label{Tab:SFR}
 \end{table}	

A linear relationship has been found between the SFR and gas surface-density in molecular clouds above a local extinction threshold of $A_{\rm K}\sim 0.8$ magnitudes, corresponding to a gas column-density of $\sim 7.4\times 10^{21}$ hydrogen molecules per cm$^{-2}$ or a gas surface-density of $\sim 116$\,\msun\,pc$^{-2}$ \citep{Lada2010}. This relationship between molecular gas surface density and SFR ($\dot{M}_{*_{\rm mol}}$) is given by:

\begin{equation}\label{Eq:MolSFR}
\dot{M}_{*_{\rm mol}}=4.6\pm2.6\times10^{-8}M_{0.8}  \; [\rm M_\odot\,yr^{-1}]
\end{equation}

\noindent where $M_{0.8}$ corresponds to the cloud mass that is above the extinction threshold of $A_{\rm K}=0.8$. We apply Equation~\ref{Eq:MolSFR} to derive the SFR for G305 based on the gas component above this threshold in two ways. First, we apply the surface density threshold ($\sum_{\rm c}\approx116$\,\msun\,pc$^{-2}$) to the CO clump catalogue and find that 59\% of the detected CO clumps exceed this threshold. The total CO derived mass above a surface density of $\sim 116$\,\msun\,pc$^{-2}$ is $M_{0.8}=2.9\pm0.2\times10^5$\,\msun\ which leads to a SFR of $\dot{M}_{*_{\rm CO}}=0.013\pm0.007$\,\msun\,yr$^{-1}$. Second, assuming that this relationship applies to densities of $n_{\rm H_2}\approx 10^4$\,\cmthree\ means that all \NH\ emission, which has a critical density of approximately $\sim10^4$\,\cmthree and is not expected to be sub-thermally excited, should be above this threshold. Therefore the SFR based on \NH, where $M_{0.8}=6.5\pm3\times10^5$\,\msun, is $\dot{M}_{*_{\rm NH^3}}=0.03\pm0.02$\,\SFR. For the SFR based on the molecular material we take the range of these two values to arrive at $\dot{M}_{*_{\rm mol}}=0.02\pm0.01$\,\msun\,yr$^{-1}$. It should be stressed that this SFR relationship is averaged over a timescale of $2\pm1$\,Myr, the age spread of low-mass YSOs (M$\oldle 0.5$\,\msun) in the local cloud sample of \cite{Lada2010} and so will overestimate the SFR in G305. This global SFR is consistent with the SFR derived by the IMF fitting method above.

Alternatively, it has been proposed that the SFR density is linearly proportional to the mean gas density divided by the free-fall time multiplied by an efficiency factor ($\eta$) estimated to be about 1\% \citep{Krumholz2012}. To derive the SFR given this relation we apply equation 10 of \cite{Longmore2013}:

\begin{equation}
\dot{M}_{*}=\eta M_{\rm g}/\tau_{\rm ff}=0.53\left ( \frac{n}{8000\,{\rm cm^{-3}}} \right )^{1/2}  \; [\rm M_\odot\,yr^{-1}]
\end{equation}

\noindent where the SFR in a given volume of gas per free-fall time ($\tau_{\rm ff}$) is determined by the mass of molecular gas ($M_{\rm mol}$) in that volume and a global efficiency of gas converted to stars per free-fall time ($\eta$) of 1\%  ($\eta=0.01$). For the average molecular density in G305 of $1.5\times10^3$\,\cmthree\ this gives an average SFR for G305 of 0.2\,\msun\,yr$^{-1}$. This is a factor of ten greater than the SFR estimated from the molecular emission and the IMF fitting techniques described above.  Using the volumetric star formation relation \cite{Longmore2013} derived a SFR of 0.4 for the central molecular zone (CMZ) of the Galaxy twice that of G305. The reason why the SFR derived by this method is so high is unclear but may be associated with the uncertainty in deriving the free-fall time. The free fall time is likely to be underestimated due to the complex nature of the structure and density in molecular clouds and assumption of a quasi-static state. In addition the virial parameter indicates much of the molecular gas in G305 is not undergoing collape and there is an inherent uncertainty in deriving the molecular gas density.

The SFR may also be independently estimated using the Lyman continuum photon rate derived from radio continuum observations.  According to \cite{Kennicutt1994} and \cite{Kennicutt1998}, a SFR of 1\,\msun\,yr$^{-1}$ produces a Lyman continuum photon rate of $N_{\rm Ly}=9.26\times10^{52}$\,photons\,s$^{-1}$ for the Salpeter IMF (assuming a mass range of 0.1--100\,\msun). It then follows that the SFR based on the Lyman continuum photon rate is:

\begin{equation}\label{Eq:IonSFR}
\dot{M}_{*_{\rm N_{ly}}}=1.08\times10^{-53}N_{\rm Ly}  \; [\rm M_\odot\,yr^{-1}]
\end{equation}

The integrated flux density of G305 at 5\,GHz, based on single dish measurements, is approximately 200\,Jy \citep{Clark2004}, therefore the corresponding total Lyman continuum rate given by Equation~\ref{Eq:Lyman} is $N_{\rm Ly}=2.61\times10^{50}$\,photons\,s$^{-1}$. Applying this Lyman continuum rate to Equation~\ref{Eq:IonSFR} yields a SFR based on the ionised gas of $\dot{M}_{*_{\rm N_{ly}}}\approx0.002$--$0.005$\,\msun\,yr$^{-1}$ similar to the SFR derived in the same way for M17 \citep{Chomiuk2011} and the CMZ \citep{Longmore2013}. The SFR derived in this way should be interpreted as the continuous SFR required to maintain a steady state population (i.e.\ number of ionising stars born equals deaths) of ionising stars that produce the ionising flux rate. This means that the timescale assumed in Eq.\,~\ref{Eq:IonSFR} is in excess of the lifetime of the ionising stars. For a late O star this implies a time scale of $\sim 8$\,Myrs whilst for an early O star the time scale is $\sim 5$\,Myrs \citep{Bertelli1994,Martins2005}. For G305 this steady state clearly does not apply, the oldest cluster Danks\,2 is only 3$^{+3}_{-1}$\,Myrs old so the SFR derived by Eq.\,~\ref{Eq:IonSFR} is an underestimate. In addition there is a high level of photon leakage towards G305 (see Section~\ref{Sect:Flux}) and dust absorption that implies the Lyman continuum derived SFR may be underestimated by as much as an order of magnitude.

\subsection{Star and clump forming efficiencies}\label{Sect:Eff}
Evidence suggests that star formation is an inherently inefficient process; in the Milky Way and normal external galaxies the percentage of molecular mass transformed into stars is observed on the order of $\sim1\%$ whilst GMCs tend to have slightly higher SFEs of 2--17\% \citep{Williams1997,Myers1986,Evans2009}. Our observations reveal the total molecular mass of G305, which in combination with estimates of the total mass in stars allows us to probe the SFE of the region.

The combined mass of stars in Danks\,1 and 2 is $1.1\times10^4$\,\msun\ \citep{Davies2012}, whilst the mass of stars associated with compact and classical HII regions is $0.67$ and $2.17\times10^4$ respectively (Section~\ref{Sect:SFR}) and UC HII regions is $0.15\times10^4$\,\msun\ \citep{Hindson2012}. Finally, we include the contribution from YSOs associated with embedded high-mass star forming region $0.8\times10^4$\,\msun\ derived from Herschel observations \citep{Faimali2012}. This gives a total mass in stars of $4.89\times10^4$\,\msun. With a  mass in molecular gas of $>3.5\times10^5$\,\msun\ we are in a position to estimate the SFE via:

\begin{equation}
SFE=\frac{M_{\rm *}}{M_{\rm GMC}+M_{\rm *}}
\end{equation}

\noindent where $M_{\rm *}$ is the total mass in stars and $M_{\rm GMC}$ is the total molecular gas mass expressed in solar mass units. We find that the SFE of G305 is 7--12\% in good agreement with the SFE of the Milky Way. 

The clump formation efficiency (CFE) is a measure of the fraction of molecular gas that has been converted into dense clumps. This quantity is analogous (or a precursor) to the SFE and provides an upper limit:

\begin{equation}
CFE=\frac{M_{\rm dense}}{M_{\rm dense}+M_{\rm diffuse}}
\end{equation}

With an effective critical density of $\sim 10^4$\,\cmthree\ \citep{Ho1983} the \NH\ emission presented in \cite{Hindson2010} is an excellent tracer of the dense gas component on the scale of molecular clouds. The CO emission presented in this work has a lower critical density $\sim 10^2$\,\cmthree\ \citep{Evans1999} and tends to freeze-out onto dust grains in the coldest and most dense regions of molecular clouds \citep{Bacmann2002} and therefore preferentially traces the diffuse gas in GMCs. Also the higher resolution of the CO observations traces emission on the scale of clumps ($<3$\,pc). We determine the positional dependant CFE for G305 by cross matching the mass of dense and diffuse gas derived in the \NH\ and CO analysis respectively. This is achieved by deriving the average mass of the \COII\ clumps within the bounds of the \NH\ clouds along the line of sight using the global \COII\ column density derived using Eq.\,~\ref{Eq:NCO}.

The CFE within G305 ranges from 13\% to 37\% with an average of 24\%. The Global CFE of G305 is derived from the CO and \NH\ mass derived for the whole complex; $3.5\times10^5$ and $6.5\times10^5$\,\msun\ respectively. This gives a CFE of $65\%$, however this is considered an upper limit due to the incomplete mapping in CO. Errors in the CFE originate from the uncertainty on the mass estimates in the CO and \NH\ studies. We suggest that the various uncertainties associated with the \NH\ and CO derived mass and complications with the comparison of the two studies leads to uncertainties in the CFE on the order of 50\%. However, the relative positional CFE within the complex should be comparable because any variations, with the exception of abundance ratios, ought to be constant across the complex.

The CFE derived for G305 is much higher than the Galactic SFE but similar to the CFEs estimated in a number of star-forming regions such as; 11--29\% for the W51 GMC \citep{Parsons2012},  20\% for the entire W43 star-forming complex \citep{Nguyen2011}, $\sim 1$--2\% in quiescent gas and $\sim 20$--$40$\% toward regions of star formation \citep{Polychroni2012,Moore2007}. It has been suggested that the CFE towards W43 shows a significant increase in the CFE towards HII regions from 3.6 up to 55\% \citep{Eden2012}.  

Analysis of the molecular clumps presented in Section~\ref{Sect:MolProp} reveals that 31\% of the CO molecular clumps in G305 are gravitationally bound with masses exceeding the virial mass. It is therefore more informative to report the clump formation efficiency of bound clumps (BCFE), as this is more likely to reflect how the SFE will evolve over time. We estimate the BCFE using:

\begin{equation}
BCFE=\frac{M_{\rm \alpha<1}}{M_{\rm dense}+M_{\rm diffuse}}
\end{equation}

where the total mass in bound CO clumps is $M_{\rm \alpha<1}=1.8\times10^5$\,\msun. We find that for the G305 complex the BCFE is $\sim20\pm10$\%. The BCFE is comparable to the CFE reported for the regions described above and comparable to the SFE of molecular clouds within the Milky Way ($\sim 2$--$17\%$ \citealt{Williams1997}). However, we have only considered thermal support when deriving the virial mass, it is likely that many clumps are experiencing significant external pressure from the ionising radiation and winds which will help them to overcome thermal support and induce collapse. We therefore consider the BCFE a lower limit.

\subsection{The star formation history}

Our analysis and discussion provides the information necessary to broadly describe the SFH of the G305 complex. Initially the dense central region of the natal G305 GMC collapsed to form Danks\,2 $\sim 3^{+3}_{-1}$\,Myr ago and then shortly after Danks\,1 $\sim 1.5^{+1.5}_{-0.5}$\,Myr ago \citep{Davies2012}. We find a diffuse field of high-mass stars surrounding Danks\,1 and 2 (Fig.~\ref{fig:SW}), which may have formed in-situ buy could also be runaway stars ejected from Danks\,1 and 2. Some time shortly after the formation of Danks\,2 the G305.2+0.2 cluster formed in the north-west. For the last $\sim$3--6\,Myr the powerful UV radiation and winds from these high-mass stars have swept up and dispersed the surrounding molecular material into the morphology we see today. The molecular material has been dispersed more rapidly in the east and to some extent the west of the complex, which suggests that the natal molecular density of gas was lower in these regions than in the north and south of the complex. Observations to date see no signs of ongoing star formation within the bounds of the central cavity driven by Danks\,1 and 2 which suggests feedback appears to have cut-off present day star formation within the direct vicinity of Danks\,1 and 2 and G305.2+0.2. Shortly after the formation of Danks\,2 a second generation of high-mass stars have formed, apparent by the presence of classical HII regions around the cavity with dynamical ages $<2$\,Myr. The inherent uncertainty in the dynamical age estimate for these HII regions leads to some ambiguity. This generation of high-mass stars could have been triggered by Danks\,1 and 2, however there is some overlap between the dynamical age of the oldest HII regions and Danks\,1 and 2. High-mass star formation has taken place away from the main complex at a projected distance from Danks\,2 of 23 to 42\,pc (G305.532+0.348 and G304.929+0.552 respectively). Star formation occurring at this distance from Danks\,1 and 2 could not have been affected by the winds and radiation of the massive clusters suggesting that some fraction of the high-mass stars in G305 formed spontaneously. It is around HII regions on the periphery of the central cavity that we find evidence of a third generation of high-mass stars in compact HII regions and around these we identify the fourth and most recent episodes of high-mass star formation traced by \mbox{UC\,HII} regions and masers $<10^5$\,yrs. Thus, the morphological evidence suggests that the star formation in G305 is multi-generational and has at least partly occurred in spatially and temporally isolated bursts that may have been triggered. However, the SFR derived by fitting the IMF to the most high-mass stars associated with these four epochs of star formation is approximately constant given the errors. Further work is required to more accurately derive the SFR of these epochs of star formation to determine if the SFR in G305 was constant or varied. 

\begin{figure*}
\begin{center}
\includegraphics[width=0.9\textwidth]{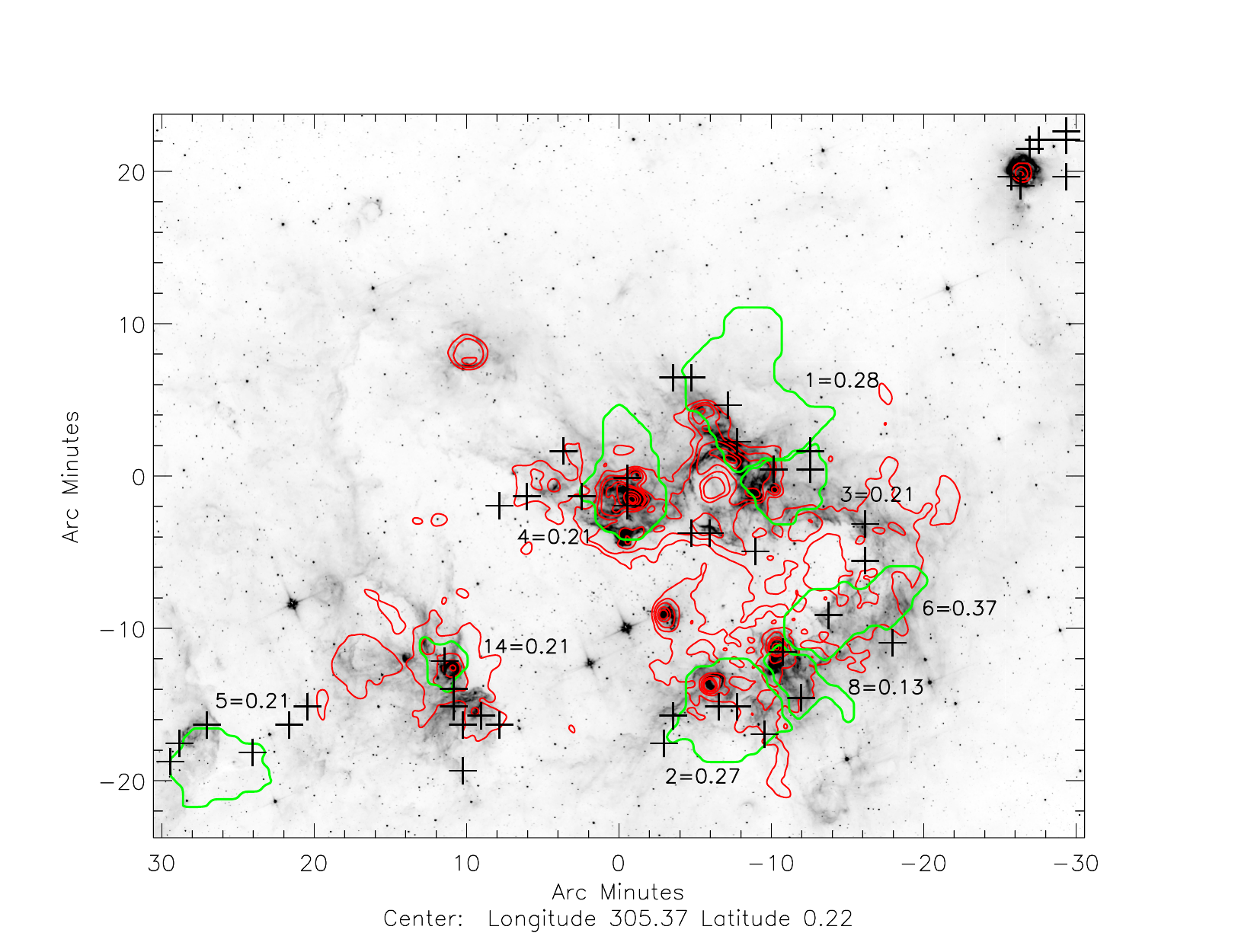}
\end{center}
\caption[CFE and bound clumps]{The 5.5\,GHz radio emission and the clump forming efficiency are shown by red and green contours with gravitationally bound CO clumps indicated by black crosses. The \NH\ cloud numbers from \cite{Hindson2010} are presented as well as the corresponding value of the CFE.}
\label{im:CFE-Bound}
\end{figure*}

\subsection{Future star formation in G305}\label{Sect:SFH}

We now speculate how the G305 complex may evolve in the future. Observations of the molecular and ionised gas suggest that the energy input into the region by Danks\,1 and 2 is no longer the sole or dominant driving force behind the evolution of the complex (Section~\ref{Sect:Flux}). Instead, the compact and classical HII regions surrounding the central cavity now dominate the observable Lyman continuum photon flux affecting the molecular gas and are the principal source of feedback driving the evolution of the complex. Despite this strong and seemingly continuous feedback from numerous high-mass stars over the last 3--6\,Myr there is still more than $10^{5}$\,\msun\ of cool ($\sim 20$\,K) molecular material within which star formation is taking place (Fig.~\ref{Fig:comb}). 

To examine where star formation may occur in the future we first consider the dynamical stability of the molecular clumps in G305. Analysis of the CO emission presented in Section~\ref{Sect:MolProp} reveals that 31\% of the molecular clumps detected in the G305 complex are gravitationally unstable and therefore likely to collapse. These clumps make up $\sim60$\% of the total CO mass ($1.8\times10^5$\,\msun). The location of these bound clumps can be seen in Figure~\ref{im:CFE-Bound} as black crosses and are found to reside on the borders of HII regions and are almost exclusively associated with ionised regions. If high-mass star formation takes place in these clumps, be it spontaneous or triggered, G305 will be host to a fifth generation of high-mass stars that will strengthen the morphological evidence for triggered star formation. 

In Section~\ref{Sect:Eff} we derive the CFE of the large molecular clouds in G305 (Fig.~\ref{im:CFE-Bound}). One might expect that if expanding HII regions are sweeping up molecular material the CFE would increase but we see no clear evidence of this being the case in G305. We do note that the \NH\ Cloud\,6 is in close proximity to the western wall of the central cavity, has the highest CFE and yet shows little sign of ongoing star formation or bound clumps. HII regions that surround the central cavity are clearly blowing away what remains of the molecular material in G305, leaving only the densest molecular clumps. This destruction of the molecular gas is apparent in a number of regions such as the southern lobe where the HII region G305.195+0.033 appears to be destroying the molecular gas associated with \NH\ Cloud\,8 lowering the CFE to 0.13 (Fig.~\ref{im:CFE-Bound}) and excavating a hole in the southern lobe of gas. 

The age and morphology of G305 suggests the complex is in an advanced stage of evolution and may be going through the final episodes of star formation before the molecular cloud is completely destroyed or rendered unrecognisable by feedback from high-mass stars. Given that the star formation efficiency in G305 is 6--12\% and the molecular mass is 3.5--$6.5\times10^5$\,\msun\ we expect approximately 0.2--$7.8\times10^4$\,\msun\ of molecular material to go on to form stars in the future. Given the mass in Danks\,1 and 2 and the mass of stars inferred from the IMF fitting technique to HII regions and embedded star-forming regions is $4.89\times10^4$ the end product of star formation in G305 may be a loosely bound OB association \citep{Zinnecker2007} with a mass of 0.5--$1.3\times10^5$\,\msun. Given the average SFR of G305 is 0.007-0.014\,\SFR\ we can compute a depletion time for the cloud (e.g \citealt{Evans2009}). If we assume that 0.2--7.8$\times10^4$\,\msun\ of gas will form stars and that star formation will continue at the average SFR (0.007-0.014\,\SFR) G305 would exhaust all its available molecular gas in 0.2-11.1\,Myr. It is unlikely that G305 will survive for a sufficient length of time for all the molecular gas to be consumed in star formation. It appears that star formation in G305 will cease in the next 2--3\,Myr when high-mass stars in Danks\,2 end their lives as supernova or when the high-mass stars embedded around the cavity blow away the remaining molecular gas via HII regions and winds.

The turbulent core model proposed by \cite{Krumholz2008} predicts that the surface density threshold necessary for the formation of high-mass stars of 10--200\,\msun\ is 1\,g\,cm$^{-2}$. Below this threshold the molecular gas is thought to fragment into much lower masses. By converting the range of CO derived surface densities presented in Table~\ref{Tab:COProp} to units of g\,cm$^{-2}$ we are able to test this model and determine if molecular gas in G305 is suitable for high-mass star formation as proposed by \cite{Krumholz2008}. We find that the CO surface density lies between 0.01 and 0.2\,g\,cm$^{-2}$ with a mean of 0.04\,g\,cm$^{-2}$. Based on the 1\,g\,cm$^{-2}$ threshold none of the CO clumps in G305 are capable of forming high-mass stars. This is clearly not the case, not only do we identify gravitationally unstable clumps, in Section~\ref{Sect:Overview} we report on the presence of 14 methanol masers that are a clear indication of ongoing high-mass star formation. It is possible that on scales smaller than the resolution of the CO and \NH\ observations the surface density exceeds this critical threshold but we are unable to resolve such dense gas due to beam dilution. Higher resolution observations of the dense gas and dust are required to probe the substructure and determine if the environment is conducive to high-mass star formation based on the surface density threshold presented by \cite{Krumholz2008}. This trend of surface densities below the 1\,g\,cm$^{-2}$ threshold has been reported in a number of studies using a range of instruments \citep{Parsons2012} which suggests either a fundamental problem with the theory or limitation of the observational approaches.

\section{Summary}\label{Sect:summary}
This paper presents the most detailed observations of the molecular and ionised emission in G305 to date. We trace molecular gas in G305 via observations of \COI, \COII\ and \COIII\ in the $J=1$--$0$ transition using Mopra and identify discrete clumps using the {\sc CLUMPFIND} algorithm. We identify 156 molecular clumps within the observed region of G305. The clumps have characteristic diameters of 1.4\,pc (0.68--3.0\,pc), excitation temperatures of 13\,K (7.1--24.6\,K), column densities of $0.9\times10^{22}$\,\cmthree\ (0.13--4.04$\times10^{22}$\,\cmthree), surface densities of 0.04\,g\,cm$^{-2}$ (0.01--0.2\,g\,cm$^{-2}$) and a total mass of 3.14$\times10^{5}$.

Our observations of the 5.5 and 8.8\,GHz observations reveal the ionised gas towards G305 and again by using {\sc CLUMPFIND} we are able to separate this complex emission into discrete sources. By comparing the detected radio sources to previous studies and three colour mid-infrared image we are able to identify nine HII regions, seven compact HII regions, one \mbox{UC\,HII} region and four extended features. All of which are projected against extended low surface brightness ionised gas. We determine the dynamical age, spectral type and Lyman continuum flux associated with UC, compact and classical HII regions. We find dynamical ages of 0.1--4.0\,Myr, spectral classes ranging from B0--O5 and Lyman fluxes of Log 46.94--49.67\,s$^{-1}$. By considering the distribution of the ionised emission we find that Danks\,1 and 2 are no longer the principal source of observable ionising radiation but rather the embedded HII regions are driving the ionisation of the molecular gas.

Comparison between the CO and radio continuum observations presented here and high-mass stars and ongoing star formation identified in the literature has allowed us to explore the role of triggering in the evolution of G305 and broadly trace the SFR, SFE and SFH of the complex. We find strong morphological evidence that the collect and collapse process is responsible for triggering at least some of the star formation in G305. We find additional evidence of triggering in G305 by applying a simple theoretical model of the fragmentation timescale involved in the collect and collapse process and exploring the velocity distribution of the molecular gas. By fitting a Salpeter IMF to the most high-mass stars inferred from the Lyman continuum flux we derive the total mass in stars corresponding to HII regions. From this we are able to trace the SFR across four epochs of star formation and find a SFR of 0.013 and 0.043 for compact and classical HII regions respectively. The total mass of stars in the complex is $4.89\times10^4$\,\msun, with a total molecular mass of 3.5--$6.5\times10^5$ we derive a SFE for G305 of 7--12\%. Finally we describe a possible star formation history of the complex in which star formation has occurred in both spontaneous and triggered modes.

\section*{Acknowledgments}
We would like to thank the director and staff of the Paul Wild Observatory for their assistance with a pleasant and productive observing run. The Mopra telescope is part of the Australia Telescope which is funded by the Commonwealth of Australia for operation as a National Facility managed by CSIRO. The University of New South Wales Digital Filter Bank used for the observations was provided with support from the Australian Research Council. This research has made use of the NASA/ IPAC Infrared Science Archive, which is operated by the Jet Propulsion Laboratory, California Institute of Technology, under contract with the National Aeronautics and Space Administration. This paper made use of information from the Red MSX Source survey database at www.ast.leeds.ac.uk/RMS, which was constructed with support from the Science and Technology Facilities Council of the UK.

\section{Appendix}

  \setcounter{table}{2}
\begin{table*}
 \caption[\COII\ emission properties]{The observed properties of the 156 detected $^{13}$CO clumps reported by {\sc CLUMPFIND}.}
\centering
\begin{tabular}{cccccccccccc}
   \hline 
	Clump & Clump & \multicolumn{3}{c}{Peak} & \multicolumn{4}{c}{Dimension}  &  Peak \\\cline{3-5} \cline{6-9}
	Number  & Name  &        \multicolumn{2}{c}{Galactic}  &$V_{\rm LSR}$         & $\Delta l$ & $\Delta b$ &$R_{\rm eff}$& $\Delta V$ &          $T_{\rm b}$       \\
	       &        &        $(l)$ & $(b)$  & (\kms)& (\arcsec) & (\arcsec) & (pc)& (\kms)  & (K)\\ 
\hline
1	&	G305.24+0.26	&	305.24	&	0.26	&	-38.7	&	123.4	&	157.46	&	2.84	&	5.66	&	15.38	\\
2	&	G304.94+0.55	&	304.94	&	0.55	&	-35	&	97.55	&	112.86	&	1.98	&	1.94	&	14.29	\\
3	&	G305.56+0.02	&	305.56	&	0.02	&	-39.6	&	112.39	&	83.02	&	1.97	&	3.04	&	13.55	\\
4	&	G305.26-0.03	&	305.26	&	-0.03	&	-32.7	&	136	&	184.61	&	3	&	4.27	&	13.17	\\
5	&	G304.93+0.54	&	304.93	&	0.54	&	-34.6	&	103.69	&	121.68	&	2.11	&	2.19	&	12.69	\\
\end{tabular}
 \end{table*}	
 
   \setcounter{table}{2}
\begin{table*}
\centering
  \begin{tabular}{cccccccccccc}
   \hline 
	Clump & Clump & \multicolumn{3}{c}{Peak} & \multicolumn{4}{c}{Dimension}  &  Peak \\\cline{3-5} \cline{6-9}
	Number  & Name  &        \multicolumn{2}{c}{Galactic}  &$V_{\rm LSR}$         & $\Delta l$ & $\Delta b$ &$R_{\rm eff}$& $\Delta V$ &          $T_{\rm b}$       \\
	       &        &        $(l)$ & $(b)$  & (\kms)& (\arcsec) & (\arcsec) & (pc)& (\kms)  & (K)\\ 
\hline
50	&	G305.18-0.02	&	305.18	&	-0.02	&	-32.3	&	71.54	&	86.29	&	1.44	&	3.14	&	6.43	\\
51	&	G305.17-0.02	&	305.17	&	-0.02	&	-30	&	76.32	&	62.27	&	1.24	&	2.65	&	6.36	\\
52	&	G305.32-0.07	&	305.32	&	-0.07	&	-33.6	&	71.97	&	50.14	&	1	&	1.58	&	6.33	\\
53	&	G305.1+0.09	&	305.1	&	0.09	&	-35.9	&	63.04	&	72.19	&	1.33	&	2.7	&	6.29	\\
54	&	G305.32+0.28	&	305.32	&	0.28	&	-42.3	&	103.22	&	98.69	&	1.97	&	2.96	&	6.29	\\
\end{tabular}
 \end{table*}	
 
   \setcounter{table}{2}
\begin{table*}
\centering
  \begin{tabular}{cccccccccccc}
   \hline 
	Clump & Clump & \multicolumn{3}{c}{Peak} & \multicolumn{4}{c}{Dimension}  &  Peak \\\cline{3-5} \cline{6-9}
	Number  & Name  &        \multicolumn{2}{c}{Galactic}  &$V_{\rm LSR}$         & $\Delta l$ & $\Delta b$ &$R_{\rm eff}$& $\Delta V$ &          $T_{\rm b}$       \\
	       &        &        $(l)$ & $(b)$  & (\kms)& (\arcsec) & (\arcsec) & (pc)& (\kms)  & (K)\\ 
\hline
101	&	G305.28+0.37	&	305.28	&	0.37	&	-33.2	&	74.69	&	47.61	&	1.07	&	3.07	&	3.95	\\
102	&	G305.54-0.05	&	305.54	&	-0.05	&	-39.1	&	49.46	&	83.47	&	1.08	&	1.16	&	3.88	\\
103	&	G305.11+0.05	&	305.11	&	0.05	&	-40.1	&	116.15	&	66.02	&	1.48	&	1.64	&	3.88	\\
104	&	G305.34+0.34	&	305.34	&	0.34	&	-35	&	64.35	&	65.96	&	1.15	&	1.74	&	3.83	\\
105	&	G305.71-0.03	&	305.71	&	-0.03	&	-38.7	&	56.9	&	58.96	&	1.05	&	1.24	&	3.83	\\
\hline
\end{tabular}
 \end{table*}

  \setcounter{table}{3}
\begin{table*}
 \caption[CO emission physical properties]{Physical properties of the156 $^{13}$CO clumps detected within G305.}
\centering
  \begin{tabular}{cccccccccc}
   \hline 
	Clump & Clump   & $\tau$ & $T_{\rm ex}$ & $N_{\rm H_{2}}$ & $n_{\rm H_2}$& $\sum_{\rm c}$ & $M$ & $M_{\rm vir}$ & $\alpha_{\rm vir}$ \\
	No.  & Name  & (\COII)    &  (K)     &  $10^{22}$(cm$^{-2}$)  &$10^3$(\cmthree)&$10^2$(\msun\,pc$^{-2}$) & $10^3$(\msun) & $10^3$(\msun)&  \\        
\hline
1	&	G305.24+0.26	&	    0.38 $\pm$     0.02	&	   16.93 $\pm$     0.13	&	    4.04 $\pm$     0.19	&	     3.6 $\pm$      1.7	&	     9.1 $\pm$      2.9	&	    23.0 $\pm$      1.1	&	    19.0 $\pm$      9.5	&	    0.83 $\pm$     0.41	\\
2	&	G304.94+0.55	&	    0.32 $\pm$     0.01	&	   24.57 $\pm$     0.09	&	    2.30 $\pm$     0.09	&	     3.0 $\pm$      1.4	&	     5.2 $\pm$      1.6	&	     6.3 $\pm$      0.2	&	     1.6 $\pm$      0.8	&	    0.25 $\pm$     0.12	\\
3	&	G305.56+0.02	&	    0.30 $\pm$     0.01	&	   17.90 $\pm$     0.13	&	    1.99 $\pm$     0.09	&	     2.6 $\pm$      1.2	&	     4.5 $\pm$      1.4	&	     5.5 $\pm$      0.3	&	     3.8 $\pm$      1.9	&	    0.70 $\pm$     0.35	\\
4	&	G305.26-0.03	&	    0.50 $\pm$     0.02	&	   16.70 $\pm$     0.13	&	    3.50 $\pm$     0.15	&	     3.0 $\pm$      1.4	&	     7.8 $\pm$      2.5	&	    22.0 $\pm$      1.0	&	    11.0 $\pm$      5.7	&	    0.51 $\pm$     0.26	\\
5	&	G304.93+0.54	&	    0.43 $\pm$     0.01	&	   20.26 $\pm$     0.11	&	    2.11 $\pm$     0.08	&	     2.5 $\pm$      1.2	&	     4.7 $\pm$      1.5	&	     6.6 $\pm$      0.2	&	     2.1 $\pm$      1.1	&	    0.32 $\pm$     0.16	\\
6	&	G305.37+0.21	&	    0.39 $\pm$     0.01	&	   18.16 $\pm$     0.12	&	    2.41 $\pm$     0.09	&	     3.3 $\pm$      1.5	&	     5.4 $\pm$      1.7	&	     6.0 $\pm$      0.2	&	     6.7 $\pm$      3.4	&	    1.13 $\pm$     0.57	\\
\end{tabular}
 \end{table*}

 \newpage

  \setcounter{table}{3}
\begin{table*}
\centering
  \begin{tabular}{cccccccccc}
   \hline 
	Clump & Clump   & $\tau$ & $T_{\rm ex}$ & $N_{\rm H_{2}}$ & $n_{\rm H_2}$& $\sum_{\rm c}$ & $M$ & $M_{\rm vir}$ & $\alpha_{\rm vir}$ \\
	No.  & Name  & (\COII)    &  (K)     &  $10^{22}$(cm$^{-2}$)  &$10^3$(\cmthree)&$10^2$(\msun\,pc$^{-2}$) & $10^3$(\msun) & $10^3$(\msun)&  \\        
\hline
41	&	G305.34+0.22	&	    0.31 $\pm$     0.02	&	   13.12 $\pm$     0.16	&	    1.16 $\pm$     0.09	&	     1.5 $\pm$      0.7	&	     2.6 $\pm$      0.8	&	     3.2 $\pm$      0.2	&	     5.0 $\pm$      2.5	&	    1.56 $\pm$     0.79	\\
42	&	G305.77-0.08	&	    0.58 $\pm$     0.03	&	   12.29 $\pm$     0.16	&	    1.24 $\pm$     0.08	&	     1.7 $\pm$      0.8	&	     2.8 $\pm$      0.9	&	     3.0 $\pm$      0.2	&	     2.5 $\pm$      1.3	&	    0.83 $\pm$     0.42	\\
43	&	G305.79-0.09	&	    0.61 $\pm$     0.03	&	   12.10 $\pm$     0.16	&	    1.40 $\pm$     0.09	&	     2.0 $\pm$      1.0	&	     3.1 $\pm$      1.0	&	     3.1 $\pm$      0.2	&	     4.1 $\pm$      2.1	&	    1.34 $\pm$     0.68	\\
44	&	G305.82-0.1	&	    0.54 $\pm$     0.03	&	   11.71 $\pm$     0.16	&	    1.29 $\pm$     0.10	&	     1.3 $\pm$      0.6	&	     2.9 $\pm$      0.9	&	     5.8 $\pm$      0.4	&	     8.2 $\pm$      4.1	&	    1.40 $\pm$     0.71	\\
45	&	G305.55-0.06	&	    0.36 $\pm$     0.02	&	   12.58 $\pm$     0.16	&	    0.76 $\pm$     0.06	&	     1.1 $\pm$      0.5	&	     1.7 $\pm$      0.6	&	     1.6 $\pm$      0.1	&	     2.5 $\pm$      1.3	&	    1.59 $\pm$     0.81	\\
46	&	G305.16+0.25	&	    0.30 $\pm$     0.01	&	   18.91 $\pm$     0.10	&	    1.47 $\pm$     0.08	&	     2.9 $\pm$      1.4	&	     3.3 $\pm$      1.1	&	     1.7 $\pm$      0.1	&	     1.1 $\pm$      0.5	&	    0.63 $\pm$     0.32	\\
\end{tabular}
 \end{table*}	

  \setcounter{table}{3}
\begin{table*}
\centering
  \begin{tabular}{cccccccccc}
   \hline 
	Clump & Clump   & $\tau$ & $T_{\rm ex}$ & $N_{\rm H_{2}}$ & $n_{\rm H_2}$& $\sum_{\rm c}$ & $M$ & $M_{\rm vir}$ & $\alpha_{\rm vir}$ \\
	No.  & Name  & (\COII)    &  (K)     &  $10^{22}$(cm$^{-2}$)  &$10^3$(\cmthree)&$10^2$(\msun\,pc$^{-2}$) & $10^3$(\msun) & $10^3$(\msun)&  \\        
\hline
81	&	G305.5-0.05	&	    0.61 $\pm$     0.04	&	   10.81 $\pm$     0.18	&	    0.72 $\pm$     0.05	&	     1.3 $\pm$      0.6	&	     1.6 $\pm$      0.5	&	     0.9 $\pm$      0.1	&	     0.6 $\pm$      0.3	&	    0.67 $\pm$     0.34	\\
82	&	G305.3+0.1	&	    0.32 $\pm$     0.03	&	   11.54 $\pm$     0.18	&	    0.44 $\pm$     0.04	&	     0.7 $\pm$      0.3	&	     1.0 $\pm$      0.3	&	     0.9 $\pm$      0.1	&	     1.1 $\pm$      0.6	&	    1.30 $\pm$     0.67	\\
83	&	G305.15+0.2	&	    0.28 $\pm$     0.02	&	   13.42 $\pm$     0.15	&	    0.86 $\pm$     0.07	&	     1.8 $\pm$      0.9	&	     1.9 $\pm$      0.6	&	     0.9 $\pm$      0.1	&	     1.3 $\pm$      0.6	&	    1.43 $\pm$     0.72	\\
84	&	G305.65-0.02	&	    0.37 $\pm$     0.03	&	   11.67 $\pm$     0.17	&	    0.78 $\pm$     0.07	&	     1.0 $\pm$      0.5	&	     1.7 $\pm$      0.6	&	     2.0 $\pm$      0.2	&	     2.7 $\pm$      1.3	&	    1.35 $\pm$     0.69	\\
85	&	G305.52-0.02	&	    0.31 $\pm$     0.02	&	   13.28 $\pm$     0.15	&	    0.58 $\pm$     0.04	&	     1.3 $\pm$      0.6	&	     1.3 $\pm$      0.4	&	     0.5 $\pm$      0.1	&	     0.7 $\pm$      0.3	&	    1.33 $\pm$     0.67	\\
86	&	G305.86-0.14	&	    0.43 $\pm$     0.03	&	   10.42 $\pm$     0.19	&	    0.40 $\pm$     0.03	&	     1.1 $\pm$      0.5	&	     0.9 $\pm$      0.3	&	     0.3 $\pm$      0.1	&	     0.6 $\pm$      0.3	&	    2.44 $\pm$     1.24	\\
\end{tabular}
 \end{table*}	

  \setcounter{table}{3}
\begin{table*}
\centering
  \begin{tabular}{cccccccccc}
   \hline 
	Clump & Clump   & $\tau$ & $T_{\rm ex}$ & $N_{\rm H_{2}}$ & $n_{\rm H_2}$& $\sum_{\rm c}$ & $M$ & $M_{\rm vir}$ & $\alpha_{\rm vir}$ \\
	No.  & Name  & (\COII)    &  (K)     &  $10^{22}$(cm$^{-2}$)  &$10^3$(\cmthree)&$10^2$(\msun\,pc$^{-2}$) & $10^3$(\msun) & $10^3$(\msun)&  \\        
\hline
121	&	G305.21+0.06	&	    0.34 $\pm$     0.03	&	   10.60 $\pm$     0.18	&	    0.23 $\pm$     0.02	&	     0.5 $\pm$      0.2	&	     0.5 $\pm$      0.2	&	     0.2 $\pm$      0.1	&	     0.3 $\pm$      0.2	&	    1.47 $\pm$     0.75	\\
122	&	G305.46+0.22	&	    0.38 $\pm$     0.03	&	   10.06 $\pm$     0.19	&	    0.24 $\pm$     0.02	&	     0.6 $\pm$      0.3	&	     0.5 $\pm$      0.2	&	     0.2 $\pm$      0.1	&	     0.2 $\pm$      0.1	&	    1.28 $\pm$     0.66	\\
123	&	G305.34+0.05	&	    0.35 $\pm$     0.04	&	    9.77 $\pm$     0.20	&	    0.34 $\pm$     0.04	&	     0.7 $\pm$      0.4	&	     0.8 $\pm$      0.3	&	     0.3 $\pm$      0.1	&	     0.6 $\pm$      0.3	&	    1.82 $\pm$     0.94	\\
124	&	G305.21+0.07	&	    0.31 $\pm$     0.03	&	   10.02 $\pm$     0.19	&	    0.24 $\pm$     0.03	&	     0.6 $\pm$      0.3	&	     0.5 $\pm$      0.2	&	     0.2 $\pm$      0.1	&	     0.3 $\pm$      0.2	&	    2.02 $\pm$     1.04	\\
125	&	G305.17+0.06	&	    0.34 $\pm$     0.03	&	    9.41 $\pm$     0.21	&	    0.22 $\pm$     0.02	&	     0.3 $\pm$      0.2	&	     0.5 $\pm$      0.2	&	     0.5 $\pm$      0.1	&	     0.6 $\pm$      0.3	&	    1.30 $\pm$     0.67	\\
126	&	G305.2+0.22	&	    0.19 $\pm$     0.02	&	   13.02 $\pm$     0.15	&	    0.28 $\pm$     0.03	&	     0.6 $\pm$      0.3	&	     0.6 $\pm$      0.2	&	     0.2 $\pm$      0.1	&	     0.4 $\pm$      0.2	&	    1.78 $\pm$     0.91	\\
\hline
\end{tabular}
 \end{table*}

\label{lastpage}

\end{document}